\documentclass[preprint,12pt]{aastex}
\usepackage{amsmath}
\usepackage{amssymb}
\usepackage{mathrsfs}

\newcommand{\ei}{\ensuremath{{\varepsilon_i}}}
\newcommand{\et}{\ensuremath{{\varepsilon_t}}}

\newcommand{\trsh}{\ensuremath{\tilde{R}_{\rm sh}}}
\newcommand{\tre}{\ensuremath{\tilde{R}_{\rm e}}}
\newcommand{\hrt}{\ensuremath{\hat{R}_t}}
\newcommand{\hrz}{\ensuremath{\hat{R}_0}}
\newcommand{\tgg}{\ensuremath{\tau_{\gamma\gamma}}}

\title{Opacity Build-up in Impulsive Relativistic Sources}

\author{Jonathan Granot\altaffilmark{1,2,\ddagger}, 
Johann Cohen-Tanugi\altaffilmark{1,\ddagger}, 
and Eduardo do Couto e Silva\altaffilmark{1}}

\altaffiltext{1}{Kavli Institute for Particle Astrophysics and Cosmology, 
Stanford Linear Accelerator Center, 
Stanford University, P.O. Box 20450, MS 29, Stanford, CA 94309}
\altaffiltext{2}{Centre for Astrophysics Research, University of
  Hertfordshire, College Lane, Hatfield, Herts, AL10 9AB, UK}
\altaffiltext{$\ddagger$}{The first two authors contributed equally to this work. 
Send reprint requests to granot@slac.stanford.edu, cohen@slac.stanford.edu}

\begin{document}

\begin{abstract}

Opacity effects in relativistic sources of high-energy gamma-rays,
such as gamma-ray bursts (GRBs) or Blazars, can probe the Lorentz
factor of the outflow as well as the distance of the emission site
from the source, and thus help constrain the composition of the
outflow (protons, pairs, magnetic field) and the emission
mechanism. Most previous works consider the opacity in steady
state. Here we study the effects of the time dependence of the opacity
to pair production ($\gamma\gamma \to e^+e^-$) in an impulsive
relativistic source, which may be relevant for the prompt gamma-ray
emission in GRBs or flares in Blazars. We present a simple, yet rich,
semi-analytic model for the time and energy dependence of the optical
depth, $\tau_{\gamma\gamma}$, in which a thin spherical shell expands
ultra-relativistically and emits isotropically in its own rest frame
over a finite range of radii, $R_0 \leq R \leq R_0 + \Delta R$. This
is particularly relevant for GRB internal shocks. We find that in an
impulsive source ($\Delta R \lesssim R_0$), while the instantaneous
spectrum (which is typically hard to measure due to poor photon
statistics) has an exponential cutoff above the photon energy
$\varepsilon_1(T)$ where $\tau_{\gamma\gamma}(\varepsilon_1) = 1$, the
time integrated spectrum (which is easier to measure) has a power-law
high-energy tail above the photon energy $\varepsilon_{1*} \sim
\varepsilon_1(\Delta T)$ where $\Delta T$ is the duration of the
emission episode. Furthermore, photons with energies $\varepsilon >
\varepsilon_{1*}$ are expected to arrive mainly near the onset of the
spike in the light curve or flare, which corresponds to the short
emission episode.  This arises since in such impulsive sources it
takes time to build-up the (target) photon field, and thus the optical
depth $\tau_{\gamma\gamma}(\varepsilon)$ initially increases with time
and $\varepsilon_1(T)$ correspondingly decreases with time, so that
photons of energy $\varepsilon > \varepsilon_{1*}$ are able to escape
the source mainly very early on while $\varepsilon_1(T) >
\varepsilon$.  As the source approaches a quasi-steady state ($\Delta
R \gg R_0$), the time integrated spectrum develops an exponential
cutoff, while the power-law tail becomes increasingly suppressed.

\end{abstract}

\keywords{gamma rays: theory --- relativity ---
  methods: analytical --- gamma rays: bursts --- galaxies: jets}

\maketitle

\section{Introduction and motivation}\label{sec:intro}

Astrophysical sources of gamma-rays that are both compact and very
luminous may be optically thick to pair production ($\gamma\gamma \to
e^+e^-$) within the source. The corresponding optical depth, $\tgg$,
is usually an increasing function of the photon energy, and therefore
a large optical depth would prevent the escape of high-energy photons
from the source, causing a high-energy cutoff in the observed
spectrum. For sufficiently high optical depths, enough $e^+e^-$ pairs
may be produced, so that the optical depth of all photons (even low
energy photons that are optically thin to pair production) to
scattering on these electrons/positrons would be much larger than
unity, in which case the photon energy spectrum would be
thermalized. The size of the gamma-ray emitting region is usually hard
to constrain directly from observations, since the angular resolution
of gamma-ray telescopes is much poorer than their counterparts in
lower energy photons (e.g. X-rays, optical, or radio). Nevertheless,
the physical properties of the emitting region can be constrained
using compactness arguments, and the observed properties of the
source. In particular, rapid flux variability of the source is often
used in order to set upper limits on the size of the emitting region,
making highly variable sources with significant non-thermal
high-energy emission a prime target for such analysis. One of the best
examples for such sources are gamma-ray bursts (GRBs), and we shall
focus on them below, although most of our analysis has a much broader
range of applicability \citep[similar opacity considerations have also
been used to constrain the properties of other sources, such as
Blazars, e.g.][]{Sikora94}.

It has been realized early on that, in GRBs, pair production within
the source is expected to cause a high-energy cutoff in the observed
photon energy spectrum \citep[see][and references therein]{Piran05}.
Naively, if the source shows significant flux variability on an
observed time scale of $\Delta T$, its size is inferred to be $R
\lesssim c\Delta T/(1+z)$ where $z$ is its cosmological redshift, and
the optical depth to pair production at a dimensionless photon energy
$\varepsilon \equiv E_{\rm ph}/m_e c^2$ is $\tgg(\varepsilon) \sim
\sigma_T L_{1/\varepsilon(1+z)} / 4\pi m_e c^3 R \gtrsim \sigma_T
L_{1/\varepsilon(1+z)}(1+z) / 4\pi m_e c^4\Delta T \sim
10^{14}(1+z)[L_{1/\epsilon(1+z)}/(10^{51}\;{\rm erg\;s^{-1}})][\Delta
T/(1\;{\rm ms})]^{-1}$, where $L_\varepsilon = F_{\varepsilon/(1+z)}
4\pi d_L^2(1+z)^{-1}$ and $F_\varepsilon$ are the source isotropic
equivalent luminosity and observed flux per unit dimensionless photon
energy, and $d_L$ is the luminosity distance
to the source. For GRBs the (observed part of the) $\varepsilon
F_\varepsilon$ spectrum typically peaks around $\varepsilon \sim 1$,
and being at cosmological distances their isotropic equivalent
luminosity is typically in the range of $10^{50}-10^{53}\;{\rm
erg\;s^{-1}}$. Furthermore, they often show significant variability
down to millisecond timescales. This implies huge values of $\tgg$, as
high as $\sim 10^{15}$, under the above naive assumptions. Such huge
optical depths are clearly inconsistent with the non-thermal GRB
spectrum, which has a significant power law high-energy tail. This is
known as the compactness problem \citep{Ruderman75}.

If the source is moving relativistically toward us with a Lorentz
factor $\Gamma \gg 1$, then in its own rest frame the photons have
smaller energies, $\varepsilon' \sim \varepsilon(1+z)/\Gamma$, while
in the lab frame (i.e. the rest frame of the central source) most of
the photons propagate at angles $\lesssim 1/\Gamma$ relative to its
direction of motion. The latter implies that in the lab frame the
typical angle between the directions of the interacting photons is
$\theta_{12} \sim 1/\Gamma$, which has two effects. First, it
increases the threshold for pair production,
$(1+z)^2\varepsilon_1\varepsilon_2>2/(1-\cos\theta_{12})$, to
$(1+z)^2\varepsilon_1\varepsilon_2 \gtrsim \Gamma^2$ (compared to
$\varepsilon'_1\varepsilon'_2 \gtrsim 1$ for the roughly isotropic
distribution of angles between the directions of the interacting
photons in the rest frame of the source, where $\theta'_{12}\sim
1$). This reduces $\tgg(\varepsilon)$ by a factor of
$\Gamma^{2(1-\alpha)}$ where $L_\varepsilon \approx L_0
\varepsilon^{1-\alpha}$ at high photon energies (corresponding to
$dN_{\rm ph}/d\varepsilon \propto \varepsilon^{-\alpha}$,
i.e. $\alpha$ is the high-energy photon index), since
$L_{1/\varepsilon(1+z)}$ needs to be replaced by
$L_{\Gamma^2/\varepsilon(1+z)} =
\Gamma^{2(1-\alpha)}L_{1/\varepsilon(1+z)}$. Second, the expression
for the optical depth includes a factor of $1-\cos\theta_{12}$ (that
represents the rate at which photons pass each other and have an
opportunity to interact) which for a stationary source is $\sim 1$,
but for a relativistic source moving toward us is $\sim \Gamma^{-2}$.
Finally, the size of the emitting region can be as large as $R \sim
\Gamma^2 c\Delta T/(1+z)$, which reduces $\tgg$ by an additional factor of
$\Gamma^{-2}$. altogether, $\tgg(\varepsilon)$ is reduced by a factor
of $\sim \Gamma^{2(\alpha+1)}$, and since typically $\alpha \sim 2-3$
this usually implies $\Gamma \gtrsim 10^2$ in order to have $\tgg < 1$
and overcome the compactness problem. Using similar arguments, the
lack of such a high-energy cutoff due to pair production in the
observed spectrum of the prompt gamma-ray emission in GRBs has been
used to place lower limits on the Lorentz factor of the outflow
\citep{KP91,Fenimore93,WL95,BH97,LS01}.

We note, however, that $\tgg$ generally depends both on the radius of
emission, $R$, and on the bulk Lorentz factor, $\Gamma\;$:
$\;\tgg(\varepsilon) \propto \Gamma^{-2\alpha} R^{-1} L_0
\varepsilon^{\alpha-1}$. Therefore, one needs to assume a relation
between $R$ and $\Gamma$ in order to obtain a lower limit on the
latter. Most works assume $R \sim \Gamma^2 c \Delta T/(1+z)$
\citep[e.g.,][]{LS01}, which gives $\tgg(\varepsilon) \propto
\Gamma^{-2(\alpha+1)}(\Delta T)^{-1} L_0 \varepsilon^{\alpha-1}$, while
the lack of a high-energy cutoff up to some photon energy
$\varepsilon$ implies $\tgg(\varepsilon) < 1$. This, in turn, provides
a lower limit on $\Gamma$ since one can directly measure the
variability time $\Delta T$, the photon index $\alpha$, and $L_0
\approx 4\pi d_L^2(1+z)^{\alpha-2} \varepsilon^{\alpha-1}
F_\varepsilon$. However, the relation $R \sim \Gamma^2 c \Delta
T/(1+z)$ does not hold for all models of the prompt GRB emission. For
example, this relation does not hold if the prompt GRB emission is
generated by relativistic magnetic reconnection events, with angular
scales $\ll 1/\Gamma$, that create local relativistic motion with
 Lorentz factor $\gamma_{rel} \sim 5-10$ relative to the average bulk value $\Gamma$ 
of the emitting shell 
\citep{LB02,LB03}. In this case $\Delta T/(1+z) \ll R/c\Gamma^2$ and
the inferred value of the Lorentz factor from standard opacity
arguments would be $\sim\gamma_{rel}\Gamma$ rather than the bulk Lorentz
factor of the shell, $\Gamma$. This allows the radius of the prompt
emission to be as large as $R \sim 10^{16}-10^{17}\;$cm, close to the
deceleration radius where most of the energy of the outflow is
transferred to the swept-up external medium, and is much larger than
the prompt emission radius that is expected in the internal shocks
model, $R \sim 10^{13}-10^{14}\;$cm. Therefore, we adopt a more
model-independent approach and do not automatically make this
assumption. Instead, we derive most of our formulas without this
assumption, as well as derive expressions for $\Gamma$ under this
assumption, which could serve in order to test its validity.

The Gamma-ray Large Area Space Telescope (GLAST) mission
\citep{glast}, to be launched in early 2008, is expected to shed light
on the high-energy emission from GRBs and other impulsive relativistic
sources. In particular, opacity effects due to the local photon field
within the source\footnote{In the present work, we will not consider
opacity effects due to interaction of high energy photons with the
extra-galactic background light. Such an attenuation, interesting in
its own right, can be added to the ``in source'' opacity in a
straightforward way. Furthermore, it is expected to become significant
(i.e. produce $\tgg > 1$) only at cosmological redshifts ($z \gtrsim
1$) and for very high photon energies \citep[$\gtrsim 56-100\;$GeV at
$z = 1$ and $\gtrsim 18-63\;$GeV at $z = 3$;][]{Kneiske04}, and is
therefore likely to significantly affect only the high end of the
GLAST energy range, where the photon statistic might
be too poor to reliably measure this effect. This source of opacity
will be independent of time (and depends only on the redshift of the
source, and on the photon energy), which would help in disentangling
it from the time dependent opacity intrinsic to the source that we
calculate in this work.}  are expected to be most relevant in the
GLAST Large Area Telescope (LAT) energy range \citep[20 MeV to more
than $300\;$GeV, see][]{glast-lat}. Thus, it represents a powerful
tool for probing the physics of these sources. GLAST is likely to
detect the high-energy cutoff due to pair production opacity which
would actually determine $\Gamma^{2\alpha}R$, rather than just provide
a lower limit for it. Furthermore, in GRBs, the outflow Lorentz factor
$\Gamma$ may be constrained by the time of the afterglow onset
\citep{PK02,LR-RG05,Molinari07}, provided that the reverse shock is
not highly relativistic, so that if GLAST detects the high-energy pair
production opacity cutoff, the radius of emission $R$ could be
directly constrained, thus helping to test the different GRB
models. In particular, this could directly test whether the relation
$R \sim \Gamma^2 c\Delta T/(1+z)$ that is expected in many models indeed
holds, since both $R$ and $\Gamma$ could be determined
separately. This, however, requires a reliable way of identifying the
observed signatures of opacity to pair production. This is one of the
main motivations for this work.

The leading model for the prompt emission in GRBs features internal
shocks \citep{RM94} due to collisions between shells that are ejected
from the source at ultra-relativistic speeds ($\Gamma \gtrsim 100$).
The shells are typically quasi-spherical, i.e. their properties do not
vary a lot over angles $\lesssim {\rm a\ few}\ \Gamma^{-1}$ around
our line of sight. Under the typical physical conditions that are
expected in the shocked shells, all electrons cool on a time scale
much shorter than the dynamical time (i.e. the time it takes the shock
to cross the shell), and most of the radiation is emitted within a
very thin cooling layer just behind the shock front.  Thus, our model
which features an emitting spherical thin shell that expands outward
ultra-relativistically is appropriate for the internal shocks model.

As this emitting ``shell'' expands outward to larger radii, it
builds up a photon field that can pair produce with high-energy
photons from the same emission component. This effect has been studied
in the past \citep[see especially][and references therein]{Baring06},
but the temporal and spatial dependences of the photon field have been
averaged out, corresponding either explicitly or implicitly to a
quasi-steady state. However, in impulsive relativistic sources, the
time scale for significant variations in the properties of the
radiation field within the source is comparable to the total duration
of the emission episode, and therefore the dependence of the opacity
to pair production on space and time cannot be ignored, and may
produce important effects that are suppressed in the steady-state
limit. Therefore, in the present work we consider the full temporal
and spatial dependence of the opacity, in order to capture all the
resulting effects.

We develop a simple, yet rich, model to investigate quantitatively the
intuitive consideration that in impulsive sources it takes time to
build up the (target) photon field, and thus the optical depth
initially increases with time, so that high energy photons might be
able to escape the source mainly at the very early part of the spike
in the light curve. This results in a power law tail for the
time-integrated spectrum at high energies, while the instantaneous
spectrum (which is hard to measure due to poor photon statistics) has
an exponential cutoff. This arises since the photon energy of the
exponential cutoff in the instantaneous spectrum decreases with time,
as the opacity increases with time at all energies.  Therefore, at
sufficiently high photon energies, most of the photons escape during
the short initial time before the optical depth increases above unity,
i.e. before the cutoff energy sweeps past their energy.

We perform detailed semi-analytic calculations of the optical depth to
pair production, which improve on previous works by first calculating
the photon field at each point in space and time, and then integrating
along the trajectory of each photon. The structure of the paper is as
follows. In \S\;\ref{sec:flux} we introduce our model and derive a
general expression for the flux that reaches an observer at
infinity. This expression includes the optical depth along the
trajectory of each photon that may reach the observer, which is
derived in \S\;\ref{sec:tau}. The calculation of the optical depth
requires the knowledge of the photon field at each point along the
trajectory of each (test) photon. This local photon field is first
expressed in terms of the source emissivity
(\S\;\ref{subsec:emissivity}). Next (\S\;\ref{sec:red}) it is
conveniently rewritten as the product of the typical optical depth
(that is approached on a dynamical time, and is similar to that
derived in previous works) and dimensionless order unity expression
(containing a few integrals) which captures the new time dependent
effects that are the focus of this work. In \S\;\ref{sec:gfield}
explicit expressions are derived for the integrands of these
dimensionless order unity integrals. In \S\;\ref{sec:analytic} we
derive the relevant analytic scalings for the resulting optical depths
and observed flux, and in \S\;\ref{sec:res} we present numerical
results (i.e. numerically evaluate the semi-analytic expressions) for
the opacity, light curves, and spectra (both the instantaneous and
time-integrated spectra are addressed in \S\S\;\ref{sec:analytic} and
\ref{sec:res}).  Our conclusions are discussed in \S\;\ref{sec:dis}.

\section{Calculating the Observed Flux}\label{sec:flux}

\subsection{Model Assumptions}\label{sec:model}

We consider an ultra-relativistic (with Lorentz factor $\Gamma\gg 1$),
thin (of width $\ll R/\Gamma^2$ in the lab frame) spherical
expanding shell, that emits over a finite range of radii, $R_0 \leq R
\leq R_0 + \Delta R$ (i.e. the emission turns on at $R_0$ and turns
off at $R_0 + \Delta R$). This model can be associated with a single
pulse or flare in the light curve. In the context of internal shocks
within the outflow, $\Delta R \sim R_0$ is typically
expected~\citep[][and references therein]{RM94,Piran05}.

The emission is assumed to be isotropic in the co-moving frame of the
emitting shell (i.e. the shell rest frame), and uniform over the
spherical shell. In this work primed quantities are always measured in
the co-moving frame, while unprimed quantities are evaluated either in
the lab frame, that is the rest frame of the central source, in which
the shell is spherical (e.g. the Lorentz factor $\Gamma$), or in the
observer frame (e.g. the observed time and photon energy which suffer
cosmological time dilation and redshift, respectively, relative to the
lab frame which is at the cosmological redshift of the source). The
observer is assumed to be located at a distance from the source that
is much larger than the source size (so that the angle subtended by
the source, as seen by the observer, is very small, and the observer
can be considered as being at ``infinity'').

For convenience, we will use dimensionless photon energies,
$\varepsilon$, in which the observed photon energy, $E_{\rm ph}$, is
normalized by the electron rest energy: $\varepsilon \equiv E_{\rm
ph}/m_e c^2$.  While general expressions will be provided when
possible, we also provide detailed semi-analytical solutions to the
model by assuming that the luminosity in the shell rest frame has a
power-law dependence on rest frame photon energy $\varepsilon'$ and
radius $R$, $L'_{\varepsilon'} \propto (\varepsilon')^{1-\alpha}R^b$,
and that the Lorentz factor scales as a power law with radius,
$\Gamma^2\propto R^{-m}$. The approximation that $\Gamma$ and
$L'_{\varepsilon'}$ scale as power laws with radius is usually
expected to hold reasonably well.  For internal shocks, the colliding
shells are expected to be in the coasting stage near the collision
radius ($R_0$), which corresponds to $m=0$~\citep[see][and references
therein]{Piran05,Meszaros06}.  In the GRB afterglow, both before and
after the deceleration radius, where most of the energy is transferred
from the ejecta to the shocked external medium, $\Gamma$ \citep{BM76}
and $L'_{\varepsilon'}$ \citep[e.g.,][]{Sari98,Granot05} are expected
to scale as power laws with radius. For GRB internal shocks, the
scaling of $L'_{\varepsilon'}$ with radius $R$ generally depends on
the details of the colliding shells.

For uniform colliding shells, where the strength of the shocks going
into the shells is constant with radius, above the peak of the $\nu
F_\nu$ spectrum, $\varepsilon'_{\rm peak}$, one expects $-0.5 \lesssim
b \lesssim 0$. This may be understood as follows. In this case the
Lorentz factor in the shocked regions of the colliding shells is
constant with radius, while the magnetic field scales as $B' \propto
R^{-1}$. Therefore, since the number of emitting electrons scales
linearly with radius, $N_e \propto R$, then $L'_{\varepsilon,{\rm
max}} \propto B' N_e \propto R^{0}$. The typical synchrotron photon
energy scales as $\epsilon'_m \propto B'\gamma_m^2 \propto R^{-1}$
since the typical Lorentz factor of the electrons, $\gamma_m$, is
constant for a constant shock strength. The energy of a photon that
cools on the dynamical time (the time since the start of the
collision) scales as $\epsilon'_c \propto R$. Therefore, above the
peak of the $\nu F_\nu$ spectrum, at $\varepsilon' > \varepsilon'_{\rm
peak} = \max(\varepsilon'_c,\varepsilon'_m)$, we have
$L'_{\varepsilon'} = L'_{\varepsilon,{\rm
max}}(\varepsilon'_m/\varepsilon'_c)^{-1/2}(\varepsilon'/\varepsilon'_m)^{-p/2}
\propto R^{(2-p)/2}$, where $p$ is the power law index of the electron
distribution, $dN_e/d\gamma_e \propto \gamma_e^{-p}$ for $\gamma_e >
\gamma_m$. Since $p \sim 2-3$ is typically inferred for the GRB prompt
emission, this corresponds to $-0.5 \lesssim b \lesssim 0$. For fast
cooling ($\varepsilon'_c < \varepsilon'_m$) below $\varepsilon'_{\rm
peak} = \varepsilon'_m$, $L'_{\varepsilon'} = L'_{\varepsilon,{\rm
max}}(\varepsilon'/\varepsilon'_c)^{-1/2} \propto R^{1/2}$. For slow
cooling ($\varepsilon'_c > \varepsilon'_m$), however, below
$\varepsilon'_{\rm peak} = \varepsilon'_c$, $L'_{\varepsilon,{\rm
max}}(\varepsilon'/\varepsilon'_m)^{(1-p)/2} \propto R^{(1-p)/2}$.

The simplifying assumption of a power law emission spectrum
[$L'_{\varepsilon'} \propto (\varepsilon')^{1-\alpha}$], however, is
not always valid~\citep[see, e.g.,][]{Baring06}. For example, in GRB
internal shocks it breaks down for photons of energy $\varepsilon
\gtrsim \Gamma^2/(1+z)^2\varepsilon_{\rm peak}$,
i.e. $\varepsilon\,m_e c^2 \gtrsim 25 (1+z)^{-2} (\Gamma/100)^2
(\varepsilon_{\rm peak}m_ec^2/100\;{\rm keV})^{-1}\;$GeV.  Indeed,
photons of such energy interact with photons below the spectral break
energy $\varepsilon_{\rm break}$ which is the peak of the $\nu F_\nu$
spectrum. A detailed treatment of the case of a more realistic
spectrum for GRB internal shocks will be provided elsewhere.  The
exact shape of the spectrum at high energies is not well
constrained. Thus, we use a fiducial value of $\alpha = 2$, which
corresponds to a flat $\nu F_\nu$ (i.e. equal energy per decade in
photon energy), in our detailed illustrative solutions, and also
explore the effects of varying the value of $\alpha$.

\subsection{The Equal Arrival Time Surface of Photons to the Observer (EATS-I)}
\label{sec:flux2}
The observed normalized flux density, $F_\varepsilon =
(m_ec^2/h)F_\nu$, is calculated as a function of time and photon
energy, closely following the derivation of \cite{Granot05}. For this
purpose, the contributions to the observed flux at any given observed
time $T$ are integrated over the ``equal arrival time surface''
(EATS-I) -- the locus of points from which photons that are emitted at
the shell reach the observer simultaneously, at the observed time
$T$. In the present work, the effects of opacity to pair production
will be added at the end of this calculation, as detailed below.

\begin{figure}[ht]
\centering
\includegraphics[scale = 0.89]{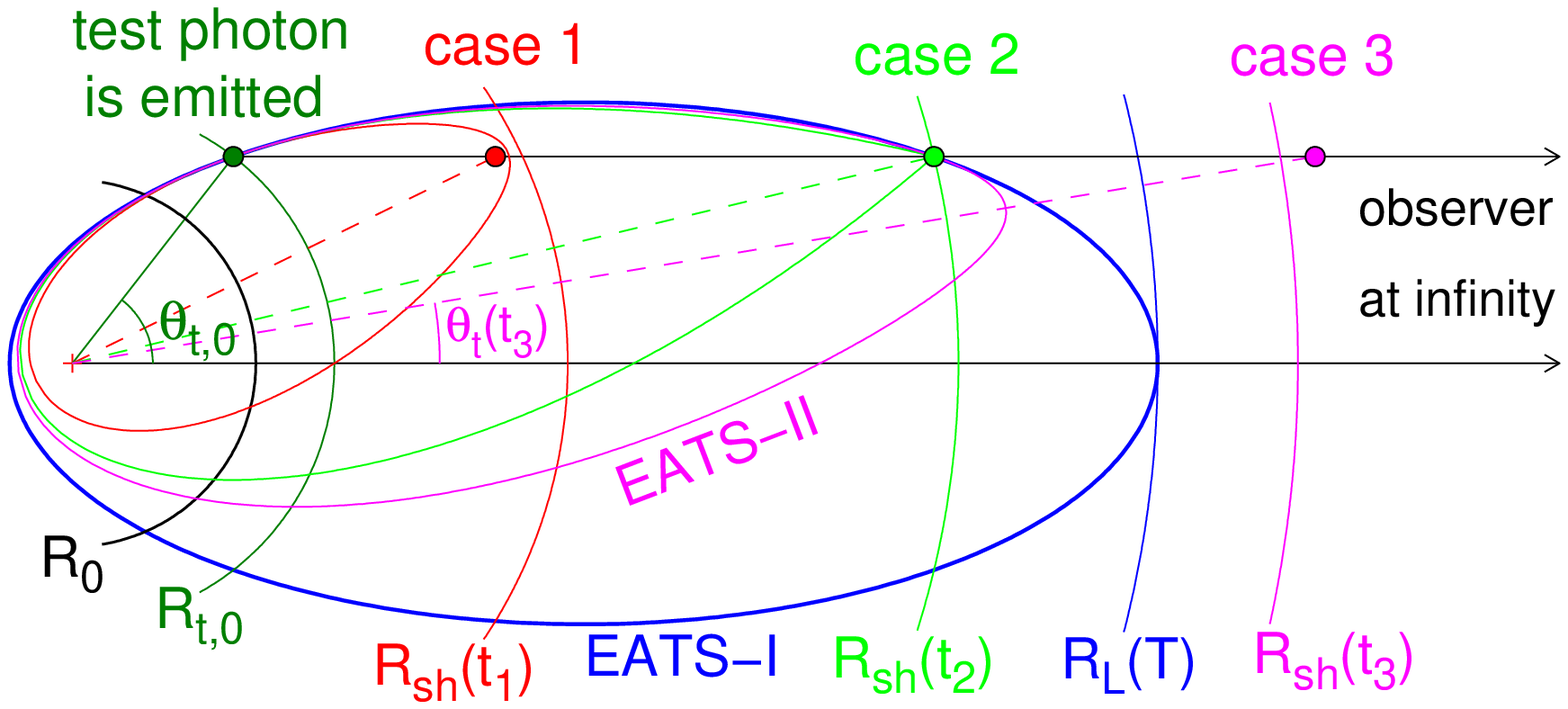}
\\
\vspace{1.4cm}
\includegraphics[scale = 0.89]{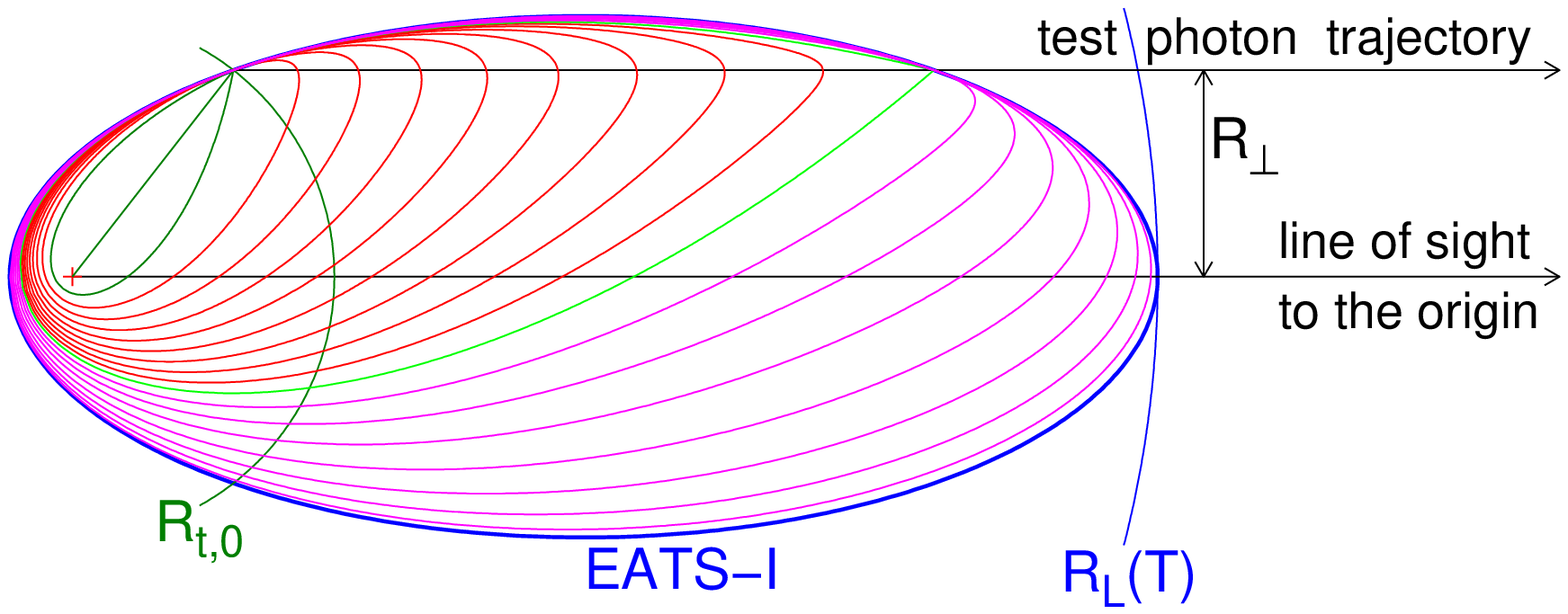}
\caption{An illustration of the two different equal arrival time
  surfaces (EATS) of photons: 1. to the observer at infinity (EATS-I,
  {\it in blue}), and 2. to the instantaneous location of a test
  photon (EATS-II, in different colors). The overall geometry as well
  as relevant radii and angles are shown in the {\it upper panel},
  along with an illustration of the three different cases that are
  discussed in the text, in which the test photon either lags behind
  the shell (case 1), coincides with the shell (case 2), or is in
  front of the shell (case 3). The {\it lower panel} shows the
  sequence of EATS-II, whose size increases with time, nested within
  the EATS-II which correspond to a larger time, and in particular
  within EATS-I which corresponds to an infinite time (i.e. an
  infinite radius for the test photon, when it reaches the observer at
  infinity).}
\label{fig:EATS12}
\end{figure}

We consider a photon initially emitted by the shell at a lab frame
time $t_0$ when the radius of the shell is $R_{t,0} \equiv R_{\rm
sh}(t_0)$ and its Lorentz factor is $\Gamma_{t,0}$, at an angle of
$\theta_{t,0}$ from our line of sight to the origin $R = 0$ (see
Fig.~\ref{fig:EATS12}). Due to the spherical symmetry of our model,
there is no dependence on the azimuthal angle. The arrival time $T$ of
the photon to a distant observer is given by the ``equal arrival
time'' formula:
\begin{equation}
\frac{T}{(1+z)} = t_0 - \frac{R_{t,0}}{c}\cos\theta_{t,0}\label{T}\ , 
\end{equation}
where the lab frame time $t$ is related to the shell radius at that
time, $R_{\rm sh}(t)$, by
\begin{equation}\label{t_beta} 
t = \int_0^{R_{\rm sh}(t)}\frac{dR}{\beta c}  ~=~ 
\frac{R_{\rm sh}(t)}{c}  
~-~\frac{1}{2c}\int_0^{R_{\rm sh}(t)}\frac{dR}{\Gamma^2(R)} ~ 
+ ~{{\cal O}(\Gamma^{-4})}\ . 
\end{equation}
In Eq.~(\ref{T}), $T = 0$ is chosen to correspond to a photon that is
emitted at the origin at $t_0 = 0$. Eq.~(\ref{t_beta}) relates $t$ and
$R_{\rm sh}(t)$, so that the locus of points $(R_{t,0},\theta_{t,0})$
that keep $T$ constant defines the EATS-I at time $T$.  For a coasting
shell ($m = 0$), it is a well-known result that the EATS-I is an
ellipse\footnote{It actually represents an ellipsoid, keeping in mind
the symmetry around the line of sight to the center of the emitting
spherical shell, and the lack of dependence on the azimuthal angle.}
of semi-major to semi-minor axis ratio $\Gamma$~\citep{Rees66}. The
flux density at the rescaled energy $\varepsilon$ is obtained by
integrating over the luminosity in the shell rest frame,
$L'_{\varepsilon'}$, along the EATS-I \citep{Granot05}:
\begin{equation}\label{F1}
F_\varepsilon(T) = \frac{(1+z)}{4\pi d_L^2}\int\delta^3 dL'_{\varepsilon'}
= \frac{(1+z)}{8\pi d_L^2}\int_{y_{\rm min}}^{y_{\rm max}} dy
\frac{d\mu_{t,0}}{dy}\delta^3(y)L'_{\varepsilon'}(y)\ ,
\end{equation}
where $\delta \equiv (1+z)\varepsilon/\varepsilon'$ is the Doppler
factor of the emitted photon (between the co-moving and lab frames),
$\mu_{t,0}\equiv\cos\theta_{t,0}$ is the cosine of its angle of
emission, and we defined the normalized radius $y \equiv R_{t,0}/R_L$,
where $R_L = R_L(T)$ is the largest radius on the EATS-I at time
$T$. The integration is performed along the EATS-I, and the boundaries
for $y$ are
\begin{equation}
y_{\rm min}(T) = \min\left[1,\frac{R_0}{R_L(T)}\right] \quad,\quad
y_{\rm max} = \min\left[1,\frac{R_0+\Delta R}{R_L(T)}\right]\ ,
\end{equation}
 since the emission turns
on at $R_0$ and turns off at $R_0 + \Delta R$. For the times $T$ relevant 
to the problem, corresponding to the arrival of photons to the observer, 
$R_0/R_L(T)$ is always smaller than 1.

In order to evaluate the integral above, we now derive expressions for
the integrand. 
Defining $\Gamma_L \equiv \Gamma(R_L)$, $\Gamma^2 \propto R^{-m}$
 can be rewritten as 
$\Gamma^2(R)R^m = \Gamma_L^2 R_L^m = {\rm constant}$, and thus
$\Gamma^2 = \Gamma_L^2\, y^{-m}$.
Eq.~(\ref{t_beta}) now becomes
\begin{equation}\label{t_0}
t_0 = \frac{R_{t,0}}{c} + \frac{R_L\,y^{m+1}}{2(m+1)\Gamma_L^2 c} 
+ {{\cal O}(\Gamma^{-4})} \ .
\end{equation}
In the limit of small angles ($\theta_{t,0} \ll 1$, which is relevant
for $\Gamma \gg 1$), Eq.~(\ref{T}) implies $t_0-R_{t,0}/c = T/(1+z) -
R_{t,0}\theta_{t,0}^2/2c$, which together with Eq.~(\ref{t_0}) yields
\begin{equation}\label{T1}
\frac{T}{(1+z)} = \frac{R_L\,y^{m+1}}{2(m+1)\Gamma_L^2 c} 
+\frac{R_{t,0}\theta_{t,0}^2}{2c}\ .
\end{equation}
As can be seen in Fig.~\ref{fig:EATS12}, a photon that is emitted at
$R_{t,0} = R_L$ (corresponding to $y = R_{t,0}/R_L(T) = 1$) remains
along the line of sight ($\theta_t = \theta_{t,0} = 0$), so that
Eq.~(\ref{T1}) yields
\begin{equation}\label{R_L}
R_L(T) = 2(m+1)\Gamma_L^2[T/(1+z)]\frac{cT}{(1+z)} =
R_0\left(\frac{T}{T_0}\right)^{1/(m+1)} \quad,\quad T_0 =
\frac{(1+z)R_0}{2(m+1)c\Gamma_0^2}\ ,
\end{equation}
where $\Gamma_0 \equiv \Gamma(R_0)$, and can be rewritten as
\begin{equation}\label{theta}
\theta_{t,0}^2 = \frac{y^{-1} - y^m}{(m+1)\Gamma_L^2}\ .
\end{equation}
We have introduced the time $T_0$ at which the first photons reach the
observer (corresponding to a photon emitted at $R_0$ along the line of
sight, $\theta = 0$): $R_L(T_0) \equiv R_0$.
Since $\mu_{t,0} \approx 1-\theta_{t,0}^2/2$, Eq.~(\ref{theta}) implies
\begin{equation}\label{dmu_dy}
\frac{d\mu_{t,0}}{dy} = \frac{y^{-2}+my^{m-1}}{2(m+1)\Gamma_L^2}\ .
\end{equation}
Finally, the Doppler factor of the emitted electron is given by
\begin{equation}\label{doppler}
\delta \equiv \frac{1}{\Gamma(1-\beta\cos\theta_{t,0})}\approx \frac{2\Gamma}{1+(\Gamma\theta_{t,0})^2} = 
\frac{2(m+1)\Gamma_L y^{-m/2}}{m+y^{-m-1}}\ ,
\end{equation}
and its value at $R_L$ (which corresponds to $y = 1$) is $\delta(R_L)
= 2\Gamma_L$. Since
\begin{equation}
L'_{\varepsilon'} = L'_{(1+z)\varepsilon/\delta(R_L)}(R_L)
\left[\frac{\varepsilon'}{\varepsilon'(R_L)}\right]^{1-\alpha}
\left(\frac{R_{t,0}}{R_L}\right)^b\ ,
\end{equation}
where $\varepsilon' = (1+z)\varepsilon/\delta$, we obtain:
\begin{equation}\label{L_int}
L'_{\varepsilon'} = L'_{(1+z)\varepsilon/2\Gamma_L}(R_L)
\left(\frac{\delta}{2\Gamma_L}\right)^{\alpha-1} y^b = 
L'_{(1+z)\varepsilon/2\Gamma_0}(R_0)
\left(\frac{\delta}{2\Gamma_L}\right)^{\alpha-1}
y^b\left(\frac{R_L}{R_0}\right)^{b-m(\alpha-1)/2}\ .
\end{equation}

The effect of pair production opacity will be treated in this work in
a somewhat simplified manner, by assuming that photons which pair
produce do not reach the observer, and ignoring the additional opacity
that is produced by the secondary pairs and the photons emitted by
these pairs. Under these simplifications, the effects of opacity to
pair production can be included by adding a term $\exp(-\tgg)$ into
the integrand in Eq.~(\ref{F1}), where $\tgg$ is a function of
$y,\varepsilon,\Delta R/R_0,\rm{ and }\ T/T_0$. Thus, by combining
eqs.~(\ref{dmu_dy}$\,-\,$\ref{L_int}) with Eq.~(\ref{F1}), we obtain:
\begin{align}\label{F2}\nonumber
F_\varepsilon(T) &=
2\Gamma_L L'_{(1+z)\varepsilon/2\Gamma_L}(R_L)\frac{(1+z)}{4\pi d_L^2}
\int_{y_{\rm min}}^{y_{\rm max}} dy 
\left(\frac{m+1}{m+y^{-m-1}}\right)^{1+\alpha}
y^{b-1-m\alpha/2}\,e^{-\tgg}
\\ \nonumber
&= 2\Gamma_0 L'_{(1+z)\varepsilon/2\Gamma_0}(R_0)\frac{(1+z)}{4\pi d_L^2}
\left(\frac{T}{T_0}\right)^{(2b-m\alpha)/[2(m+1)]}\\
&\quad\times \int_{y_{\rm min}}^{y_{\rm max}} dy 
\left(\frac{m+1}{m+y^{-m-1}}\right)^{1+\alpha}
y^{b-1-m\alpha/2}\,e^{-\tgg}\ ,\quad\quad
\end{align}
where Eq.~(\ref{R_L}) is used to derive the scaling $R_L(T)/R_0 =
 (T/T_0)^{1/(m+1)}$, and
\begin{equation}\label{tau_gen}
\tgg = \tgg\left(y,\varepsilon,\frac{\Delta R}{R_0},\frac{T}{T_0},
\frac{L_0}{\Gamma_0^{2\alpha}R_0}\right)\ ,
\end{equation}
as is shown later on, where $\Gamma_0 \equiv \Gamma(R_0)$, and
 $L_\varepsilon \approx L_0\varepsilon^{1-\alpha}$ is the observed isotropic
 equivalent luminosity. Unless specified otherwise, the derivations
 throughout this work are valid for a general value of $m$.  For a
 coasting shell ($m = 0$), which is a case of special interest (as it
 is expected, e.g., for internal shocks), Eq.~(\ref{F2}) simplifies to
\begin{equation}\label{F2m0}\nonumber
F_\varepsilon(T) = 2\Gamma_0 L'_{(1+z)\varepsilon/2\Gamma_0}(R_0)
\frac{(1+z)}{4\pi d_L^2}\left(\frac{T}{T_0}\right)^b 
\int_{y_{\rm min}}^{y_{\rm max}} dy\, y^{\alpha+b}\,e^{-\tgg}\ .
\end{equation}
We have expressed the observed flux density for our model as a
function of the observed time $T$, and we now need to derive the
expression of the optical depth $\tgg$. We gather here the dependence
on $y$ of two quantities that will be needed later on:
\begin{equation}\label{Rhat_x}
\hat{R}_0 \equiv \frac{R_0}{R_{t,0}} = \frac{y_{\rm min}}{y} = 
\frac{R_0}{\Delta R}\,\frac{\Delta R}{R_{t,0}} = 
\frac{1}{y}\left(\frac{T}{T_0}\right)^{-1/(m+1)} \quad,\quad
x \equiv (\Gamma_{t,0}\theta_{t,0})^2 = \frac{y^{-(m+1)}-1}{(m+1)}\ .
\end{equation}
In order to facilitate reading, we include in
Table~\ref{tab:basic} the most common quantities used 
throughout this work.

\section{Computation of the optical depth}\label{sec:tau}

\begin{figure}[t]
\centering \includegraphics[scale=0.3]{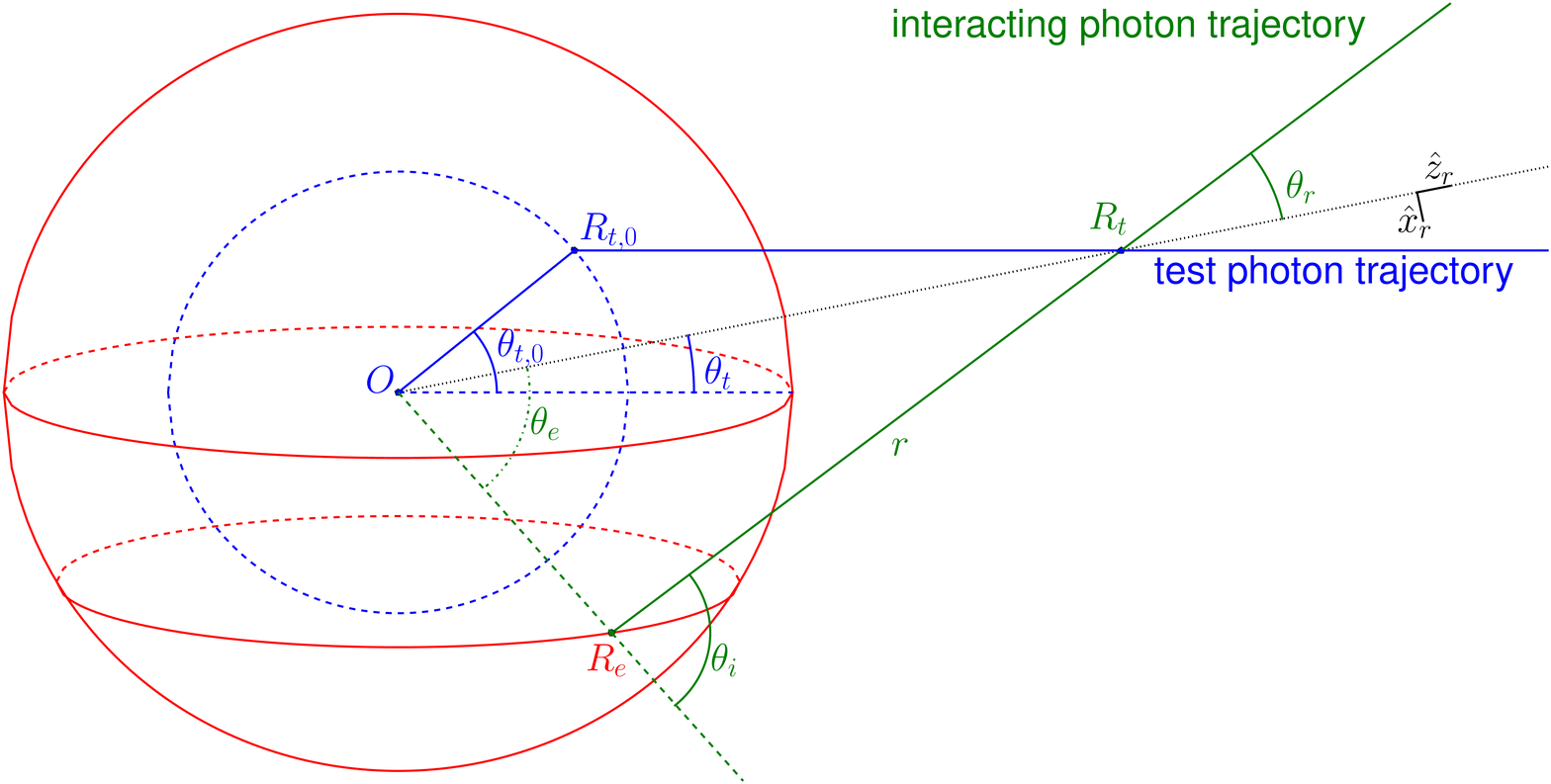}
\caption{Geometry of the interaction between two photons, for a
spherically symmetric shell.  A test photon emitted at $R_{t,0}$
reaches $R_t > R_{t,0}$ at $t_t > t_0$ and can interact with a photon
emitted at $R_e$ that reaches the location $R_t$ at the exact same
time $t_t$ as the test photon.  Note that $O$, $R_{t,0}$ and $R_t$ are
coplanar (and in the plane of the figure), whereas $R_e$ is not in the
same plane, nor is the interacting photon trajectory that goes from
$R_e$ to $R_t$.  The observer is to the right, at infinity.
The other symbols are defined in the text.}
\label{fig:geom2}
\end{figure}

As in the previous section, we consider a ``test'' photon emitted by
the shell at radius $R_{t,0}$ and angle $\theta_{t,0}$ with respect to
the line of sight (see Fig.~\ref{fig:EATS12}). All the quantities with
a subscript `$t$' will always refer to such a test photon. We wish to
calculate its optical depth to pair production with all the other
photons which are emitted by the same source and denoted by a
subscript `i' (for potentially ``interacting'').  The differential of
the optical depth to pair production is given by \citep{Weaver76}
\begin{equation}
d\tgg = \sigma^\star[\chi(\varepsilon_t,\ei,\mu_{ti})](1-\mu_{ti})
\frac{dn_i}{d\Omega_i d\ei} d\Omega_i d\ei ds\ \label{dtau_gg}\ .
\end{equation}
In this equation, $ds$ is the differential of the path length along
the trajectory of the test photon; $n_i$, $\Omega_i$ and $E_i \equiv
\ei m_e c^2$ are the number density, solid angle, and photon energy of
the photon field along the path of the test photon with which it might
interact.~\footnote{ We do not add a factor of $1/2$ due to double
counting \citep[as was done by, e.g.,][]{BH97,DS94}, as it should not
appear in the expression for the optical depth. We discuss this point
in more details in annex \ref{factor_onehalf}.} For convenience,
$\varepsilon_t$ and $\ei$ denote the values of the corresponding
dimensionless photon energies in the lab frame, rather than in the
observer frame (as is the case for $\varepsilon$), i.e. without the
cosmological redshift, so that $\varepsilon_t = (1+z)\varepsilon$
should eventually be used in order to evaluate the optical depth at an
observed value of $\varepsilon$. The Lorentz invariant cross section
for pair production $\sigma^\star(\chi)$ is
\begin{align}
\sigma^\star(\chi) &= \frac{\pi r_e^2}{\chi^6}
\left[(2\chi^4+2\chi-1){\rm ln}(\chi+\sqrt{\chi^2-1})
-\chi(1+\chi^2)\sqrt{\chi^2-1}\right] \label{sigma}\ ,\\
\chi &= \sqrt{\frac{\varepsilon_t\ei(1-\mu_{ti})}{2}}\ , \label{chi}
\end{align}
where $\chi$ is the center of momentum energy (in units of $m_e c^2$,
of each particle -- each of the two interacting photons, and the
produced electron and positron), and $\mu_{ti} = \hat{n}_t \cdot
\hat{n}_i$ is the cosine of the angle between the directions of motion
of the test photon ($\hat{n}_t$) and a potentially interacting photon
($\hat{n}_i$). In order to evaluate $\mu_{ti}$, we need to specify the
geometry for our model: a spherical emitting shell, whose emission
depends only on its radius $R_{\rm sh}$ (i.e. at any given radius its
local emission does not depend on the location within the shell) and
is isotropic in its own rest-frame. Under these assumptions, the
radiation field will depend only on the radius $R$ and the (lab frame)
time $t$, and at any given place and time it will be symmetric around
the radial direction  (see Fig.~\ref{fig:geom2}).
Therefore, at any point along the trajectory of the test photon, we
can use a local coordinate system, $S_r$, whose $z$-axis is aligned
with the radial direction (from the center of the shell to that
point), $\hat{z}_r$, and such that the direction of motion of the test
photon is in the $x$-$z$ plane. In this frame the polar angles are
denoted by $(\theta_r,\phi_r)$, and
\begin{align}
\hat{n}_t &= \hat{x}_r\sin\theta_t + \hat{z}_r \cos\theta_t\ ,
\\
\hat{n}_i &= \hat{x}_r\sin\theta_r\cos\phi_r 
             +\hat{y}_r\sin\theta_r\sin\phi_r + \hat{z}_r\cos\theta_r\ ,
\\
\mu_{ti} &= \hat{n}_i\cdot \hat{n}_t
 = \sin\theta_t\sin\theta_r\cos\phi_r 
+ \cos\theta_t\cos\theta_r\ \label{mu_ti}.
\end{align}
Note that $\theta_t$ varies only with $s$. The integration over the
solid angle in the lab frame in Eq.~(\ref{dtau_gg}) can conveniently
use the frame $S_r$ which is at rest in the lab frame, i.e. $d\Omega_i
= d\Omega_r = d\phi_r d\mu_r$.  The optical depth of the 
test photon is then given by:
\begin{equation}\label{tau_gg_1}
\tgg(\varepsilon_t,\theta_{t,0},R_{t,0}) = 
\int ds \int d\ei \int d\Omega_r
\sigma^\star[\chi(\varepsilon_t,\ei,\mu_{ti})](1-\mu_{ti})
\frac{dn_i}{d\Omega_r d\ei}\ .
\end{equation}
Next, we express the derivative in the integrand of
Eq.~(\ref{tau_gg_1}), which represents the photon field along the
trajectory of the test photon, in terms of the source emissivity. In
addition, we make a series of changes of variable in order to simplify
the expression for the optical depth.

\subsection{Expressing the photon field in terms of the source emissivity}
\label{subsec:emissivity}

In \S\ \ref{sec:flux2}, we expressed the observed flux as an
integral over the EATS-I of photons to the observer at an observed
time $T$. These photons travel along straight line trajectories that
pass through the photon field. As a result, we integrate the
contribution to the optical depth at each point along the path of each
photon, treating it as a test photon. This is the integration over
$ds$ in Eq.~(\ref{tau_gg_1}) which, as we show below, can be replaced
by an integration over $dR_t$. In the other two inner integrations
$R_t$ is kept fixed, and the photon field, $dn_i/d\Omega_rd\ei$, needs
to be evaluated as a function of $\varepsilon_i$, $\mu_r$ and
$R_t$. For a given test photon that is emitted at ($R_{t,0}$,
$\mu_{t,0}$), the value of $R_t$ also determines the value of the lab
frame time $t_t$. We remind the reader that $R_t$ and $t_t$ are always
computed in the lab frame, and that $R_t$ is in general different than
$R_{\rm sh}(t_t)$, i.e. at a general time the position of the test
photon does not coincide with that of the shell. We proceed first to
relate the photon field at $(t_t,R_t)$ to the emissivity in the local
frame of the emitting shell, which is easier to specify, and simpler.
The Doppler factor of the emitted photon is given by
\begin{equation}\label{delta}
\delta \equiv \frac{\ei}{\varepsilon'_i}
 = \frac{1}{\Gamma(1-\beta\mu_i)} = \Gamma(1+\beta\mu'_i)\ ,
\end{equation}
where $\mu_i \equiv \cos\theta_i = \hat{\beta}\cdot\hat{n}_i$ and
$\mu'_i \equiv \cos\theta'_i = \hat{\beta}\cdot\hat{n}'_i$ are the
cosines of the angle between the bulk velocity of the emitting fluid
($\vec{\beta}$) and the direction of the interacting photon in the lab
frame ($\hat{n}_i$) and in the comoving frame of the emitting fluid
($\hat{n}'_i$), respectively. Furthermore,
\begin{equation}
\mu'_i = \frac{\mu_i-\beta}{1-\beta\mu_i} \ \Longrightarrow\ 
\frac{d\Omega'_i}{d\Omega_r} = \frac{d\Omega'_i}{d\Omega_i} 
= \frac{d\mu'}{d\mu} = \delta^2\ ,
\end{equation}
since $d\Omega_i = d\phi_i d\mu_i$ and $\phi'_i = \phi_i$.  We are
interested in the differential density of photons of energy $\ei$ and
direction of motion in the solid angle $d\Omega_r$ around the
direction $\hat{n}_i$, which is at an angle $\theta_r$ from the radial
direction, at a radius $R_t$ and time $t_t$. This density is related
to the specific intensity of the photon field by:
\begin{equation}\label{I_eps}
I_\ei(\hat{n}_i) \equiv \frac{dE}{dS dt d\ei d\Omega_i } =
\ei m_e c^3 \frac{dn_i}{d\ei d\Omega_r}(\hat{n}_i)\ ,
\end{equation}
where the ({\it normalized}) specific intensity $I_\ei$ is the energy
($dE$) per unit normal area ($dS$ where $d\vec{S}/dS = \hat{n}$), per
unit time ($dt$), per unit ({\it normalized}) photon energy ($d\ei$),
per solid angle ($d\Omega_i = d\Omega_r$) around some direction
$\hat{n}_i$ of the (potentially interacting) photons.

The differential ({\it normalized}) specific luminosity (in our case,
from a small part of the emitting shell) is defined as $dL_\varepsilon
= dE/d\ei dt$, while the isotropic equivalent ({\it normalized})
specific luminosity is defined by:
\begin{equation}\label{L_iso_def}
 dL_{\ei,iso} \equiv 4\pi\frac{dL_\ei}{d\Omega_r}\ .
\end{equation}
The contribution of an emitting element with $dL_{\ei,iso}$ to the
({\it normalized}) flux density $dF_\ei\equiv dE/dSdtd\ei$ and to the
({\it normalized}) specific intensity $I_\ei$ at a point located at a
distance $r$ from it is
\begin{equation}\label{dF_eps}
 dF_\ei = \frac{dL_{\ei,iso}}{4\pi r^2}
=I_\ei(\hat{n}) d\Omega_r\ ,
\end{equation}
and is along the direction $\hat{n}$ from the emitting element to that
point (i.e. here $dS$ is the differential of the area normal to
$\hat{n}$, $dS = \hat{n} \cdot d\vec{S}$). Finally, we can
conveniently express $dL_{\ei,iso}$ in the comoving frame (i.e. the
local rest frame of the emitting shell),
\begin{equation}\label{dL_eps_iso}
dL_{\ei,iso} = 4\pi\frac{dL_\ei}{d\Omega_r}
                  = 4\pi\frac{dE}{d\ei dt d\Omega_i} 
                  = \delta^3 4\pi\frac{dE'}{d\varepsilon' dt' d\Omega'_i} 
                  = \delta^3 dL'_{\varepsilon'}\ ,
\end{equation}
\noindent where the last equality follows from the assumption that the
 emission is isotropic in the comoving frame. Because the emission is
 assumed to be uniform throughout the shell, $dL'_{\varepsilon'}$
 depends only on the radius of emission of the potentially interacting photon, $R_e$, 
and not on the location within the shell.  Apart from the emission radius,
 $R_e$, the position of an emitting point on the shell is also
 specified by the polar angle, $\theta_e$, which for convenience is
 measured with respect to the direction from the center of the sphere
 to the location of the test photon (at a radius $R_t$) where the flux (or
 some other property of the photon field) is calculated (see
 Fig.~\ref{fig:geom2}). As a result, we can write $dL'_{\varepsilon'}
 = L'_{\varepsilon'}(R_e)d\mu_e/2$, where $\mu_e = \cos\theta_e$ and
 $L'_{\varepsilon'}(R_e)$ has been defined and discussed in
 \S\;\ref{sec:flux}.  We finally combine eqs.~(\ref{I_eps}),
 (\ref{dF_eps}) and (\ref{dL_eps_iso}) to obtain the expressions for
 the normalized specific intensity,
\begin{equation}\label{spec_intensity}
 I_{\varepsilon} = \frac{L'_{\varepsilon'}(R_e)}{4\pi} 
\frac{\delta^3}{4\pi r^2}
\left|\frac{d\mu_e}{d\mu_r}\right|\ ,
\end{equation}
and the expression for the photon field which appears in the integrand
in Eq.~(\ref{tau_gg_1}),
\begin{equation}\label{photon_field}
\frac{dn_i}{d\ei d\Omega_r} = 
\frac{L'_{\varepsilon'}(R_e)}{(4\pi)^2 \varepsilon m_e c^3} 
\frac{\delta^3}{r^2}
\left|\frac{d\mu_e}{d\mu_r}\right| \ .
\end{equation}
The derivative in the last term to the right of these equations must
be computed along the equal arrival time surface (EATS-II) of photons
to $R_t$ at $t_t$, where $I_\varepsilon$ or $dn_i/d\ei d\Omega_r$ are
to be calculated. We can now rewrite Eq.~(\ref{tau_gg_1}) as:
\begin{align}
\tgg(\varepsilon_t,\theta_{t,0},R_{t,0}) &= 
\frac{\sigma_T}{(4\pi)^2 m_e c^3} 
\int ds \int d\ei \int d\Omega_r
\frac{\sigma^\star[\chi(\varepsilon_t,\ei,\mu_{ti})]}{\sigma_T}(1-\mu_{ti})
\frac{L'_{\varepsilon'_i}(R_e)}{\ei} 
\frac{\delta^3}{r^2}
\left|\frac{d\mu_e}{d\mu_r}\right|
\ . \label{tau_gg_2}
\end{align}
We have thus replaced the photon field by the specific emissivity in
the expression for the optical depth. The boundaries of integration
will be specified explicitly later on. We now want to simplify this
triple integration in order to make it easier to evaluate.

\subsection{Analytical reduction}\label{sec:red}

In the remainder of this work, we will make use of various
dimensionless radii, which are gathered in Table~\ref{tab:basic} and
greatly simplify the analysis.  Furthermore, it is much more
convenient to work with such quantities of order unity inside the
integrand. We thus rescale $R_e$ and $R_t$ by introducing $\tilde{R_e}
\equiv R_e/R_t$ and $\hat{R_t} \equiv R_t/R_{t,0}$. Furthermore, the
notations $\tilde{R} \equiv R/R_t$ and $\hat{R} \equiv R/R_{t,0}$ will
be used for other rescaled dimensionless radii as well. While clearly
$1 \leq \hrt < \infty$, the range of $\tre$ is much more complex and
will be extensively discussed in \S\;\ref{sec:gfield}.  For now, we
want to simplify Eq.~(\ref{tau_gg_2}) by changing integration
variables. We give here the main results and leave the details of the
derivations for Annex~\ref{ap:varchanges}.

As has been mentioned above, the integration over $ds$ can be replaced
by an integration over $\hrt$. Under the approximation of large
Lorentz factors ($\Gamma \gg 1$), and thus small emission angles
($\theta_{t,0} \ll 1$), one obtains $ds = R_{t,0}d\hrt$ (see
the discussion following Eq.~[\ref{ds_to_dR_t}] for more details). Besides,
since we integrate over $d\Omega_r = d\phi_r d\mu_r$ and the integrand
contains $|d\mu_e/d\mu_r|$ we can conveniently change the integration
over $\mu_r$ to an integration over $\tilde{R}_e$. We show in the
Annex~\ref{ap:varchanges} that
\begin{equation*}
\left|\frac{d\mu_e}{d\mu_r}\right|d\mu_r = 
\frac{d\mu_e}{d\tilde{R}_e}\;d\tilde{R}_e\ , 
\end{equation*}
\noindent since $d\mu_e/d\tilde{R}_e > 0$, where the limit of
integration over $\tre$ should be in increasing order (i.e. the
integration should be from small to large values of $\tre$).  The
optical depth now reads:
\begin{align}\label{tau_gg_2bis}
\tgg(\varepsilon_t,\theta_{t,0},R_{t,0}) &\approx 
\frac{\sigma_T}{(4\pi)^2 m_e c^3 R_{t,0}}
\int_1^\infty \frac{d\hat{R}_t}{\hat{R}_t^2} 
\int_{2/\varepsilon_t}^\infty d\ei 
\int_0^{2\pi} d\phi_r 
\int_{R_0/R_t} d\tilde{R}_e 
 \nonumber \\
&\quad\quad\quad\quad\quad\quad\quad\quad\quad\quad\quad\quad
\times\,\frac{\sigma^\star[\chi(\varepsilon_t,\ei,\mu_{ti})]}{\sigma_T}
(1-\mu_{ti})\frac{L'_{\varepsilon'_i}(R_e)}
{\ei}\frac{\delta^3}{\tilde{r}^2}\frac{d\mu_e}{d\tilde{R}_e}\ .\quad
\end{align}

Next, we can follow the hind-sights of \citet{Stepney83} and
\citet{Baring94} in order to cast the integrations over
$(d\phi_r,d\varepsilon_i)$ into a much more practical form. In order
to perform this change of variables, it is necessary to specialize the
specific luminosity to the dependence discussed in \S\;\ref{sec:flux}:
$L'_{\varepsilon'_i}(R_e) = L'_0\,h(R_e/R_0)(\ei')^{1-\alpha} \equiv
\Gamma_0^{-\alpha}L_0(\ei')^{1-\alpha}\times h(\tre\hrt/\hrz)$, where
$h$ is a general function of $R_e/R_0$ that satisfies $h(1) = 1$ (for
details see Appendix~\ref{local_luminosity}) and $L'_0 \equiv
L'_{\varepsilon'=1}(R_0)$. Note that $L_0 \equiv \Gamma_0^\alpha L'_0$
is approximately the observed isotropic equivalent luminosity at a
photon energy of $m_e c^2 \approx 511\;$keV, near the peak of the
spike in the light curve which corresponds to the emission episode
that we model for $\Delta R \sim R_0$. 

For convenience, we rescale all the quantities in the integrand of
Eq.~(\ref{tau_gg_2bis}) which are not of order unity by the relevant
power of the Lorentz factor at radius $R_t$, $\Gamma_t =
\Gamma(R_t)$, so that the rescaled quantities (which are denoted by a
bar) will be of order unity. We rescale $\bar{\delta} \equiv
\delta/\Gamma_t$ and $d\bar{\mu}_e \equiv \Gamma_t^2 d\mu_e$, but do
not rescale $\tilde{r}$ which is already of order unity. Thus,
\begin{equation}\label{rescaled1}
\frac{\delta^{2+\alpha}}{\tilde{r}^2} \cdot \frac{d\mu_e}{d\tre} =
\Gamma_0^\alpha\left(\frac{\hrt}{\hrz}\right)^{-m\alpha/2}
\frac{\bar{\delta}^{\,2+\alpha}}{\tilde{r}^{\,2}}
\cdot \frac{d\bar{\mu}_e}{d\tre}\ ,
\end{equation}
and the expression for the optical depth becomes:
\begin{equation} \label{tau_gg_5}
\tgg(\varepsilon_t,\theta_{t,0},R_{t,0}) = \tau_\star
\varepsilon_t^{\alpha-1}  \hrz^{1-m\alpha/2} 
\int_1^\infty \frac{d\hrt}{\hrt^{2-m\alpha/2}} 
\int d\tre \frac{\bar{\delta}^{\,2+\alpha}}{\tilde{r}^{\,2}}
\,\frac{d\bar{\mu}_e}{d\tre}\;h\left(\tre\frac{\hrt}{\hrz}\right) 
\bar{\zeta}_-^{\,\alpha} H_\alpha(\zeta) \ , 
\end{equation}
where
\begin{equation} \label{tau_star}
\tau_\star\left(\Gamma_0^{2\alpha}R_0,\alpha,L_0\right) = 
\frac{7\sigma_T}{48\pi^3m_e c^3}
\frac{\Gamma_0^{-2\alpha}L_0}{\alpha^{5/3}R_0} =
0.402\left(\frac{\alpha}{2}\right)^{-5/3}10^{4(2-\alpha)}
\frac{L_{0,52}}{(\Gamma_{0,2})^{2\alpha}R_{0,13}}\ , 
\end{equation}
and $L_{0,52} = L_0/(10^{52}\;{\rm erg\;s^{-1}})$, $R_{0,13} =
R_0/(10^{13}\;$cm), $\Gamma_{0,2} = \Gamma_0/100$, $\zeta =
(\zeta_+-\zeta_-)/\zeta_-$, $\zeta_+ =
\left[1-\cos{(\theta_r+\theta_t)}\right]/2$, $\zeta_- =
\left[1-\cos{(\theta_r-\theta_t)}\right]/2$, and $H_\alpha(z)$ is a
function discussed in Annex~\ref{local_luminosity}. In
Eq.~(\ref{tau_star}), $\tau_\star$, the only quantity requiring
astrophysical input, is a constant of the order of the optical depth
to pair production at a photon energy of $m_e c^2$ at $R_0$ in
quasi-steady state (near the peak of the spike in the light curve for
$\Delta R \sim R_0$). Note that since both the photon index $\alpha$
and $L_0$ (roughly the isotropic equivalent luminosity) are observable
quantities (the latter requiring knowledge of the source redshift),
the observation of a high-energy spectral cutoff due to pair
production opacity can enable the determination of
$\Gamma_0^{2\alpha}R_0$. In the limit of small angles that is
appropriate for large Lorentz factors, $\zeta_-$ is of order
$\Gamma^{-2}$, so we define $\bar{\zeta}_-\equiv\Gamma_t^2\zeta_-$,
where $\Gamma_t = \Gamma(R_t) = \Gamma_0\hat{R}_t^{-m/2}$. Thus,
\begin{equation}\label{rescaled2}
 \zeta_-^{\alpha} = \Gamma_0^{-2\alpha}
\left(\frac{\hrt}{\hrz}\right)^{m\alpha}
\left(\bar{\zeta}_-\right)^{\alpha}\ .
\end{equation}

Under the assumption that $h$ is also a power-law of index $b$, $h(R_e) =
(R_e/R_0)^b = (\tre\hrt/\hrz)^b$, the expression for the optical depth
in our model simplifies to:
\begin{align}
\tgg(\varepsilon_t,\theta_{t,0},R_{t,0}) &=
\tau_0(\varepsilon_t,R_{t,0}){\cal F}(x)\ , \label{tau_gg_6}
\\
\tau_0(\varepsilon_t,R_{t,0}) &= \tau_\star\;\varepsilon_t^{\alpha-1}\;
 \hrz^{1-b-m\alpha/2}\ , \label{tau_0}
 \\
{\cal F}(x) &= \int_1^\infty d\hrt\,\hrt^{b-2+m\alpha/2}
\int d\tre \frac{\bar{\delta}^{\,2+\alpha}}
{\tilde{r}^{\,2}}\cdot\frac{d\bar{\mu}_e}{d\tre} \tre^b\; 
\bar{\zeta}_-^{\,\alpha} H_\alpha(\zeta)\ . \label{cal_F}
\end{align}

In order to proceed further, we need to obtain an explicit expression
for the innermost integrand of $\cal F$, by a detailed examination of the
geometry of the photon field. The next section will be devoted to this
analysis, which constitutes the main novelty of this work. We will
evaluate the optical depth (Eq.~[\ref{tau_gg_6}]), taking into account
that the photon field is not homogeneous along the test photon
trajectory, but the contribution to the photon field is actually built
up in time.

\section{Calculating the Photon Field}\label{sec:gfield}

\subsection{Equal Arrival Time Surface of Photons to the Test Photon (EATS-II)}

In this section we calculate the photon field at a general radius
$R_t$ and time $t_t$, along the trajectory of a test photon. For this
purpose we need to consider the contribution from all photons that
arrive at the instantaneous location of the test photon, ($R_t,t_t$),
simultaneously. The locus of points where all such photons are
emitted, taking into account that the emission occurs only in the
shell, forms a two dimensional surface referred to as the equal
arrival time surface (EATS-II) of photons to the instantaneous
location of the test photon. The local photon field at ($R_t,t_t$) is
calculated by integrating the contributions over this surface.  We
stress that this surface (EATS-II), is different from the equal
arrival time surface of photons to the observer at infinity (EATS-I).

Fig.~\ref{fig:EATS12} shows the basic configuration for our
calculations and illustrates the relation between
the two different equal arrival time surfaces (EATS) of photons: 1. to
the observer at infinity (EATS-I), 2. to the instantaneous location of
a test photon (EATS-II). It can be seen that the EATS-II grows with
the lab frame time $t$, and therefore also with the radius of the test
photon $R_t$. Furthermore, each EATS-II encompasses all other EATS-II
corresponding to smaller times, and is encompassed within all the
EATS-II which correspond to larger times. In particular, all EATS-II
are within the EATS-I, which corresponds to the limit of the EATS-II
for an infinite time (when the test photon reaches the observer at
infinity). All of the EATS-II and EATS-I pass through the emission
point of the test photon, and for case 2 and 3, also through the place
where the photon crosses the shell (i.e. its location in case 2). These
are general properties of the EATS-II.

We now proceed to calculate the EATS-II and the expressions for
relevant quantities along this surface, which are needed in order to
calculate the local radiation field. From the geometry of our problem
(see Fig.~\ref{fig:geom2}), we can immediately derive the two
following equations:
\begin{align}\label{r_sq}
\tilde{r}^2 &= 1 + \tilde{R}_e^2 - 2\tilde{R}_e\mu_e 
= (1-\tilde{R}_e)^2 +2\tilde{R}_e(1-\mu_e)\ , 
\\ \label{R_sq}
\tilde{R}_e^2 &= 1 + \tilde{r}^2 - 2\tilde{r}\mu_r\ ,
\end{align}
where $\tilde{R}_e \equiv R_e/R_t$ and $\tilde{r} \equiv r/R_t$. The
equal arrival time surface (EATS-II) of photons to $(R_t,t_t)$ is
determined by the condition that $r = c (t_t-t_e) = c[t_t-t_{\rm
sh}(R_e)]$, where the photons are emitted at a previous time $t_e$ when the
shell is at a radius $R_e = R_{\rm sh}(t_e)$. The EATS-II equation is thus
given by
\begin{equation}\label{r_tilde}
\tilde{r} = \frac{c}{R_t}\left[t_t - t_{\rm sh}(R_e)\right] 
= \sqrt{(1-\tilde{R}_e)^2+2\tilde{R}_e(1-\mu_e)}\ ,
\end{equation}
which relates the radius ($R_e$) and angle ($\theta_e =
\arccos\mu_e$) of emission along this surface.

The expression for $t_{\rm sh}(R_e)$ depends on our assumption about
the expansion of the shell. If the latter occurs at constant speed,
then $t_{\rm sh}(R_e) = R_e/\beta c$ and in the limit of $R_t \to
\infty$ Eq.~(\ref{r_tilde}) reduces to $\beta(ct_t-R_t) =
R_e(1-\beta\mu_e)$, which is the usual polar equation of an ellipse
(setting $cT = ct_t - R_t$). In this simple case, using the short
notation $\tilde{R}_{\rm sh} = \tilde{R}_{\rm sh}(t_t)$, we have
$\tilde{r} = (\tilde{R}_{\rm sh}-\tre)/\beta$, and the EATS-II is
given by
\begin{equation}
\mu_e = 1 - \frac{1}{2\beta^2\tre}\left[
\left(\tilde{R}_{\rm sh}-\tre\right)^2
-\beta^2\left(1-\tre\right)^2\right]\ ,
\end{equation}
while the lower and upper limits for the range of $\tre$ values along
the EATS-II, which correspond to $\mu_e = -1$ and $\mu_e = 1$,
respectively, are given by
\begin{equation}
\tilde{R}_{e,{\rm min}} = \frac{\tilde{R}_{\rm sh}-\beta}{1+\beta}
\quad,\quad\quad
\tilde{R}_{e,{\rm max}} = \left\{\begin{matrix}
(\tilde{R}_{\rm sh}+\beta)/(1+\beta) & \quad \tilde{R}_{\rm sh}
\geq 1\ , \cr\cr
(\tilde{R}_{\rm sh}-\beta)/(1-\beta) & \quad \tilde{R}_{\rm sh}
\leq 1\ .
\end{matrix}\right.
\end{equation}
Note that we have not assumed $\Gamma = (1-\beta^2)^{-1/2} \gg 1$, so
these results are valid for an arbitrary velocity, as long as it is
constant with radius.

Combining eqs.~(\ref{r_sq}) and (\ref{R_sq}) we also obtain
\begin{equation}
\mu_r =
\frac{1-\tilde{R}_e\mu_e}{\sqrt{(1-\tilde{R}_e)^2+2\tilde{R}_e(1-\mu_e)}} =
\frac{R_t}{c}\left[\frac{1-\tilde{R}_e\mu_e}{t_t-t_{\rm sh}(R_e)}\right]\ ,
\end{equation}
where in the last equality we have also used Eq.~(\ref{r_tilde}), so
that it is valid only along the EATS-II (while the first equality is
valid more generally, as it is derived directly from the geometrical
setup).

\subsection{Radial Dependence of Relevant Angles,
$\mu_e(\tilde{R}_e)$ and $\mu_r(\tilde{R}_e)$, along EATS-II}

Specifying for $1 \ll \Gamma^2 =\Gamma_t^2\tilde{R}^{-m}$, we can
rewrite Eq.~(\ref{t_0}) as
\begin{equation}\label{t_sh}
t_{\rm sh}(R) = 
\frac{R_t}{c}\left[\tilde{R}+\frac{\tilde{R}^{m+1}}
{2(m+1)\Gamma^2_t}\right]+ {\cal O}\left(\Gamma_t^{-4}\right) \ .
\end{equation}
Thus, Eq.~(\ref{r_tilde}) implies
\begin{align}\label{eq:tre}
\tilde{r} &= \tilde{R}_{\rm sh}-\tilde{R}_e + 
\frac{\left(\tilde{R}_{\rm sh}^{m+1}-\tilde{R}_e^{m+1}\right)}
{2(m+1)\Gamma^2_t} + {\cal O}\left(\Gamma_t^{-4}\right)\ ,
\\ \label{eq4bis}
 \tilde{r}^2 &= \left(\tilde{R}_{\rm sh}-\tilde{R}_e\right)^2 
+\frac{\left(\tilde{R}_{\rm sh}-\tilde{R}_e\right)
\left(\tilde{R}_{\rm sh}^{m+1}-\tilde{R}_e^{m+1}\right)}
{(m+1)\Gamma^2_t} + {\cal O}\left(\Gamma_t^{-4}\right)\ .
\end{align}
Note that $\tilde{R}_e \leq \tilde{R}_{\rm sh}$, because $t_e \leq
t_t$ (due to causality) and $R_{\rm sh}(t)$ is an increasing function
of $t$. The equality only holds when $R_t = R_{\rm sh}(t_t)$,
i.e. when $\tilde{R}_{\rm sh} = 1$ (case 2 below). Thus,
eqs.~(\ref{r_sq}) and (\ref{eq4bis}) give (to the order of
$\Gamma_t^{-2}$),
\begin{align}\label{mu_e}
2\Gamma_t^2(1-\mu_e) &= (\Gamma_t\theta_e)^2 
\nonumber \\ 
&=\frac{1}{\tilde{R}_e}\left\{
\Gamma_t^2\left[\left(\tilde{R}_{\rm sh}-\tilde{R}_e\right)^2
-\left(1-\tilde{R}_e\right)^2\right]
+\frac{\left(\tilde{R}_{\rm sh}-\tilde{R}_e\right)
\left(\tilde{R}_{\rm sh}^{m+1}-\tilde{R}_e^{m+1}\right)}{(m+1)}\right\}\ .
\end{align}
The two terms on the right hand side of the equation are typically of
the same order since $|\tilde{R}_{\rm sh}-1| \lesssim {\rm a\ few\
}\Gamma_t^{-2}$, i.e. $\tilde{R}_{\rm sh} \cong 1 \Leftrightarrow
R_{\rm sh}(t_t) \cong R_t$. This immediately implies
\begin{equation}\label{dmu_e_dR}
\frac{d\mu_e}{d\tilde{R}_e} = \frac{1}{2\Gamma^2_t\tilde{R}_e^2}
\left[ \Gamma^2_t\left(\tilde{R}_{\rm sh}^2-1\right)
+\frac{\tilde{R}_{\rm sh}}{m+1}\left(\tilde{R}_{\rm
sh}^{m+1} + m\tilde{R}_e^{m+1}\right) - \tilde{R}_e^{m+2} \right]\ .
\end{equation}

Now we turn to $\mu_r$. From eqs.~(\ref{R_sq}) and (\ref{eq4bis}) we
obtain
\begin{align}\label{mu_r}
\mu_r &=
\frac{\left(\tilde{R}_{\rm sh}-\tilde{R}_e\right)^2
+\left(1-\tilde{R}_e^2\right)}
{2\left(\tilde{R}_{\rm sh}-\tilde{R}_e\right)}
+\frac{\left(\tilde{R}_{\rm sh}^{m+1}-\tilde{R}_e^{m+1}\right)}
      {4(m+1)\Gamma_t^2}\left[1-\frac{1-\tilde{R}_e^2}
{\left(\tilde{R}_{\rm sh}-\tilde{R}_e\right)^2}\right]
+ {\cal O}\left(\Gamma_t^{-4}\right)\, ,\quad\ \ \
\\ \nonumber
\frac{d\mu_r}{d\tilde{R}_e} &= -\frac{d\tilde{r}}{d\tilde{R}_e}
\left[\frac{1-\tilde{R}_e^2-\tilde{r}^2}{2\tilde{r}^2}\right]
-\frac{\tilde{R}_e}{\tilde{r}}
\\ \nonumber
&= \frac{1-\tilde{R}_{\rm sh}^2}
{2\left(\tilde{R}_{\rm sh}-\tilde{R}_e\right)^2}
+ \frac{\tilde{R}_e^m}{4\Gamma_t^2}\left[\frac{1-\tilde{R}_e^2}
{\left(\tilde{R}_{\rm sh}-\tilde{R}_e\right)^2}-1\right]
\\ \label{dmu_r_dR}
&\quad+\frac{\left(\tilde{R}_{\rm sh}^{m+1}-\tilde{R}_e^{m+1}\right)
\left(\tilde{R}_{\rm sh}\tilde{R}_e-1\right)}
{2(m+1)\Gamma_t^2\left(\tilde{R}_{\rm sh}-\tilde{R}_e\right)^3}
+ {\cal O}\left(\Gamma_t^{-4}\right)\ ,
\end{align}
\noindent where
\begin{equation}
-\frac{d\tilde{r}}{d\tilde{R}_e} = 1+\frac{\tilde{R}_e^m}{2\Gamma_t^2}
= 1+\frac{1}{2\Gamma^2(\tilde{R}_e)} = \frac{1}{\beta(\tilde{R}_e)}\ .
\end{equation}
This can easily be understood since $r = c(t_t-t_e)$ along the equal
arrival time surface, so that $dr = -cdt_e$ and $d\tilde{r}/d\tilde{R}_e =
dr/dR_e = -cdt_e/dR_e = -c/(dR_e/dt_e) = -1/\beta(\tilde{R}_e)$.

The maximal radius of emission, $R_{e,{\rm max}}$, from which a photon
reaches a point at radius $R_t$ at the time $t_t$ is determined by the
photon that is emitted at $\theta_e = 0$ (i.e. $\mu_e = 1$), along the
line connecting that point to the center of the sphere. Thus,
\begin{align}
\tilde{r}_{\rm min} &= \left|1-\tilde{R}_{e,{\rm max}}\right| = 
\frac{c}{R_t}\left[t_t-t_{\rm sh}(\tilde{R}_{e,{\rm max}})\right] \\
&=\tilde{R}_{\rm sh}-\tilde{R}_{e,{\rm max}} + 
\frac{\left(\tilde{R}_{\rm sh}^{m+1}-\tilde{R}_{e,{\rm max}}^{m+1}\right)}
{2(m+1)\Gamma^2_t} + {\cal O}\left(\Gamma_t^{-4}\right)\ \label{remax_cond},
\end{align}
and the problem naturally divides into three cases.

\subsection{Properties of EATS-II According to Relative Location of
  Test Photon and Shell}
\label{sec:cases}

\begin{figure}[t]
\centering
\includegraphics[scale = 0.61]{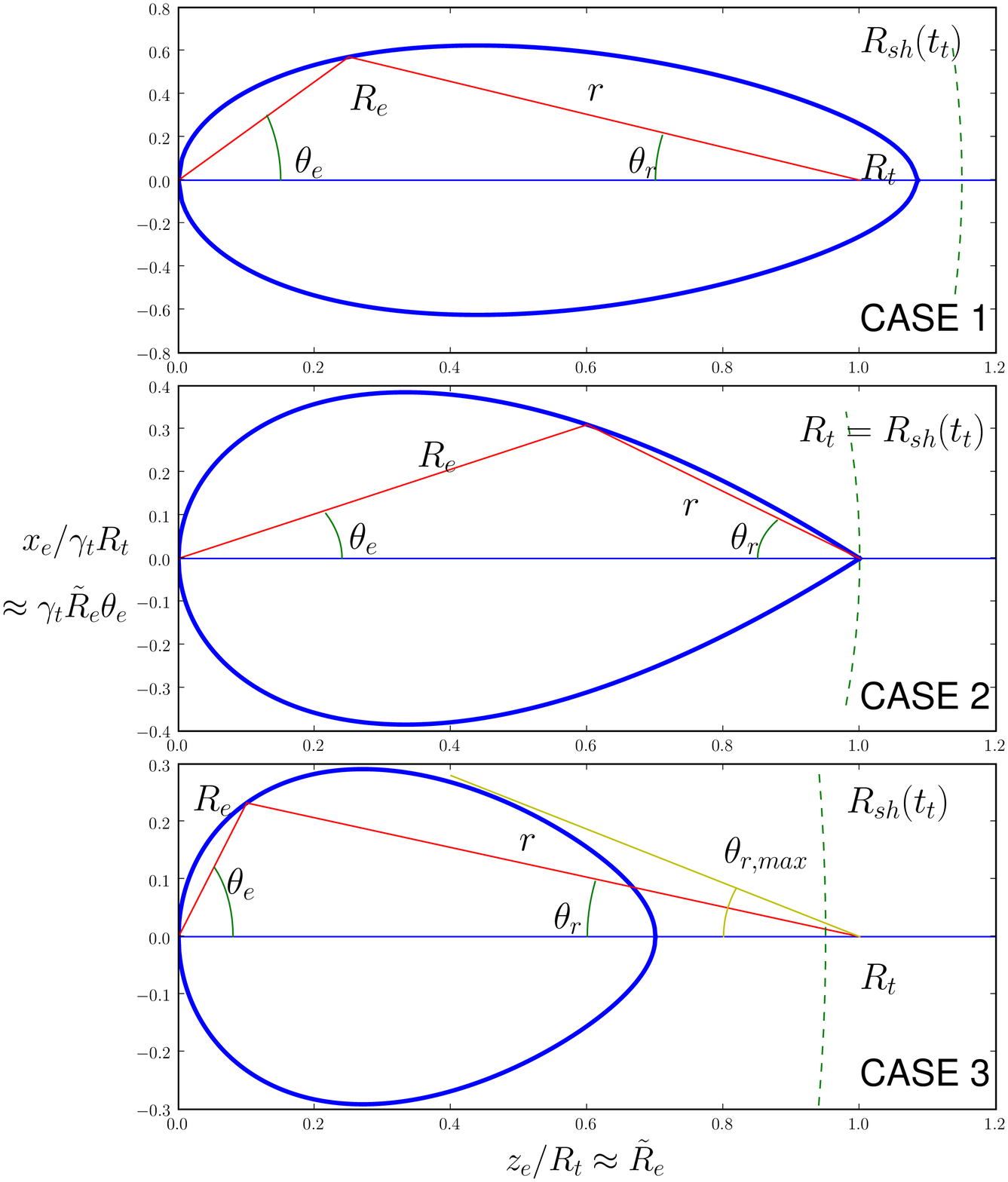}
\caption{The equal arrival time surface (EATS-II) of photons to
$(R_t,t_t)$, which represents a general point along the trajectory of
a test photon, is shown by the thick blue line. It naturally divides
into three cases: 1. the test photon is behind the shell ($R_t <
R_{\rm sh}(t_t)$ -- {\it upper panel}), 2. the test photon coincides
with the shell ($R_t = R_{\rm sh}(t_t)$ -- {\it middle panel}), and
3. the test photon is ahead of the shell ($R_t > R_{\rm sh}(t_t)$ --
{\it lower panel}). There are qualitative difference in the
properties of the EATS-II between these different cases, that are
discussed in the text.}
\label{fig:EATS2}
\end{figure}

The properties of the EATS-II qualitatively change according to the
location of the test photon relative to the shell at the same lab
frame time, $t_t$. Thus the problem naturally divides into three
cases, as illustrated in Figs.~\ref{fig:EATS12} and \ref{fig:EATS2}. If the photon is
emitted at an angle\footnote{More generally, the condition is
$\cos\theta_{t,0} < \beta$, but for $\Gamma_{t,0} \gg 1$ and
$\theta_{t,0} \ll 1$ this reduces to $\theta_{t,0} > 1/\Gamma_{t,0}$.}
$\theta_{t,0} > 1/\Gamma_{t,0}$, i.e. $x \equiv
(\Gamma_{t,0}\theta_{t,0})^2 > 1$, it initially lags behind the shell
(case 1), since due to the aberration of light (also referred to as
relativistic beaming) this corresponds to an angle greater than
$90^\circ$ from the radial direction in the co-moving frame of the
shell. The photon eventually catches-up with the shell and crosses it
(case 2), since the latter is moving at a velocity slight smaller than
the speed of light. After it crosses the shell, it remains ahead of
the shell (case 3). A photon that is emitted at $\theta_{t,0} \leq
1/\Gamma_{t,0}$, corresponding to $x \leq 1$, immediately gets ahead
of the shell (case 3). All photons are always emitted at the shell, so
the point of emission is considered case 2.  Like the later shell
crossing for photons with $x > 1$, case 2 corresponds to a single
point along that trajectory of the test photon, unlike cases 1
corresponds to a finite path along the trajectory, and case 3
corresponds to a (practically) semi-infinite interval (as far as the
observer is considered to be at ``infinity''; the contribution to the
opacity at large distances from the source, however, becomes
negligible). The three different cases are discussed in detail below,
and the relevant expressions for each case are derived. We start by
defining some useful quantities for this purpose, which will be very
helpful later on.

In the limit of small angles, Eq.~(\ref{eq:rperp}) yields
\begin{equation}\label{eq:rperp2}
 (\Gamma_t\theta_t)^2 \approx x\hrt^{-m-2}\ ,
\end{equation}
where $x \equiv (\Gamma_{t,0}\theta_{t,0})^2$ is the square of the
normalized emission angle of the test photon.  Evaluating
Eq.~(\ref{t_sh}) at $\tilde{R}_{\rm sh} = \tilde{R}_{\rm sh}(t_t)$
gives
\begin{equation}\label{R_sh}
\frac{ct_t}{R_t} = \tilde{R}_{\rm sh} 
+ \frac{\tilde{R}_{\rm sh}^{m+1}}{2(m+1)\Gamma^2_t} 
+ {\cal O}\left(\Gamma_t^{-4}\right)\ ,
\end{equation}
which can be rewritten in terms of the quantity
\begin{equation}\label{ctr}
f_m \equiv 2(m+1)\Gamma^2_t\left(\frac{ct_t}{R_t}-1\right) 
= 2(m+1)\Gamma^2_t(\tilde{R}_{\rm sh} - 1) + \tilde{R}_{\rm sh}^{m+1} 
+ {\cal O}\left(\Gamma_t^{-2}\right)\ ,
\end{equation}
that plays a major role in the following derivations.

For an emission episode starting at
 $R_0=0$, the inequality $ct_t > R_t$ is required in 
order to have a non-vanishing radiation field at the point $(R_t,t_t)$. 
If the emission turns on at a non-zero radius $R_0$, this condition generalizes to
\begin{equation}\label{R0_order}
\frac{ct_t}{R_t}-1 \geq \frac{\tilde{R}_0}{2(m+1)\Gamma^2(R_0)}
= \frac{\tilde{R}_0^{m+1}}{2(m+1)\Gamma_t^2}\ .
\end{equation}
This implies that $f_m > 0$ (for $m > -1$, which is assumed in this
 work, and is typically the case for the astrophysical sources of
 interest).

\begin{figure}[t]
\centering
\includegraphics[scale=0.88]{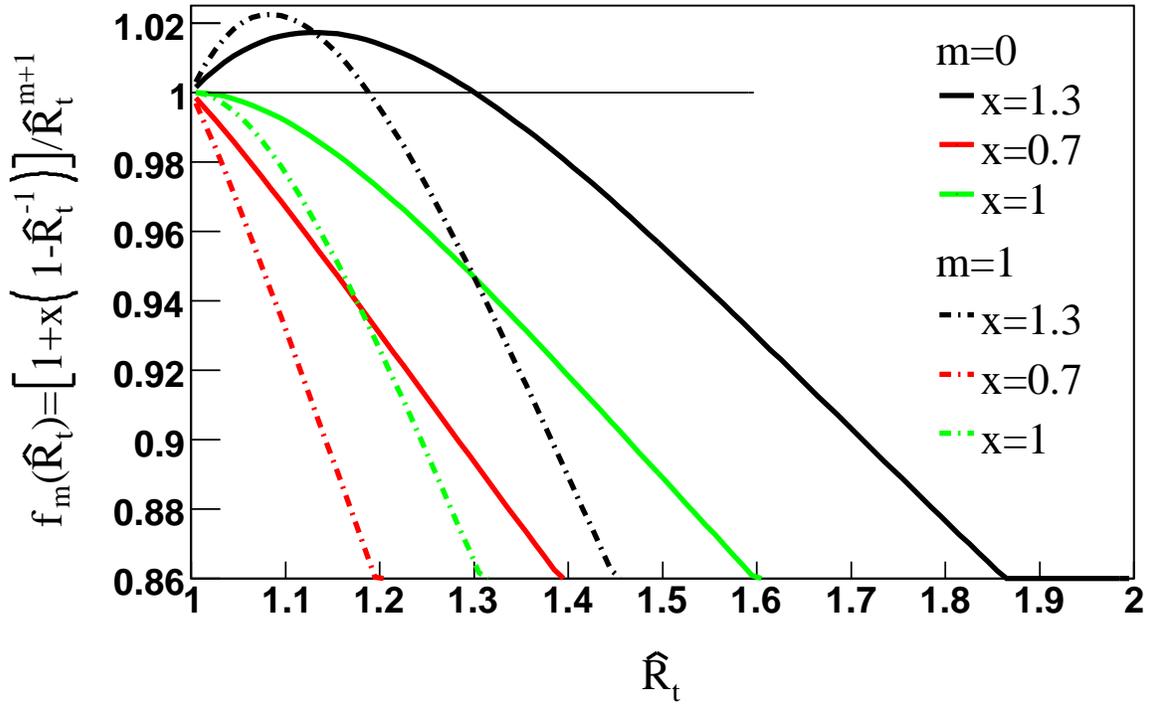}
\caption{$f_m(\hrt)$, defined in Eq.~(\ref{f_m}), as a function of
$\hrt$, for $m = 0$ and $m = 1$.  The test photon is necessarily on the
shell at the time of its emission, so that all the curves meet at
$f_m(1)=1$.}
\label{fig:x}
\end{figure}

We note that $f_m < 1$ for $\tilde{R}_{\rm sh} < 1$ (when the test
photon is traveling in front of the shell), $f_m > 1$ for
$\tilde{R}_{\rm sh} > 1$ (when the test photon is traveling behind the
shell), and $f_m = 1$ for $\tilde{R}_{\rm sh} = 1$ (when the test
photon is at the shell). It is convenient to express $f_m$ as a
function of our primary variables. Using
\begin{equation}
R_t^2 = R_\perp^2+z^2 = R^2_{t,0}\sin^2{\theta_{t,0}} +
\left[R_{t,0}\cos{\theta_{t,0} + c(t_t-t_0)}\right]^2\ ,
\end{equation}
where $R_\perp$ is the distance between the line of sight to the
origin and the trajectory of the test photon (see
Fig.~\ref{fig:EATS12} and Eq.~[\ref{eq:rperp}]) and solving this
second order equation, one obtains
\begin{equation}
\frac{c\left(t_t-t_0\right)}{R_{t,0}} = \left(\hat{R}_t-1\right)
\left(1+\frac{\theta^2_{t,0}}{2\hat{R}_t}\right) 
+ {\cal O}(\theta^4_{t,0})\ .  
\end{equation}
Recalling
that $ct_0/R_{t,0} = 1+1/2(m+1)\Gamma^2_{t,0} + {\cal
O}(\Gamma^{-4}_{t,0})$, we finally obtain:
\begin{equation}\label{f_m}
f_m(\hrt) \equiv 2(m+1)\Gamma_t^2\left(\frac{ct_t}{R_t}-1\right) = 
\frac{1+x(m+1)\left(1-\hrt^{\,-1}\right)}{\hrt^{m+1}} 
+ {\cal O}\left(\Gamma^{-4}\right)\ .
\end{equation}
Fig.~\ref{fig:x} shows the dependence of $f_m(\hrt)$ on the parameter
$x \equiv (\Gamma_{t,0}\theta_{t,0})^2$.
For $\hrt = 1$ we always have $f_m = 1$ since the test photon is
emitted at the shell.  For $x > 1$ the photon initially lags
behind the shell (case 1), and the equation $f_m = 1$ that can be
expressed as $\hrt^{\,m+2} - \left[1+(m+1)x\right]\hrt + (m+1)x = 0$
has an additional non-trivial solution, $\hat{R}_2$, which corresponds
the the point where the photon crosses the shell.  For $m = 0$ and $m
= 1$, it is given by $\hat{R}_2 = x$ and $\hat{R}_2 =
(\sqrt{1+8x}-1)/2$, respectively.

\subsubsection{Case 1: Test Photon Behind the Shell,
$R_t < R_{\rm sh}(t_t)$}\label{subsec:case1}

In this case
\begin{equation}\label{R_order2}
R_t < R_{e,{\rm max}} < R_{\rm sh}(t_t) \lesssim R_t\left(1 +
 \frac{\rm a\ few}{\Gamma_t^2}\right)\Leftrightarrow 
1 < \tilde{R}_{e,{\rm max}} < \tilde{R}_{\rm sh}(t_t) \lesssim 
1 + \frac{\rm a\ few}{\Gamma_t^2}\ ,
\end{equation}
where the last approximate inequality holds for emission angles
$(\Gamma_t\theta_e)^2 \lesssim{\rm a\ few}$, from which most of the
contribution to the observed flux arises, and are therefore the ones
of relevance. An expression for $\tilde{R}_{\rm sh}(t_t)$ may readily
be obtained through (see Eq.~[\ref{t_sh}])
\begin{equation}\label{R_sh2}
\tilde{R}_{\rm sh}(t_t) = \frac{ct_t}{R_t}\left\{1-
\frac{\left[\tilde{R}_{\rm
      sh}(t_t)\right]^m}{2(m+1)\Gamma_t^2}\right\}
+ {\cal O}\left(\Gamma_t^{-4}\right)
= \frac{ct_t}{R_t} - \frac{1}{2(m+1)\Gamma_t^2} 
+ {\cal O}\left(\Gamma_t^{-4}\right) \ ,
\end{equation}
while $\tilde{R}_{e,{\rm max}}$ is obtained by equating the two
expressions for $\tilde{r}$, from Eq.~(\ref{eq:tre}) and
Eq.~(\ref{r_sq}) for $\mu_e = 1$,
\begin{equation}
\tilde{r}_{\rm min} = \tilde{R}_{e,{\rm max}} - 1 =  
\tilde{R}_{\rm sh}-\tilde{R}_{e,{\rm max}} + 
\frac{\left(\tilde{R}_{\rm sh}^{m+1}-\tilde{R}_{e,{\rm max}}^{m+1}\right)}
{2(m+1)\Gamma^2_t} + {\cal O}\left(\Gamma_t^{-4}\right)
= \tilde{R}_{\rm sh}-\tilde{R}_{e,{\rm max}} 
+ {\cal O}\left(\Gamma_t^{-4}\right)\ ,
\end{equation}
which implies
\begin{align}
&\phantom{\Rightarrow}2\left(\tilde{R}_{e,{\rm max}}-1\right) \approx
\left(\tilde{R}_{\rm sh} - 1\right) \approx 
\left(\frac{ct_t}{R_t} - 1\right) - \frac{1}{2(m+1)\Gamma_t^2}\ ,
\\ \label{R_emax2}
&\Rightarrow\tilde{R}_{e,{\rm max}} = \frac{1}{2}\left(\frac{ct_t}{R_t}+1\right)
- \frac{1}{4(m+1)\Gamma_t^2} + {\cal O}\left(\Gamma_t^{-4}\right)
\approx \frac{\tilde{R}_{\rm sh} + 1}{2}\ .
\end{align}

While $\theta_e$ is always small, $\left(\Gamma_t\theta_e\right)^2
\lesssim {\rm a\ few}$, in the case studied in this subsection
$\theta_r$ can range from zero to $\pi$ and it is not obvious {\it a
priori} whether it can be taken to be either large,
$\left(\Gamma_t\theta_r\right)^2 \gg 1$, or small,
$\left(\Gamma_t\theta_r\right)^2 \lesssim {\rm a\ few}$.  We argue
that when $\theta_r$ is large, the photons must be emitted at a large
angle relative to the direction of motion of the emitting shell
($\theta_i = \theta_r + \theta_e$), and are therefore significantly
suppressed by relativistic beaming. This effect wins over the increase
in the reaction rate due to the larger angle between the test photon
and the interacting photons, that is manifested by the factor of
$(1-\mu_{ti})$ in the integrand for the optical depth.  Therefore, the
dominant contribution to the optical depth occur from small $\theta_r$
values, and we can therefore make the approximations that are
appropriate for $\left(\Gamma_t\theta_r\right)^2 \lesssim {\rm a\
few}$. We express these considerations more quantitatively in 
Annex~\ref{ap:case1}. Thus, we obtain:
\begin{align}\nonumber
\left(\Gamma_t\theta_e\right)^2 &= 2\Gamma_t^2(1-\mu_e) = 
\frac{\left(\tilde{R}_{e,{\rm max}}-\tilde{R}_e\right)}{(m+1)\tilde{R}_e}
\left[\tilde{R}_{e,{\rm max}}^{m+1}-\tilde{R}_e^{m+1}
+4(m+1)\Gamma_t^2\left(\tilde{R}_{e,{\rm max}}-1\right)\right]
+ {\cal O}\left(\Gamma_t^{-2}\right)
\\ \label{mu_e2}
&=
\frac{\left(1-\tilde{R}_e\right)}{(m+1)\tilde{R}_e}
\left[f_m(\hrt)-\tilde{R}_e^{m+1}\right] 
+ {\cal O}\left(\Gamma_t^{-2}\right) \ ,
\\ \nonumber
\\ \nonumber
\left(\Gamma_t\theta_r\right)^2 &= 2\Gamma_t^2(1-\mu_r) = 
\frac{\tilde{R}_e\left[\tilde{R}_{\rm sh}^{m+1} - \tilde{R}_e^{m+1} 
+ 2(m+1)\Gamma_t^2\left(\tilde{R}_{\rm sh}-1\right)\right]}{(m+1)\left(\tilde{R}_{\rm sh}-\tilde{R}_e\right)}
+ {\cal O}\left(\Gamma_t^{-2}\right)
\\ \label{mu_r2}
&= \frac{\tilde{R}_e\left[f_m(\hrt)-\tilde{R}_e^{m+1}\right]}
{(m+1)\left(1-\tilde{R}_e\right)}
+ {\cal O}\left(\Gamma_t^{-2}\right)\ ,
\end{align}
\begin{align}\nonumber
\frac{d\mu_e}{d\tilde{R}_e} &= 
\frac{1}{2(m+1)\Gamma_t^2}\left\{\frac{\tilde{R}_{e,{\rm max}}}
{\tilde{R}_e^2}\left[\tilde{R}_{e,{\rm max}}^{m+1}-\tilde{R}_e^{m+1}
+4(m+1)\Gamma_t^2\left(\tilde{R}_{e,{\rm max}}-1\right)\right]\right.
\\ \nonumber
& \left. \quad\quad\quad\quad\quad\quad\quad\quad
\quad\quad\quad\quad\quad\quad\quad\quad
+(m+1)\tilde{R}_e^{m-1}
\left(\tilde{R}_{e,{\rm max}}-\tilde{R}_e\right)\right\}
\\ \label{dmu_e_dRe2} 
&=  
\frac{\left[f_m(\hrt)-\tilde{R}_e^{m+1}
+(m+1)\tilde{R}_e^{m+1}\left(1-\tilde{R}_e\right)\right]}{2(m+1)\Gamma_t^2\tilde{R}_e^2} \ ,
\\ \nonumber
\\ \nonumber
\frac{d\mu_r}{d\tilde{R}_e} &= \frac{(m+1)\tilde{R}_e^{m+1}
\left(\tilde{R}_{\rm sh}-\tilde{R}_e\right) - \tilde{R}_{\rm sh}
\left[\tilde{R}_{\rm sh}^{m+1} - \tilde{R}_e^{m+1} 
+ 2(m+1)\Gamma_t^2\left(\tilde{R}_{\rm sh}-1\right)\right]}
{2(m+1)\Gamma_t^2\left(\tilde{R}_{\rm sh}-\tilde{R}_e\right)^2}
\\ \label{dmu_r_dRe2} 
&\approx  
\frac{\left[(m+1)\tilde{R}_e^{m+1}\left(1-\tilde{R}_e\right)
-f_m(\hrt)+\tilde{R}_e^{m+1}
\right]}{2(m+1)\Gamma_t^2\left(1-\tilde{R}_e\right)^2} \ .
\end{align}
We note that, as expected, $\mu_e(\tilde{R}_{e,{\rm max}}) = 1$, since
$\left(1-\tilde{R}_e\right) = \left(\tilde{R}_{e,{\rm
max}}-\tilde{R}_e\right) + {\cal O}\left(\Gamma_t^{-2}\right)$ while
$d\mu_e/d\tilde{R}_e > 0$.
The Doppler factor is given by
\begin{equation}\label{delta2}
\delta \approx \frac{2\Gamma}{1+\Gamma^2(\theta_e+\theta_r)^2}
= \frac{2(m+1)\Gamma_t\tilde{R}_e^{(m+2)/2}\left(1-\tilde{R}_e\right)}
{(m+1)\tilde{R}_e^{m+1}\left(1-\tilde{R}_e\right)
+f_m(\hrt)-\tilde{R}_e^{m+1}}\ ,
\end{equation}
where we have used eqs.~(\ref{mu_e2}) and (\ref{mu_r2}) as well as
$\Gamma^2 = \Gamma_t^2\tilde{R}_e^{-m} \gg 1$ and $\theta_e+\theta_r
\ll 1$.  Finally, $\tilde{r} \approx 1-\tilde{R}_e$, and thus
\begin{equation}\label{eps2}
\frac{\delta^{\alpha+2}}{\tilde{r}_e^2}\cdot\frac{d\mu_e}{d\tre}
= \Gamma_t^\alpha\,\frac{\bar{\delta}^{\alpha+2}}{\tilde{r}_e^2}
\cdot\frac{d\bar{\mu}_e}{d\tre} 
\approx
\frac{2(2\Gamma_t)^{\alpha}(m+1)^{1+\alpha}
\tre^{\alpha+\frac{m}{2}(2+\alpha)}(1-\tre)^{\alpha}}
{\left[(m+1)\tre^{m+1}\left(1-\tilde{R}_e\right) + 
f_m(\hrt)-\tre^{m+1}\right]^{1+\alpha}}\ .
\end{equation}

\subsubsection{Case 2: Test photon at the shell, 
$R_t = R_{\rm sh}(t_t)$}

This is a limiting case between case 1 and case 3, when the test photon
is located on the shell:
$t_t = t_{\rm sh}(\tilde{R}_{e,{\rm max}})$,
$\tilde{r}_{\rm min} = 0$, and $\tilde{R}_{e,{\rm max}} = 1$, i.e.
\begin{equation}
R_{e,{\rm max}} = R_{\rm sh}(t_t) = R_t\quad,\quad
\tilde{R}_{e,{\rm max}} = \tilde{R}_{\rm sh}(t_t) = 1\ .
\end{equation}
This means that the last emitted photons that still reach the point
$(R_t,t_t)$ are emitted at that same point in space and time, i.e. the
equal arrival time surface ends at that point. Therefore,
\begin{align}
(\Gamma_t\theta_e)^2  &=  2\Gamma_t^2(1-\mu_e) 
= \frac{\left(1-\tilde{R}_e\right)
\left(1-\tilde{R}_e^{m+1}\right)}{(m+1)\tilde{R}_e}\ ,
\\ \nonumber
\\
(\Gamma_t\theta_r)^2  &=  2\Gamma_t^2(1-\mu_r) 
= \frac{\tilde{R}_e\left(1-\tilde{R}_e^{m+1}\right)
}{(m+1)\left(1-\tilde{R}_e\right)}\ ,
\\ \nonumber
\\
\frac{d\mu_e}{d\tilde{R}_e} &= 
\frac{m\tilde{R}_e^{m+1}\left(1-\tilde{R}_e\right)
+\left(1-\tilde{R}_e^{m+2}\right)}
{2(m+1)\Gamma^2_t\tilde{R}_e^2}\ ,
\\ \nonumber
\\
\frac{d\mu_r}{d\tilde{R}_e} &= 
-\frac{\left(1-\tilde{R}_e^{m+2}\right)-(m+2)\tilde{R}_e^{m+1}
\left(1-\tilde{R}_e\right)}
{2(m+1)\Gamma_t^2\left(1-\tilde{R}_e\right)^2}\ .
\end{align}
In the limit where $\tilde{R}_e \approx 1$ (i.e. $1-\tilde{R}_e \ll
1$) we have:
\begin{align}
&\left(\Gamma_t\theta_e\right)^2  \approx  
\left(1-\tilde{R}_e\right)^2 \ ,\ \quad
\left(\Gamma_t\theta_r\right)^2  \approx  
1 -\frac{m+2}{2}\left(1-\tilde{R}_e\right)\ ,
&\frac{d\mu_e}{d\tilde{R}} \approx \frac{1-\tilde{R}_e}
{\Gamma^2_t} \approx \frac{\theta_e}{\Gamma_t} \ ,\ \quad
\frac{d\mu_r}{d\tilde{R}} \approx -\frac{(m+2)}{4\Gamma_t^2}\ ,\ \quad
\frac{d\mu_e}{d\mu_r} \approx -\frac{4}{(m+2)}\left(1-\tilde{R}_e\right)\ .
\end{align}
In this limit $\tilde{r} \approx 1-\tilde{R}_e$, which implies
[see Eq.~(\ref{photon_field})] that
$I_\varepsilon \propto |d\mu_e/d\mu_r|/\tilde{r}^2
\propto (1-\tilde{R}_e)^{-1}$,
i.e. the specific intensity diverges at the angle $\theta_r =
\theta_{r,{\rm max}} = 1/\Gamma_t$, and vanishes above this
angle. This can be understood as follows. In this limit $\tilde{r} \ll
1$, i.e. $r \ll R_t = R_{\rm sh}(t_t)$ and the curvature of the shock
front becomes unimportant, so that in order for a photon to reach the
point $(R_t,t_t)$ together with the shock front it must propagate
along the shock front, which corresponds locally to an angle of
$1/\Gamma$ (or more generally $\cos\theta = \beta$) from the normal to
the shock front, i.e. the radial direction in our case.

\subsubsection{Case 3: Test Photon Ahead of the Shell, 
$R_t > R_{\rm sh}(t_t)$}

With the causality condition $R_{e,{\rm max}} < R_{\rm sh}(t_t)$ and
Eq.~(\ref{R0_order}), we now have:
\begin{align}\label{R_order3}
 &R_0 \leq R_{e,{\rm max}} < R_{\rm sh}(t_t) < R_t < ct_t - 
\frac{R_t\tilde{R}_0^{m+1}}{2(m+1)\Gamma_t^2},\\
\Longleftrightarrow &\tilde{R}_0 \leq 
\tilde{R}_{e,{\rm max}} < \tilde{R}_{\rm sh} < 1
\leq \frac{ct_t}{R_t} - \frac{\tilde{R}_0^{m+1}}{2(m+1)\Gamma_t^2}\ . 
\end{align}
As a result, Eq.~(\ref{remax_cond}) yields:
\begin{align}
1- \tilde{R}_{e,{\rm max}} &=  
\tilde{R}_{\rm sh}-\tilde{R}_{e,{\rm max}} + 
\frac{\left(\tilde{R}_{\rm sh}^{m+1}-\tilde{R}_{e,{\rm max}}^{m+1}\right)}
{2(m+1)\Gamma^2_t} + {\cal O}\left(\Gamma^{-4}\right)\ ,
\\
\Longleftrightarrow\ \  1-\tilde{R}_{\rm sh} &= \frac{
\tilde{R}_{\rm sh}^{m+1}-\tilde{R}_{e,{\rm max}}^{m+1}}{2(m+1)\Gamma_t^2} 
\leq \frac{\tilde{R}_{\rm sh}^{m+1}-\tilde{R}_0^{m+1}}{2(m+1)\Gamma_t^2}
< \frac{\tilde{R}_{\rm sh}^{m+1}}{2(m+1)\Gamma_t^2}
< \frac{1}{2(m+1)\Gamma_t^2},
\\ \nonumber
\text{ and }\quad \tilde{R}_{e,{\rm max}} &= \tilde{R}_{\rm sh}
\left[1-\frac{2(m+1)\Gamma_t^2}{\tilde{R}_{\rm sh}^{m+1}}
(1-\trsh)\right]^{1/(m+1)}
\\
& =
\left[2(m+1)\Gamma_t^2\left(\frac{ct_t}{R_t}-1\right)\right]^{1/(m+1)}
= \left[f_m(\hrt)\right]^{1/(m+1)}\ .
\end{align}

Taking these results into account, we now derive the relevant
expressions from eqs.~(\ref{mu_e}$\,-\,$\ref{dmu_r_dR}). The leading terms
for $1-\mu_e$, $1-\mu_r$, and their derivatives with respect to
$\tilde{R}_e$ are all of the order of ${\cal
O}\left(\Gamma_t^{-2}\right)$.  Thus, we obtain:

\begin{align}\nonumber
(\Gamma_t\theta_e)^2  = 2\Gamma_t^2(1-\mu_e) &= 
\frac{\left(1-\tilde{R}_e\right)
\left(\tilde{R}_{e,{\rm max}}^{m+1}-\tilde{R}_e^{m+1}\right)}
{(m+1)\tilde{R}_e} + {\cal O}\left(\Gamma_t^{-2}\right)
\\ \label{mu_e3}
&=
\frac{\left(1-\tilde{R}_e\right)\left[f_m(\hrt)-\tilde{R}_e^{m+1}\right]}
{(m+1)\tilde{R}_e} + {\cal O}\left(\Gamma_t^{-2}\right)\ ,
\\ \nonumber
(\Gamma_t\theta_r)^2  = 2\Gamma_t^2(1-\mu_r) &= \frac{\tilde{R}_e
\left(\tilde{R}_{e,{\rm max}}^{m+1}-\tilde{R}_e^{m+1}\right)}
{(m+1)\left(1-\tilde{R}_e\right)} + {\cal O}\left(\Gamma_t^{-2}\right)
\\ \label{mu_r3}
&= \frac{\tilde{R}_e\left[f_m(\hrt)-\tilde{R}_e^{m+1}\right]}
{(m+1)\left(1-\tilde{R}_e\right)}+{\cal O}\left(\Gamma_t^{-2}\right)\ ,
\\ \nonumber
\end{align}
and for the derivatives
\begin{align}
\frac{d\mu_e}{d\tilde{R}_e} &= 
\frac{(m+1)\tilde{R}_e^{m+1}\left(1-\tilde{R}_e\right)
+\left[f_m(\hrt)-\tilde{R}_e^{m+1}\right]}
{2(m+1)\Gamma_t^2\tilde{R}_e^2}
\ + \ {\cal O}\left(\Gamma_t^{-4}\right) \ ,
\\ \nonumber
\\ \label{dmu_r3_dR}
\frac{d\mu_r}{d\tilde{R}_e} &=
\frac{(m+1)\tilde{R}_e^{m+1}\left(1-\tilde{R}_e\right)
-\left[f_m(\hrt)-\tilde{R}_e^{m+1}\right]}
{2(m+1)\Gamma_t^2\left(1-\tilde{R}_e\right)^2}
\ + \ {\cal O}\left(\Gamma_t^{-4}\right) \ .
\end{align}

The Doppler factor is given by
\begin{equation}\label{delta3}
\delta \approx \frac{2\Gamma}{1+\Gamma^2(\theta_e+\theta_r)^2}
= \frac{2(m+1)\Gamma_t\tilde{R}_e^{(m+2)/2}\left(1-\tilde{R}_e\right)}
{(m+1)\tilde{R}_e^{m+1}\left(1-\tilde{R}_e\right)
+\left[f_m(\hrt)-\tilde{R}_e^{m+1}\right]}\ ,
\end{equation}
where we have used eqs.~(\ref{mu_e3}) and (\ref{mu_r3}) as well as
$\Gamma^2 = \Gamma_t^2\tilde{R}_e^{-m} \gg 1$ and $\theta_e+\theta_r
\ll 1$.  Finally, $\tilde{r} \approx 1-\tilde{R}_e$, and thus
\begin{align}\nonumber
\frac{\delta^{\alpha+2}}{\tilde{r}_e^2}\cdot\frac{d\mu_e}{d\tre} 
= \Gamma_t^\alpha\,\frac{\bar{\delta}^{\alpha+2}}{\tilde{r}_e^2}
\cdot\frac{d\bar{\mu}_e}{d\tre} 
&\approx \frac{2(2\Gamma_t)^{\alpha}}
{(1-\tre)\tre^{1+m\alpha/2}}
\left[1+\frac{f_m(\hrt)-\tre^{m+1}}
{(m+1)\tre^{m+1}(1-\tre)}\right]^{-(1+\alpha)} 
\\ \nonumber
\\ \label{x_3}
&\approx \frac{2(2\Gamma_t)^{\alpha}(m+1)^{1+\alpha}
\tre^{\alpha+m(2+\alpha)/2}(1-\tre)^{\alpha}}
{\left[(m+1)\tre^{m+1}\left(1-\tilde{R}_e\right)
+f_m(\hrt)
-\tre^{m+1}\right]^{1+\alpha}}\ .
\end{align}
We note that the expressions above for case 3 are identical to those
for case 1 -- Eq.~(\ref{eps2}).  While $dn_i/d\ei d\Omega_r$ diverges
as $|\tilde{R}_e-\tilde{R}_e(\theta_{r,{\rm max}})|^{-1} \propto
|\theta_r-\theta_{r,{\rm max}}|^{-1/2}$ at $\theta_{r,{\rm max}}$,
$dn_i/d\ei = \int d\Omega_r(dn_i/d\ei d\Omega_r) \approx
2\pi\int\theta_r d\theta_r(dn_i/d\ei d\Omega_r)$ remains finite
(i.e. both the energy density and the energy flux of the radiation
field remain finite). This has been noticed in the context of the
diverging surface brightness of the afterglow image at its outer edge,
when the emission comes from an infinitely thin shell
\citep{Sari98,GL01}. In that context, it has also been shown
\citep{Waxman97,GPS99a,GPS99b,GL01} that when the emission comes from
a shell of finite width, the surface brightness (i.e. the specific
intensity $I_\varepsilon$) does not diverge.

\subsection{Putting it all together}

Analytical expressions for our model have now been fully derived, and
are reported for convenience here.  The scaled spectral flux density,
Eq.~(\ref{F2}) is rewritten as:
\begin{equation}
\frac{F_\varepsilon(T)}{F_{\varepsilon,0}} =
\left(\frac{T}{T_0}\right)^{\frac{2b-m\alpha}{2(m+1)}}
\int_{y_{\rm min}}^{y_{\rm max}} dy 
\left(\frac{m+1}{m+y^{-m-1}}\right)^{1+\alpha}
y^{b-1-m\alpha/2}\exp\left[-\tgg\left(
y,\varepsilon_t,\frac{\Delta R}{R_0},
\frac{T}{T_0}\right)\right]\ ,
\end{equation}
where $\varepsilon_t = (1+z)\varepsilon$, $y_{\rm min} =
\min[1,R_0/R_L(T)]$ and $y_{\rm max} = \min[1,(R_0+\Delta R)/R_L(T)]$,
while the flux normalization is given by
\begin{align}
F_{\varepsilon,0} &= 2\Gamma_0 L'_{(1+z)\varepsilon/2\Gamma_0}(R_0)
\frac{(1+z)}{4\pi d_L^2}
= \frac{2^\alpha L_0\varepsilon^{1-\alpha}(1+z)^{2-\alpha}}{4\pi
  d_L^2}\ ,
\\ \label{F_0}
F_0 &\equiv \varepsilon^{\alpha-1} F_{\varepsilon,0} = (\varepsilon
F_{\varepsilon,0})|_{\varepsilon=1} =
\frac{L_0}{\pi d_L^2}
\left(\frac{1+z}{2}\right)^{2-\alpha} 
\\  \nonumber
&= 
7.6\times 10^{-6}\left(\frac{1+z}{2}\right)^{2-\alpha}
L_{0,52}d_{L,28}^{-2}\;{\rm erg\;cm^{-2}\;s^{-1}}\ ,
\end{align}
where $d_L = 10^{28}d_{L,28}\;$cm, and may be used in order to infer
the value of $L_0$ from the observed flux level. The optical depth in
the integrand above is:
\begin{align}
\tgg(\varepsilon_t,\theta_{t,0},R_{t,0}) &=
\tau_\star\;\varepsilon_t^{\alpha-1}\;\hrz^{1-b-m\alpha/2} {\cal F}(x)\ ,
\end{align}
where $\hrz =y^{-1}(T/T_0)^{-1/(m+1)}$ and $x=(y^{-(m+1)}-1)/(m+1)$.
The function ${\cal F}$ is the following double integral:

\begin{equation}\label{2int}
{\cal F}(x) = \int_1^{\hat{R}_{2}} d\hrt
\int_{\hrz/\hrt}^{\tilde{R}_{e,2}} d\tre \ {\cal I}(\hrt,\tre) 
 \; + \; \int_{\hat{R}_{2}}^\infty d\hrt
\int_{\hrz/\hrt}^{\tilde{R}_{e,3}} d\tre \ {\cal I}(\hrt,\tre)\ . 
\end{equation}
The two integrals above correspond to cases 1 and 3 respectively, as
discussed in \S\ \ref{sec:cases}.  When $x>1$, the test photon
lags behind the shell and $f_m(\hrt)>1$, with
$f_m(\hrt)\equiv[1+x(m+1)(1-\hrt^{\,-1})]/\hrt^{m+1}$; $\hat{R}_{2}$
is then defined by the implicit equation $f_m(\hat{R}_{2}) \equiv 1$,
and $\tilde{R}_{e,2} \equiv {\rm min}
[(\hrz+\Delta\hat{R})/\hrt,\;1]$. For $m = 0$, $\hat{R}_{2} =
\max(1,x)$, while for $m = 1$, $\hat{R}_{2} =
\max[1,(\sqrt{1+8x}-1)/2]$. The test photon eventually overtakes the
shell at radius $R_{2}$, and will travel ahead of the shell ever
after, which corresponds to the second integral, where
$\tilde{R}_{e,3}\equiv {\rm min}[(\hrz+\Delta\hat{R})/\hrt,\;
f_m(\hrt)^{1/(m+1)}]$.  When $x \leq 1$, only the second integral
contributes, as the photon is emitted on the shell and immediately
travels ahead of it. Thus, $\hat{R}_{2} = 1$ so that the first
integral vanishes.  Note that, for all practical purposes,
Eq.~(\ref{2int}) can be cast into a single integral:
\begin{equation}\label{1int}
{\cal F}(x) = \int_1^{\infty} d\hrt
\int_{\hrz/\hrt}^{\tilde{R}_{e,M}} d\tre \ {\cal I}(\hrt,\tre) \ ,
\end{equation}
where $\tilde{R}_{e,M}=\tilde{R}_{e,2}$ when $x>1$ and $f_m(\hrt)>1$,
and $\tilde{R}_{e,M}=\tilde{R}_{e,3}$ in all other cases,
i.e. $\tilde{R}_{e,M} = \min[(\hrz+\Delta\hat{R})/\hrt,\;
f_m(\hrt)^{1/(m+1)},\;1]$.
\noindent Finally, the integrand is equal to:
\begin{equation}\label{I}
{\cal I}(\hrt,\tre) = \frac{1}{\hat{R}_t^2} 
\left(\frac{\delta^3}{\tilde{r}^2}\cdot\frac{d\mu_e}{d\tre}\right) \; 
\times\int_{\zeta_-}^{\zeta_+} \frac{\zeta d\zeta}
{\sqrt{(\zeta_+-\zeta)(\zeta-\zeta_-)}} 
 \int_1^{+\infty} \frac{d\chi}{\chi}\frac{\sigma^\star(\chi)}{\sigma_T}
L'_{\chi^2/\varepsilon_t\zeta\delta}(\tre) \ .
\end{equation}
Specializing to $\displaystyle{L'_{\varepsilon'_i}(R) =
L'\ei'^{1-\alpha}\times (\tre\hrt/\hrz)^b}$, Eq.~(\ref{1int}) becomes:
\begin{equation}
{\cal F}(x) = \int_1^{\infty}d\hrt\,\hrt^{b-2+m\alpha/2}
\int_{\hrz/\hrt}^{\tilde{R}_{e,M}} d\tre \ \frac{\bar{\delta}^{\,2+\alpha}}
 {\tilde{r}^{\,2}}\cdot\frac{d\bar{\mu}_e}{d\tre} \tre^b\; 
 \bar{\zeta}_-^{\,\alpha} H_\alpha(\zeta) \ ,
\end{equation}
where the integrands are further expressed as:
\begin{align}
\frac{\bar{\delta}^{2+\alpha}}{\tilde{r}^2}
\cdot\frac{d\bar{\mu}_e}{d\tre}\tre^b & 
\approx \nonumber \frac{[2(m+1)]^{1+\alpha}
\tre^{b+\alpha+\frac{m}{2}(2+\alpha)}(1-\tre)^{\alpha}}
{\left[(m+1)\tre^{m+1}\left(1-\tilde{R}_e\right)
+f_m(\hrt) -\tre^{m+1}\right]^{1+\alpha}}\ ,\\
\bar{\zeta}_- = \frac{(\Gamma_t\theta_r -\Gamma_t\theta_t)^2}{4}\ , \quad
\zeta &= \frac{4(\Gamma_t\theta_r)(\Gamma_t\theta_t)}
{(\Gamma_t\theta_r -\Gamma_t\theta_t)^2}\ ,\quad
H_\alpha(\zeta) = {}_2F_1(-\alpha,0.5;1;-\zeta)\ ,\\
(\Gamma_t\theta_r)^2=
\frac{\tre\left[f_m(\hrt)-\tre^{m+1}\right]}{(m+1)(1-\tre)}&
\ ,\quad (\Gamma_t\theta_t)^2=\frac{x}{\hrt^{m+2}}\ .\label{Hvar}
\end{align}
This concludes the set of general equations that have been obtained. 
For reference, the hypergeometric expressions for $\alpha=1,2,3$ respectively
read
\begin{align}
\bar{\zeta}_-^1H_1(\zeta) &= \bar{\zeta}_-\left(1+\frac{\zeta}{2}\right)=\frac{1}{4}\left[(\Gamma_t\theta_r)^2 + (\Gamma_t\theta_t)^2\right]\ ,
\label{H_alpha1}\\
\bar{\zeta}_-^2H_2(\zeta) &= \bar{\zeta}_-^2\left(1+\zeta+\frac{3}{8}\zeta^2\right)
=\frac{(\Gamma_t\theta_t)^4+(\Gamma_t\theta_r)^4}{16} + 
\frac{(\Gamma_t\theta_t)^2(\Gamma_t\theta_r)^2}{4}\ ,
\label{H_alpha2}\\
\bar{\zeta}_-^3H_3(\zeta) &= \bar{\zeta}_-^3\left(1+\frac{3}{2}\zeta+\frac{9}{8}\zeta^2+\frac{5}{16}\zeta^3\right)\nonumber\\
&=\frac{1}{64}\left[(\Gamma_t\theta_r)^6 + 
9(\Gamma_t\theta_r)^4(\Gamma_t\theta_t)^2 + 
9(\Gamma_t\theta_r)^2(\Gamma_t\theta_t)^4 +
(\Gamma_t\theta_t)^6\right]\ .
\label{H_alpha3}
\end{align}
For our fiducial case, $\alpha=2$, we also explicitly write the relevant expressions : 
\begin{align}
\tau_0(\varepsilon_t,R_{t,0}) &= \tau_\star\; \varepsilon_t\;\hrz^{-m-b+1}\ , 
\quad 
\tau_\star = 0.402\left(\frac{\Gamma_0}{100}\right)^{-4}
\frac{L_{0,52}}{R_{0,13}}\ ,
\\
{\cal F}(x) &= \int_1^\infty \hrt^{b+m-2}\; d\hrt 
\int d\tre\; \frac{\bar{\delta}^{4}}{\tilde{r}^2}   
\frac{d\bar{\mu}_e}{d\tre}\tre^b\; \bar{\zeta}_-^{2}
H_2(\zeta)\ ,\\
\frac{\bar{\delta}^{4}}{\tilde{r}^2}  	
\frac{d\bar{\mu}_e}{d\tre}\; &= \frac{[2(m+1)]^{3}\tre^{2+2m}(1-\tre)^{2}}
{\left[(m+1)\tre^{m+1}\left(1-\tilde{R}_e\right)
+f_m(\hrt)-\tre^{m+1}\right]^{3}} \ ,
\\
\bar{\zeta}_-^2H_2(\zeta) &=\frac{x^2}{16\hrt^{2(m+2)}}+\frac{x}{4\hrt^{m+2}}
\frac{\tre\left[f_m(\hrt)-\tre^{m+1}\right]}{(m+1)(1-\tre)}+\frac{1}{16}
\frac{\tre^2\left[f_m(\hrt)-\tre^{m+1}\right]^2}{(m+1)^2(1-\tre)^2} \ .
\end{align}

\section{Analytic Scalings of the Flux and Optical Depth}
\label{sec:analytic}

Before showing our results for the lightcurves and spectra, it is
useful to first analytically derive some of the relevant scaling laws
(from the equations obtained in the preceding sections), and discuss
the qualitative behavior of the system in different regimes. It is
convenient to define a normalized time $\bar{T} \equiv (T/T_0)-1$, 
which is zero when the first photon from $R_{t,0} = R_0$
and $\theta_{t,0} = 0$ reaches the observer, and is $\sim 1$ about a
dynamical time later, when the system starts to approach a
quasi-steady state. It is also useful to define the time $T_f =
T_0(1+\Delta R/R_0)^{m+1}$, where $R_L(T_f) \equiv R_0 + \Delta R$,
when the lack of emission from outside the outer edge of the emitting
region ($R > R_0+\Delta R$) starts being noticed by the observer, and
the corresponding normalized time
\begin{equation}\label{T_bar_f}
\bar{T}_f = \frac{T_f}{T_0}-1 = 
\left(1+\frac{\Delta R}{R_0}\right)^{m+1} - 1
\ \approx \left\{\begin{matrix}
(m+1)\Delta R/R_0 \ll 1 & \quad (\Delta R \ll R_0)\ , \cr\cr
(\Delta R/R_0)^{m+1} \gg 1 & \quad (\Delta R \gg R_0)\ .
\end{matrix}\right.
\end{equation}

Note that at $T \leq T_f$ the outer boundary of the emission region
does not affect either the emission, since the outer edge of the
EATS-I is still fully within the emission region, or the opacity of
the emitted photons, since the maximal radius of the EATS-II
($R_{e,{\rm max}}$) at all points along the trajectory of any photon
is always smaller than that of the EATS-I [$R_L(T)$]: 
$R_{e,{\rm max}}(T,y,R_t) < R_L(T)$. 
As shown in Fig.~\ref{fig:EATS12}, the two radii become nearly equal for
$R_t \gg R_L(T)$. In fact, for $R_t \gg R_L(T)$, not only does
$R_{e,{\rm max}}(T,y,R_t)$ approach $R_L(T)$, but the EATS-II approaches
the EATS-I (the two must become identical when the test photon reaches the
observer, which corresponds to $R_t \to \infty$ for a distant
observer, at ``infinity''). This immediately implies that for $T \leq
T_f$ the observed flux and the opacity along the trajectory of all
photons (which reach the observer at time $T$) are independent of
$\Delta R$. Thus, in order to calculate the light curves for a family
of model parameters that differ only in their $\Delta R$ values, it is
sufficient to calculate the observed flux and opacity for $\Delta R
\to \infty$ and use them for $T \leq T_f$, and do the full calculation
for each specific value of $\Delta R$ only for $T > T_f$.

The temporal scaling of the unattenuated flux, at sufficiently low
photon energies $\varepsilon$, can be understood as follows. For $1
\gg \bar{T} < \bar{T}_f$, $y_{\rm min} = R_0/R_L(T) =
(T/T_0)^{-1/(m+1)} = (1+\bar{T})^{-1/(m+1)}$ and $y_{\rm max} = 1$, so
that $\Delta y = y_{\rm max} - y_{\rm min} \approx \bar{T}/(m+1)$
while $y \approx 1$ and the integrand in Eq.~(\ref{F2}) is also
$\approx 1$, implying that $F_\varepsilon \propto \bar{T}$. For
$\bar{T}_f < \bar{T} \ll 1$ the emission is from $R \approx R_0$ from
angles $\theta$ which satisfy $(\Gamma_0\theta)^2 \approx
\bar{T}/(m+1) \ll 1$, while the Doppler factor at this stage is almost
constant, $\delta \approx 2\Gamma_0/[1+(\Gamma_0\theta)^2] \approx
2\Gamma_0/[1+\bar{T}/(m+1)] \approx 2\Gamma_0$, and therefore the flux
is approximately constant in time, $F_\varepsilon \propto
\delta^{-1-\alpha} \propto \bar{T}^0$. For $\bar{T}_f \ll 1 \ll T$,
$\delta \approx 2\Gamma_0/(\Gamma_0\theta)^2 \propto \bar{T}^{-1}$ and
$F_\varepsilon \propto \bar{T}^{-1-\alpha}$.
For $1 \ll \bar{T} < \bar{T}_f$, $y_{\rm
min} = R_0/R_L(T) = (1+\bar{T})^{-1/(m+1)} \approx \bar{T}^{-1/(m+1)}
\ll 1$ and $y_{\rm max} = 1$, so that the integral over $y$ in
Eq.~(\ref{F2}) approaches a constant (corresponding to its value for
$\int_0^1dy$), and $F_\epsilon \propto
\bar{T}^{(2b-m\alpha)/[2(m+1)]}$. Finally, for $\bar{T} \gg
\bar{T}_f$, the emission is dominated by $R \sim R_0+\Delta R$ and
angles $\theta_{t,0} \gg 1/\Gamma_{t,0}$ (i.e. $x \gg 1$) and we
obtain the familiar result for ``high latitude'' emission
\citep{KP00}, $F_\epsilon \propto \bar{T}^{-1-\alpha}$.  Altogether,
\begin{equation}\label{F_eps_T}
F_{\varepsilon<\varepsilon_1(\bar{T})} \propto
\left\{\begin{matrix}
\bar{T} & \quad 
(1 \gg \bar{T}<\bar{T}_f)\ , \cr\cr
\bar{T}^{0} & \quad 
(\bar{T}_f < \bar{T} \ll 1)\ , \cr\cr
\bar{T}^{(2b-m\alpha)/[2(m+1)]} & \quad 
(1 \ll \bar{T} < \bar{T}_f)\ , \cr\cr
\bar{T}^{-1-\alpha} & \quad (\bar{T} \gg \max[1,\bar{T}_f])\ .
\end{matrix}\right.
\end{equation}

\begin{figure}[ht]
\centering
\includegraphics[scale=0.6]{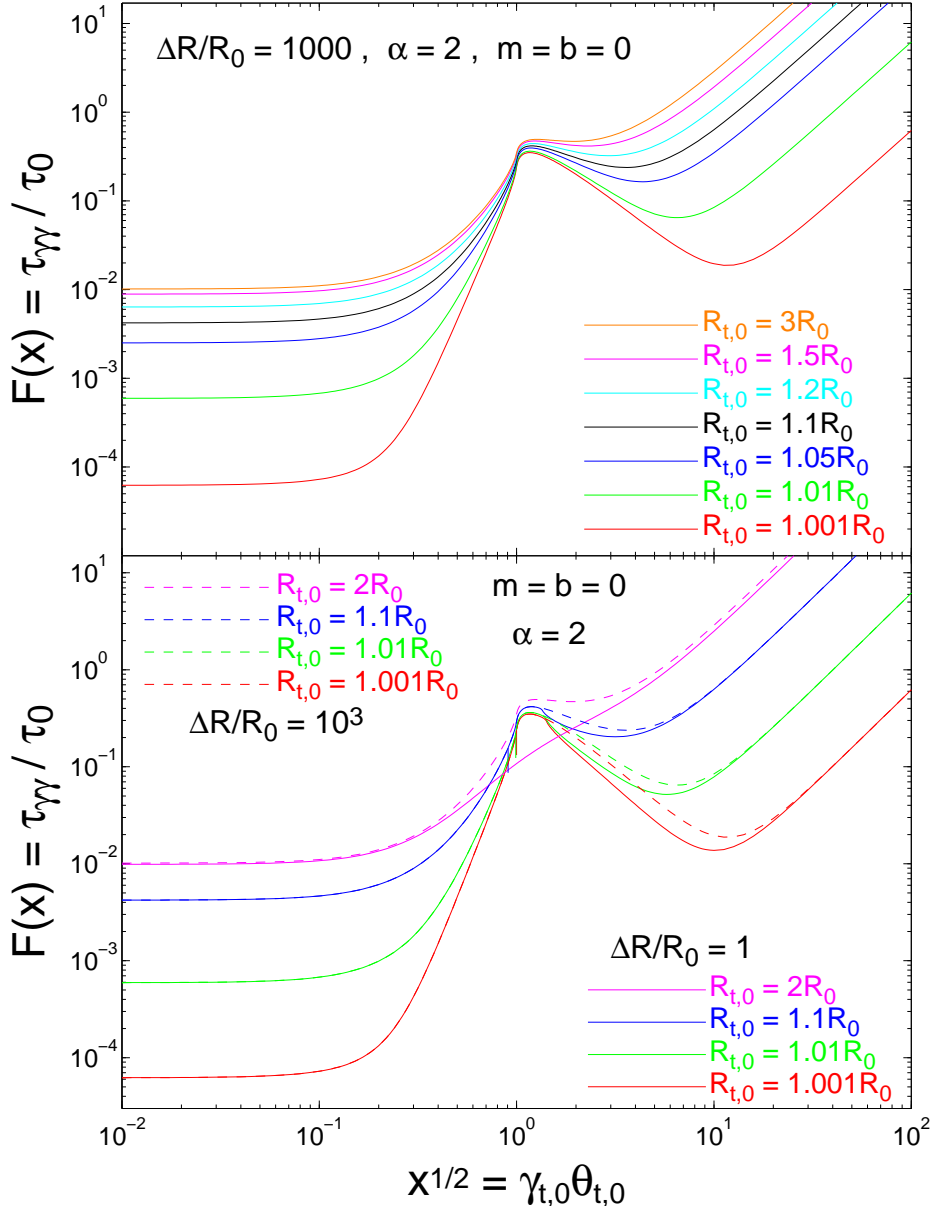}
\caption{The normalized optical depth ${\cal F}(x) = \tgg/\tau_0$, as
a function of the renormalized emission angle, $x^{1/2} =
\gamma_{t,0}\theta_{t,0}$, for several different emission radii
$R_{t,0}$. The {\it upper panel} is for $\Delta R/R_0 = 1000$ while
the {\it lower panel} shows the results for $\Delta R/R_0 = 1$ ({\it
solid lines}) and for $\Delta R/R_0 = 1000$ ({\it dashed lines})
overlaid on each other. The small vertical lines in the {\it lower
panel} indicate the angle that corresponds to $\bar{T} = \bar{T}_f$,
outside of which the contributions to the opacity from $R > R_0+\Delta
R$ for $\Delta R/R_0 = 1$ start being missed (this effect becomes
significant only at somewhat larger angles; see discussion in the
text). For $R_{t,0} = 2R_0 = R_0 + \Delta R$, this corresponds to
$x^{1/2} = \gamma_{t,0}\theta_{t,0} = 0$, which is outside the range
shown in the figure. In both panels the photon index is $\alpha = 2$
while the Lorentz factor and the total luminosity in the comoving
frame are independent of radius ($m = b = 0$).}
\label{fig:tau_th}
\end{figure}

\begin{figure}[ht]
\centering
\includegraphics[scale=0.62]{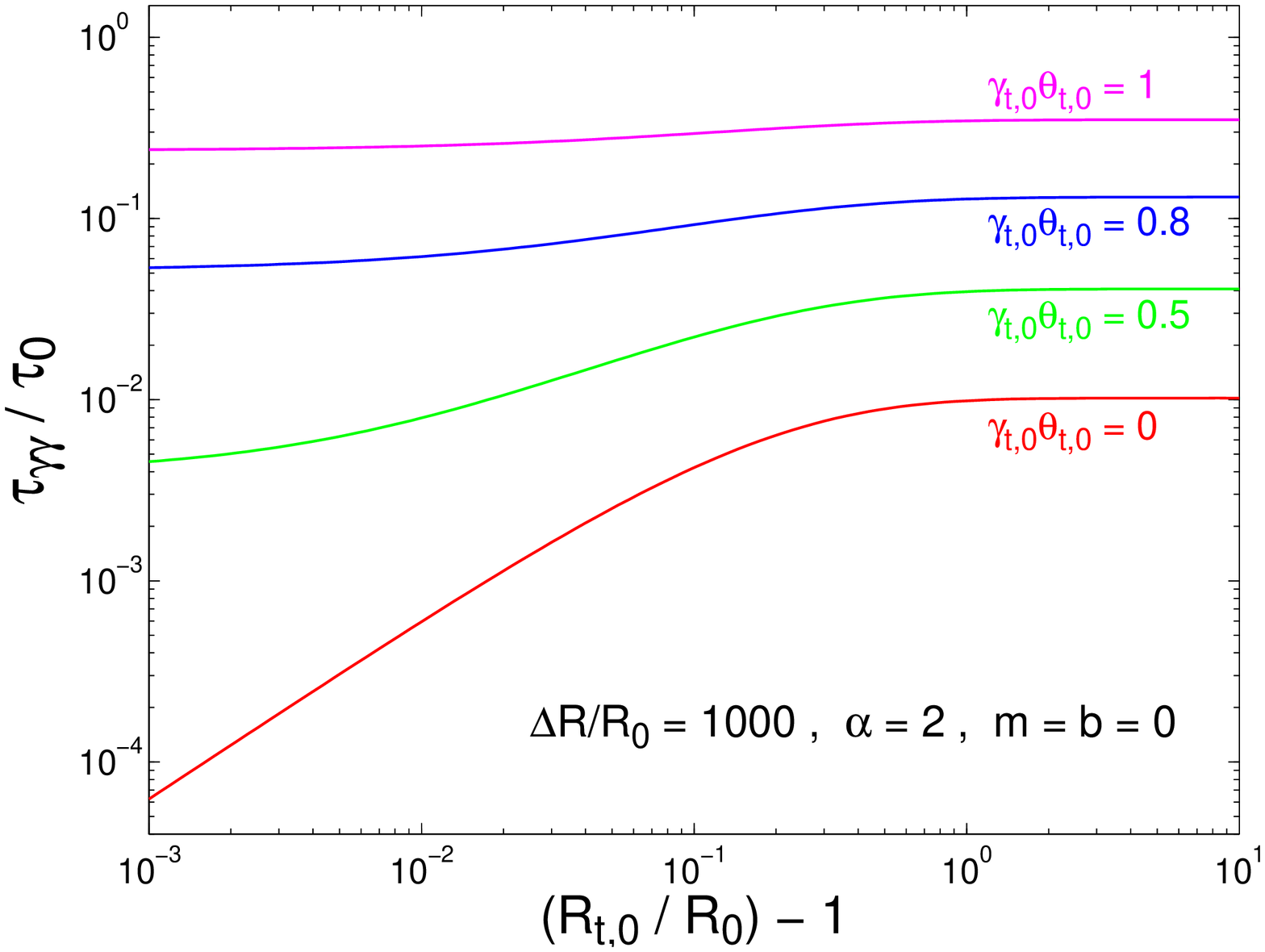}
\vspace{0.5cm}\\
\includegraphics[scale=0.62]{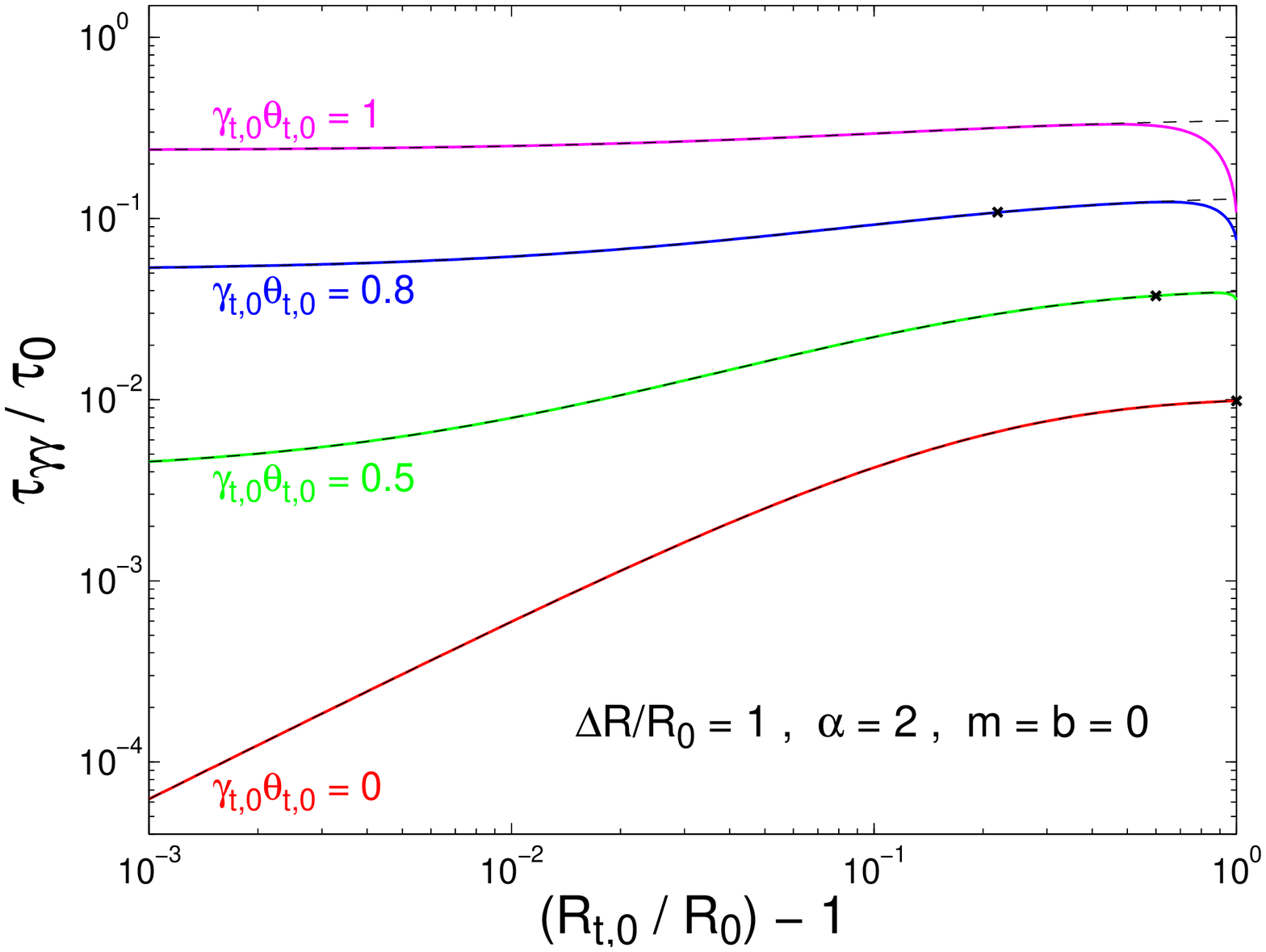}
\caption{The normalized optical depth ${\cal F}(x) = \tgg/\tau_0$, as
a function of the renormalized emission radius, $(R_{t,0}/R_0)-1$, for
different values of the normalized emission angle $x^{1/2} =
\gamma_{t,0}\theta_{t,0}$. The upper panel is for $\Delta R/R_0 =
1000$. The lower panel is for $\Delta R/R_0 = 1$ but also shows the
corresponding result for $\Delta R/R_0 = 1000$ in dashed lines, where
the x-symbols show the value of the emission radius corresponding to
an observed time of $T = T_f$ (for $\gamma_{t,0}\theta_{t,0} = 1$ this
corresponds to $(R_{t,0}/R_0)-1 = 0$ which is outside the range shown
in the figure). Note the deviation near $(R_{t,0}/R_0)-1 \sim 1$ and
see the text for discussion of its origin.}
\label{fig:tau_R}
\end{figure}

Now we move on to discuss the opacity effects in some detail. As can
be seen in Fig.~\ref{fig:tau_th}, at a given emission radius the
optical depth is smallest for small emission angles (i.e. small values
of $x$). There is a local maximum near $x = \gamma_{t,0}\theta_{t,0}
\approx 1$ since for such emission angles the photon is emitted almost
parallel to the shell in the comoving frame, and a relatively large
part of its trajectory (also in the lab frame) is close to the
emitting shell, which enhances the optical depth. For a given
normalized emission angle, $x^{1/2} = \gamma_{t,0}\theta_{t,0}$, the
normalized optical depth increases with emission radius, as can be
seen in Fig.~\ref{fig:tau_R}, where the increase is largest for small
emission angles. The optical depth generally increases with $\Delta
R/R_0$ when all other model parameters are held fixed, due to the
larger range of emission radii which enhances the photon field that
can potentially interact with test photons. However, as pointed out
above, for $T \leq T_f$ the optical depth in this case is independent
of $\Delta R/R_0$. This is demonstrated in the lower panel of
Fig.~\ref{fig:tau_R}, where it can be seen that in practice a
noticeable increase in the optical depth due to the increase in
$\Delta R/R_0$ does not occur immediately after $T_f$ but takes some
time to come into effect. This is since for $0 < T-T_f \ll T_f$ the
added contribution to the opacity from $R > R_0+\Delta R$ for the
smaller $\Delta R$ is very small, since the additional photons can
interact with the test photon only at very large radii ($R_t \gg
R_L(T_f) = R_0+\Delta R$) where the intensity of the photon field is
very small, and at very small angles between the directions of the
photons which are very unfavorable for interaction.

\begin{figure}[ht]
\centering
\includegraphics[scale=0.63]{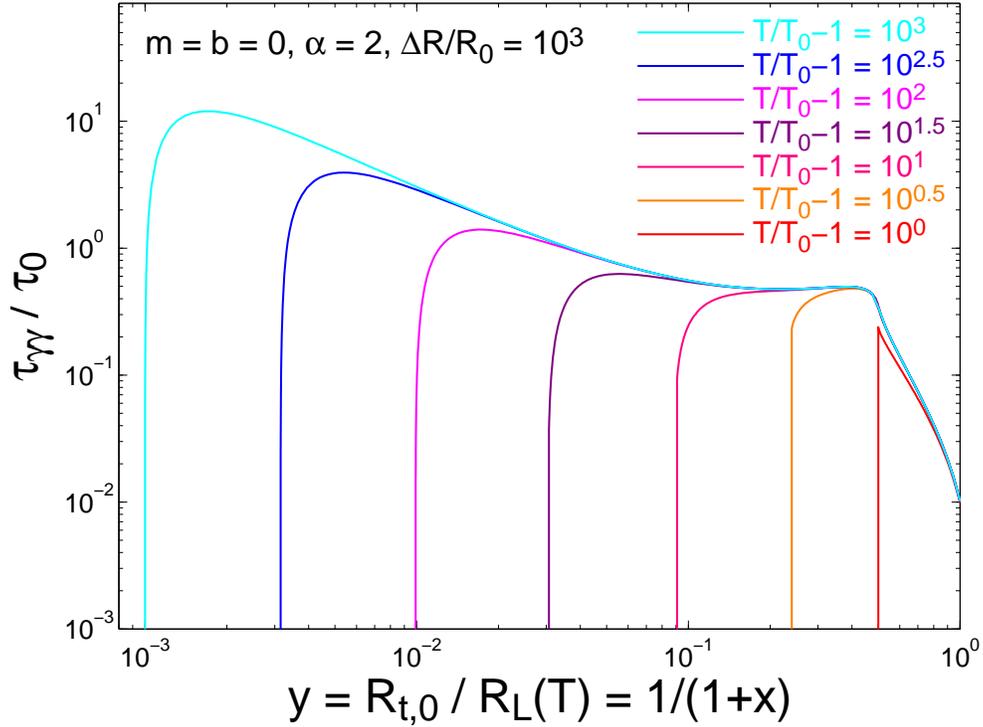}
\vspace{0.5cm}\\
\includegraphics[scale=0.63]{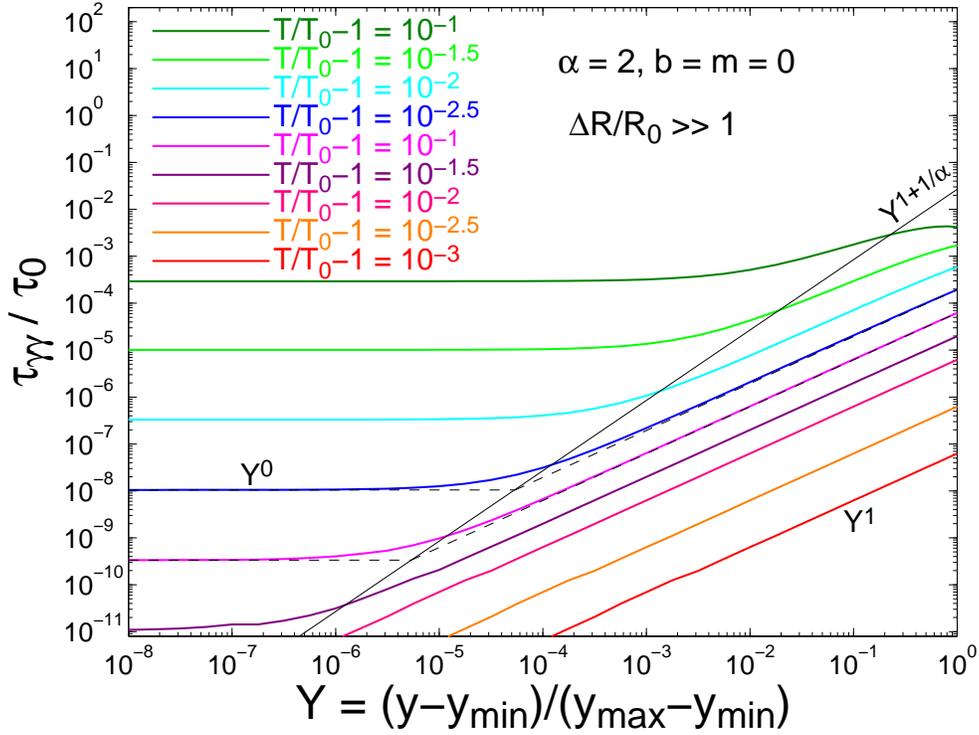}
\caption{The normalized optical depth ${\cal F}(x) = \tgg/\tau_0$,
along the equal arrival time surface of photons to the observer
(EATS-I), for several different values of the normalized time $\bar{T}
= (T/T_0)-1$: the {\it upper panel} shows ${\cal F}(x)$ as a function
of the normalized emission radius $y = R_{t,0}/R_L(T)$ for several
values of $1 \geq \bar{T} < \bar{T}_f$, while the {\it lower panel}
shows ${\cal F}(x)$ as a function of $Y \equiv (y-y_{\rm min})/(y_{\rm
max}-y_{\rm min}) \approx (x_{\rm max}-x)/x_{\rm max}$ for several
values of $1 \ll \bar{T} < \bar{T}_f$.}
\label{fig:tauEATS}
\end{figure}

One would also like to define $\varepsilon_1$ as the photon energy at
which the optical depth becomes unity: $\tgg(\varepsilon_1) \equiv 1$.
However, this definition gives a different value along the
trajectories of different (test) photons, making it hard to define a
unique value for $\varepsilon_1(\bar{T})$, since its value varies
along the EATS-I (see Fig.~\ref{fig:tauEATS}). For $1 \ll \bar{T} <
\bar{T}_f$, ${\cal F}(x)$ becomes independent of $\bar{T}$ and depends
only on $x$ (see {\it upper panel} of Fig.~\ref{fig:tauEATS}). Most of
the contributions to the observed flux come from $x \lesssim 1$, since
for $x \gg 1$ the radiation is strongly beamed away from the
observer. The {\it upper panel} of Fig.~\ref{fig:tauEATS} shows that
${\cal F}(x \lesssim 1)$ varies over a factor of $\sim 50$ for $1 \ll
\bar{T} < \bar{T}_f$, and therefore in this regime it still makes some
sense to define a single typical value of $\varepsilon_1(\bar{T})$,
and derive its scaling. It is good to keep in mind, however, both the
spectral transitions around $\epsilon_1(\bar{T})$ in the instantaneous
spectrum and around $\epsilon_1(\bar{T}_f)$ in the time integrated
spectrum, as well as the transition in the light curve when
$\varepsilon_1(\bar{T})$ sweeps past the observed photon energy
$\varepsilon$, are all expected to be somewhat smoothed due to this
relatively large range of opacity values across the (unresolved)
observed image of the GRB
projected in the sky. According to Eq.~(\ref{tau_0}), $\tgg \propto \tau_0
\propto \et^{\alpha-1} \hrz^{1-b-m\alpha/2} \propto \et^{\alpha-1}
\bar{T}^{-(1-b-m\alpha/2)/(m+1)}$, and therefore $\epsilon_1(1 \ll
\bar{T} < \bar{T}_f) \propto
\bar{T}^{(1-b-m\alpha/2)/[(m+1)(\alpha-1)]}$.

The {\it lower panel} of Fig.~\ref{fig:tauEATS} shows ${\cal F}(x)$ as
a function of $Y \equiv (y-y_{\rm min})/(y_{\rm max}-y_{\rm min})
\approx (x_{\rm max}-x)/x_{\rm max}$ for several values of $1 \gg
\bar{T} < \bar{T}_f$, along the equal arrival time surface of photons
to the observer (EATS-I). In this limit $R_{t,0} \approx R_0$, $y
\approx 1-x$ and $x_{\rm max} \approx \bar{T}/(m+1) \approx [\Gamma_0
R_{\perp,{\rm max}}(\bar{T})/R_0]^2$ where $R_{\perp,{\rm
max}}(\bar{T})$ is the radius of the GRB observed image, projected in
the sky, at a normalized observed time $\bar{T}$. As is shown
analytically in Appendix~\ref{tau_Tbar} and is apparent in the {\it
lower panel} of Fig.~\ref{fig:tauEATS}, in this limit
\begin{equation}
{\cal F}(x = 0, 1 \gg \bar{T} < \bar{T}_f) \propto \bar{T}\ .
\end{equation}
The {\it lower panel} of Fig.~\ref{fig:tauEATS} also shows that
\begin{equation}\label{Y_Ystar}
{\cal F}(x, 1 \gg \bar{T} < \bar{T}_f) \approx
{\cal F}(x = 0, 1 \gg \bar{T} < \bar{T}_f) \times
\left\{\begin{matrix}
Y & \quad 
Y > Y_*(\bar{T})\ , \cr\cr
Y_*(\bar{T}) & \quad 
Y < Y_*(\bar{T})\ ,
\end{matrix}\right.
\end{equation}
where $Y_* \approx (x_{\rm max} - x_*)/x_{\rm max} \propto
\bar{T}^\alpha$ is the value of $Y$ where the scaling of ${\cal F}(x)$
changes from $\propto Y^{0}$ to $\propto Y^{1}$. The corresponding
value of $x$ is $x_*$, and ${\cal F}(x_*) \approx {\cal F}(x_{\rm
max}) \propto \bar{T}^{\alpha+1} \propto Y_*^{(\alpha+1)/\alpha}$.

The scaling of ${\cal F}(x)$ with $Y$ for $1 \gg \bar{T} < \bar{T}_f$
can be understood as follows. For a given emission radius $R_{t,0}$,
the dependence of the optical depth and ${\cal F}(x)$ on the emission
angle $\theta_{t,0}$ is very weak for $x =
(\Gamma_{t,0}\theta_{t,0})^2 \ll 1$, and becomes significant only for
$x \gtrsim 1$ (see Fig.~\ref{fig:tau_th}). Therefore, as $x$ starts
increasing from $x = 0$ at the line of sight, along the EATS-I, ${\cal
F}(x)$ initially varies following its dominant radial dependence. The
latter may be derived from that along the line of sight where ${\cal
F}(x) \propto \bar{T} \propto (R_{t,0}/R_0)-1$, where for a general
value of $x < x_{\rm max} \ll 1$ we have $(R_{t,0}/R_0)-1 \approx
x_{\rm max}-x \propto Y$ and therefore ${\cal F}(x) \propto Y$. As $x$
approaches $x_{\rm max}$, $Y$ approaches zero, until eventually the
optical depth becomes dominated by the small angular dependence on
$\theta_{t,0}$ at a fixed emission radius $R_{t,0}$, and ${\cal F}(x)$
approaches a constant value of ${\cal F}(x_{\rm max})$ which
corresponds to $R_{t,0} = R_0$ and $x = (\Gamma_0\theta_{t,0})^2 =
x_{\rm max} = \bar{T}/(m+1)$. We find numerically that ${\cal
F}(x_{\rm max}) \propto \bar{T}^{\alpha+1} \propto x_{\rm
max}^{\alpha+1} = (\Gamma_0\theta_{t,0})^{2(\alpha+1)}$. This may be
understood as follows, starting from the expression for the optical
depth in Eq.~(\ref{tau_gg_2}). In this regime $\tilde{R}_{e,{\rm max}}
= f_m^{1/(m+1)}$ and for $R_{t,0} = R_0$ we have 
\begin{equation}
\frac{\tilde{R}_{e,{\rm max}}}{R_0}-1 =
\left[1+x(m+1)(1-\hrt^{-1})\right]^{1/(m-1)}-1 \approx 
x\left(1-\hrt^{-1}\right) \ll 1\ .
\end{equation}
This means that the contribution to the local photon field at each
point along the trajectory of the test photon is always from a very
narrow range of radii near $R_0$. This implies that
$r^{-2}|d\mu_r/d\mu_r|$ which appears in Eq.~(\ref{tau_gg_2}) remains
approximately constant, since the geometry of the problem implies
$R_0\theta_e \approx r\theta_r$ so that $r^{-2}|d\mu_r/d\mu_r| \approx
R_0^{-2} = {\rm const}$. At any given point along the test photon
trajectory $\theta_r \leq \theta_t \leq \theta_{t,0}$, simply because
in this regime $\theta_{r,{\rm max}}$ is obtained where EATS-II is
truncated at $R_0$, which must correspond to $\theta_r = \theta_t$ for
a test photons that is emitted at $R_{t,0} = R_0$ (the test photon is
always on its own EATS-I and EATS-II, by definition). This implies
that $\delta \approx 2\Gamma_0 = {\rm const}$ since
$\Gamma(\theta_e+\theta_r) \leq 2\Gamma_0\theta_t \leq
2\Gamma_0\theta_{t,0} = 2 x^{1/2} \ll 1$. Furthermore,
$L'_{\varepsilon'_i}(\tre)$ is approximately constant since $\tre
\approx R_0 = {\rm const}$. Since $\theta_t/\theta_{t,0} \approx
R_{t,0}/R_t = \hrt$, the effective solid angle that contributes to
interaction at $R_t$ is $\sim \theta_t^2 \propto R_t^{-2}$ and there
is also a factor of $1-\mu_{ti} \sim \theta_t^2 \propto R_t^{-2}$ in
the integrand of Eq.~(\ref{tau_gg_2}), most of the contribution to the
total optical depth is from $R_t \sim R_{t,0}$ (i.e. $\hrt \lesssim
2$). Therefore, the integration over the solid angle effectively
introduces a factor of $\sim \theta_{t,0}^2$, while the factor of
$1-\mu_{ti}$ in the integrand introduces a similar factor, together
giving a factor of $\sim \theta_{t,0}^4$. The integration over energy
$\varepsilon_i$ together with the threshold
$\varepsilon_t\varepsilon_i > 2/(1-\mu_{ti})$ for pair production give
$L'_{\varepsilon'_i} \propto \varepsilon_i^{1-\alpha} \propto
\varepsilon_t^{\alpha-1}(1-\mu_{ti})^{\alpha-1} \propto
\theta_{t,0}^{2(\alpha-1)}$. Altogether, with the previous factor of
$\theta_{t,0}^4$, the optical depth in this regime scales as
$\theta_{t,0}^{2(\alpha+1)}$.

Thus, for fixed values of $\tau_\star$ and $\varepsilon$, $\tgg$ first
becomes larger than unity at the center of the image ($x \ll 1$ and $Y
\approx 1$) at $\bar{T}_{1i} \sim \alpha\,2^{2\alpha-1} / \tau_\star
\epsilon^{\alpha-1}$. From this time on the central part of the image
is opaque, at $x < x_1$ which corresponds to $Y_1 \approx
\bar{T}_{1i}/\bar{T}$, so that photons of energy $\varepsilon$ can
escape mainly from a thin ring in the outer part of the image, that
corresponds to $x_1 < x < x_{\rm max}$ and occupies a fraction $Y_1
\approx \bar{T}_{1i}/\bar{T} \propto \bar{T}^{-1}$ of the image area
(since that area is linear in $x$). Thus, the observed flux is
suppressed by a similar factor and turns from $\propto \bar{T}$ at
$\bar{T} < \bar{T}_{1i}$ to $\propto \bar{T}^0$ at $\bar{T} >
\bar{T}_{1i}$. Eventually, at a later time $\bar{T}_{1f} \sim
\bar{T}_{1i}^{1/(\alpha+1)} \propto
\varepsilon^{-(\alpha-1)/(\alpha+1)}$ when $x_1 = x_*$, the whole
image becomes opaque, i.e. $\tgg > 1$ for all $0 \leq x \leq x_{\rm
max}$, and the observed flux starts to drop exponentially with
time. This behavior can be seen e.g. in Fig.~\ref{fig:6panel_DR}. In summary,
\begin{equation}\label{F_eps_high}
F_{\varepsilon\gg\varepsilon_{1*}}(\bar{T}\ll 1) 
\sim F_{\varepsilon<\varepsilon_{1*}}(\bar{T}=1) \times
\left\{\begin{matrix}
\bar{T} & \quad 
\bar{T} < \bar{T}_{1i}(\varepsilon)\ , \cr\cr
\bar{T}_{1i} & \quad 
\bar{T}_{1i}(\varepsilon) < \bar{T} < \bar{T}_{1f}(\varepsilon)\ , \cr\cr
\bar{T}^{\alpha+1}\exp[-(\bar{T}/\bar{T}_{1f})^{\alpha+1}] 
& \quad \bar{T} > \bar{T}_{1f}(\varepsilon)\ .
\end{matrix}\right.
\end{equation}

Similarly, for $1 \gg \bar{T} < \bar{T}_f$ it is natural to
define $\varepsilon_{1i}(\bar{T})$ and $\varepsilon_{1f}(\bar{T})$
as the two photon energies above which the center and outer
edge of the observed image, respectively, become optically thick to
pair production: by definition, $\bar{T}_{1i,f}[\varepsilon_{1i,f}(\bar{T})] \equiv
\bar{T}$. This implies that $\varepsilon_{1i} \sim
(\alpha\,2^{2\alpha-1} / \tau_\star\bar{T})^{1/(\alpha-1)} \propto
\bar{T}^{-1/(\alpha-1)}$, and since $\bar{T}_{1f} \sim
\bar{T}_{1i}^{1/(\alpha+1)}$, we have $\varepsilon_{1i}/\varepsilon_{1f}
\sim \bar{T}^{\alpha/(\alpha-1)}$ and $\varepsilon_{1f} \propto
\bar{T}^{-(\alpha+1)/(\alpha-1)}$. Eq.~(\ref{F_eps_high}) determines
the instantaneous spectrum in this regime,
\begin{equation}\label{F_eps_smallTbar}
F_{\varepsilon\gg\varepsilon_{1*}}(\bar{T}\ll 1) 
\sim \bar{T}F_{\varepsilon=1<\varepsilon_{1*}}(\bar{T}=1) \times
\left\{\begin{matrix}
\varepsilon^{-(\alpha-1)} & \quad 
\varepsilon < \varepsilon_{1i}(\bar{T})\ , \cr\cr
\varepsilon_{1i}^{\alpha-1}\varepsilon^{-2(\alpha-1)} & \quad 
\varepsilon_{1i}(\bar{T}) < \varepsilon < \varepsilon_{1f}(\bar{T})\ , \cr\cr
 \varepsilon_{1i}^{\alpha-1}\varepsilon_{1f}^{-2(\alpha-1)}
\exp[-(\varepsilon/\varepsilon_{1f})^{\alpha-1}]
& \quad \varepsilon > \varepsilon_{1f}(\bar{T})\ .
\end{matrix}\right.
\end{equation}
At $\bar{T} \sim 1$ the opacity becomes more uniform across the image,
$\bar{T}_{1i} \sim \bar{T}_{1f} \sim 1$, and $\varepsilon_{1i} \sim
\varepsilon_{1f} \sim \varepsilon_1(\bar{T}=1) \sim \varepsilon_{1*}$.

For $\varepsilon \gg \varepsilon_1(\bar{T} = 1)$, the time integrated
flux $f_\varepsilon = \int dT F_\varepsilon(T)$ is approximately given
by $\sim T_0
F_{\varepsilon<\varepsilon_1}(\bar{T}=1)\bar{T}_{1i}\bar{T}_{1f}$
where $\bar{T}_{1i}\bar{T}_{1f} \propto
\bar{T}_{1i}^{(\alpha+2)/(\alpha+1)} \propto
\varepsilon^{-(\alpha-1)(\alpha+2)/(\alpha+1)}$, since $\bar{T}_{1i}
\propto \varepsilon^{1-\alpha}$. Therefore, the spectral slope of the
time integrated spectrum, $f_\varepsilon$, for impulsive sources
($\Delta R \lesssim R_0$ and $\bar{T}_f \lesssim 1$) where the total
time integrated flux is comparable to that from the rising phase,
steepens by $\Delta \alpha = (\alpha-1)(\alpha+2)/(\alpha+1)$ above
$\epsilon_1(\bar{T}_f)$,
\begin{equation}\label{spectral_break}
f_\varepsilon(\Delta R \sim R_0) \propto
\left\{\begin{matrix}
\varepsilon^{-(\alpha-1)} & \quad 
[\varepsilon \ll \varepsilon_1(\bar{T}_f)]\ , \cr\cr
\varepsilon^{-(\alpha-1)(2\alpha+3)/(\alpha+1)} & \quad 
[\varepsilon \gg \varepsilon_1(\bar{T}_f)]\ .
\end{matrix}\right.
\end{equation}
This can be seen in Fig.~\ref{fig:int_spectra}.  For a quasi-steady
source ($\Delta R \gg R_0$ and $\bar{T}_f \gg 1$), a similar time
integrated spectrum is obtained only if the flux at $1 < \bar{T} <
\bar{T}_f$ decays faster than $\bar{T}^{-1}$, i.e. if $m(\alpha-2) >
2(b+1)$ [see Eq.~(\ref{F_eps_T})], so that $f_\varepsilon$ is dominated
by contributions near $\bar{T} \sim 1$. For a slower decay or a rising
flux at $1 < \bar{T} < \bar{T}_f$, $f_\varepsilon$ is dominated by
contributions from $\bar{T} \sim \bar{T}_f$ and there is an
exponential cutoff above $\varepsilon_1(\bar{T}_f)$, while the power
law high energy tail from the rising phase is encountered only after a
significant (exponential in $\varepsilon$) flux drop. For extremely
impulsive sources, where $\bar{T}_f \ll 1$ (i.e. $\Delta R \ll R_0$),
there is also an intermediate power law segment in the time integrated
spectrum:
\begin{equation}\label{time_int_spec2}
f_\varepsilon(\Delta R \ll R_0) \propto \left\{\begin{matrix}
\varepsilon^{-(\alpha-1)} & \quad [\varepsilon <
\varepsilon_{1i}(\bar{T}_f)]\ , \cr\cr 
\varepsilon^{-2(\alpha-1)} & \quad
[\varepsilon_{1i}(\bar{T}_f) < \varepsilon <
\varepsilon_{1f}(\bar{T}_f)]\ , \cr\cr
\varepsilon^{-(\alpha-1)(2\alpha+3)/(\alpha+1)} & \quad 
[\varepsilon > \varepsilon_{1f}(\bar{T}_f)]\ .
\end{matrix}\right.
\end{equation}

\section{Results: Semi-Analytic Light Curves and Spectra}
\label{sec:res}

\begin{figure}[ht]
\centering
\includegraphics[scale=0.8]{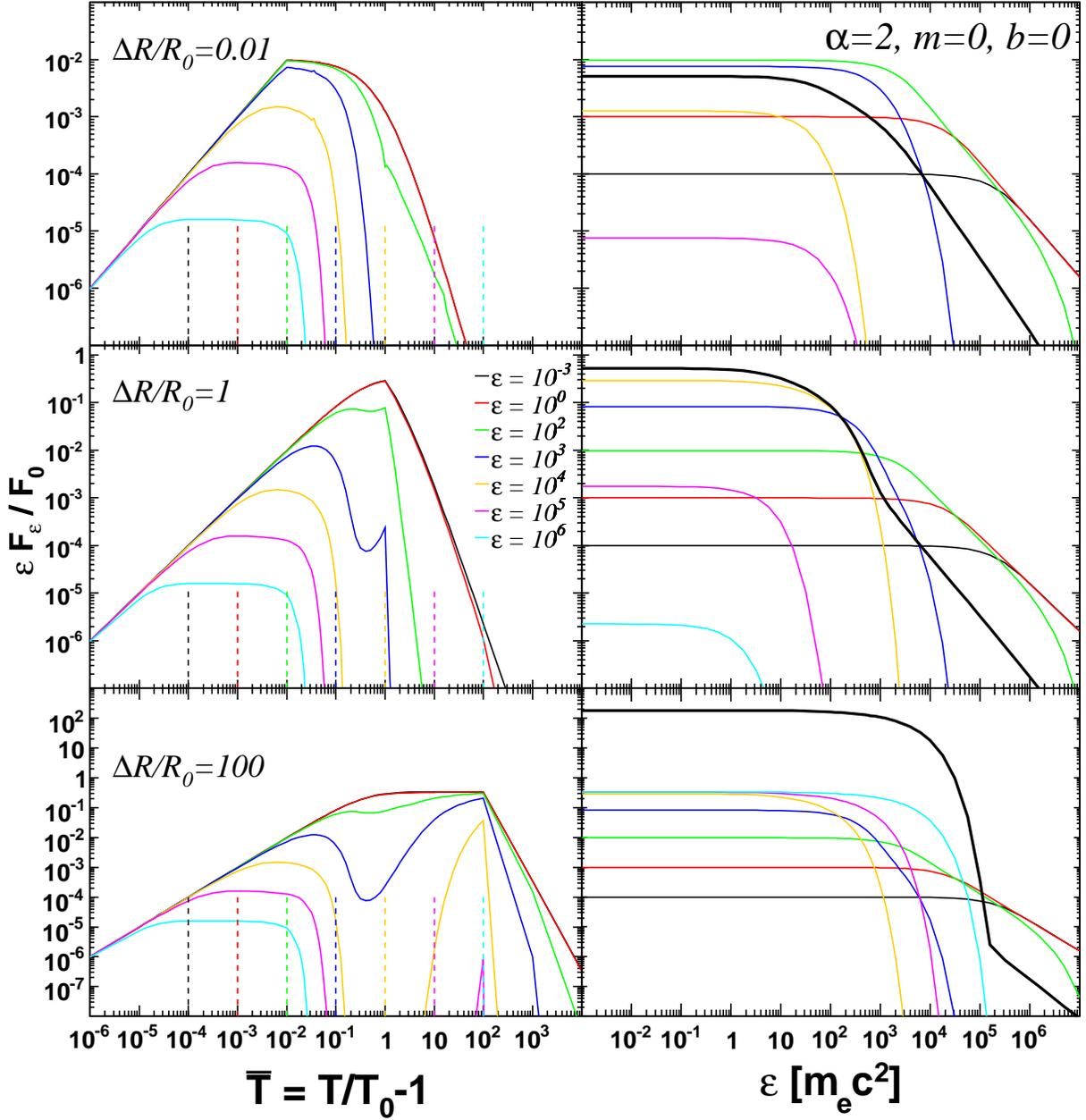}
\caption{Lightcurves ({\it left panels}), instantaneous
  ({\it thin lines}) and time integrated ({\it thick line}) spectra
  ({\it right panels}), calculated using our semi-analytic model, for
  a constant Lorentz factor ($m = 0$) and a comoving emissivity
  independent of radius ($b = 0$) with equal energy per decade of
  photon energy (corresponding to a photon index of $\alpha = 2$). We
  show results for three different radial extents of the emission
  region, $\Delta R / R_0 = 0.01$, 1, and 100, from top to bottom. We
  also use $\tau_\star = 1$ (see Eq.~[\ref{tau_star}]).}
\label{fig:6panel_DR}
\end{figure}

\begin{figure}[ht]
\centering
\includegraphics[scale=0.8]{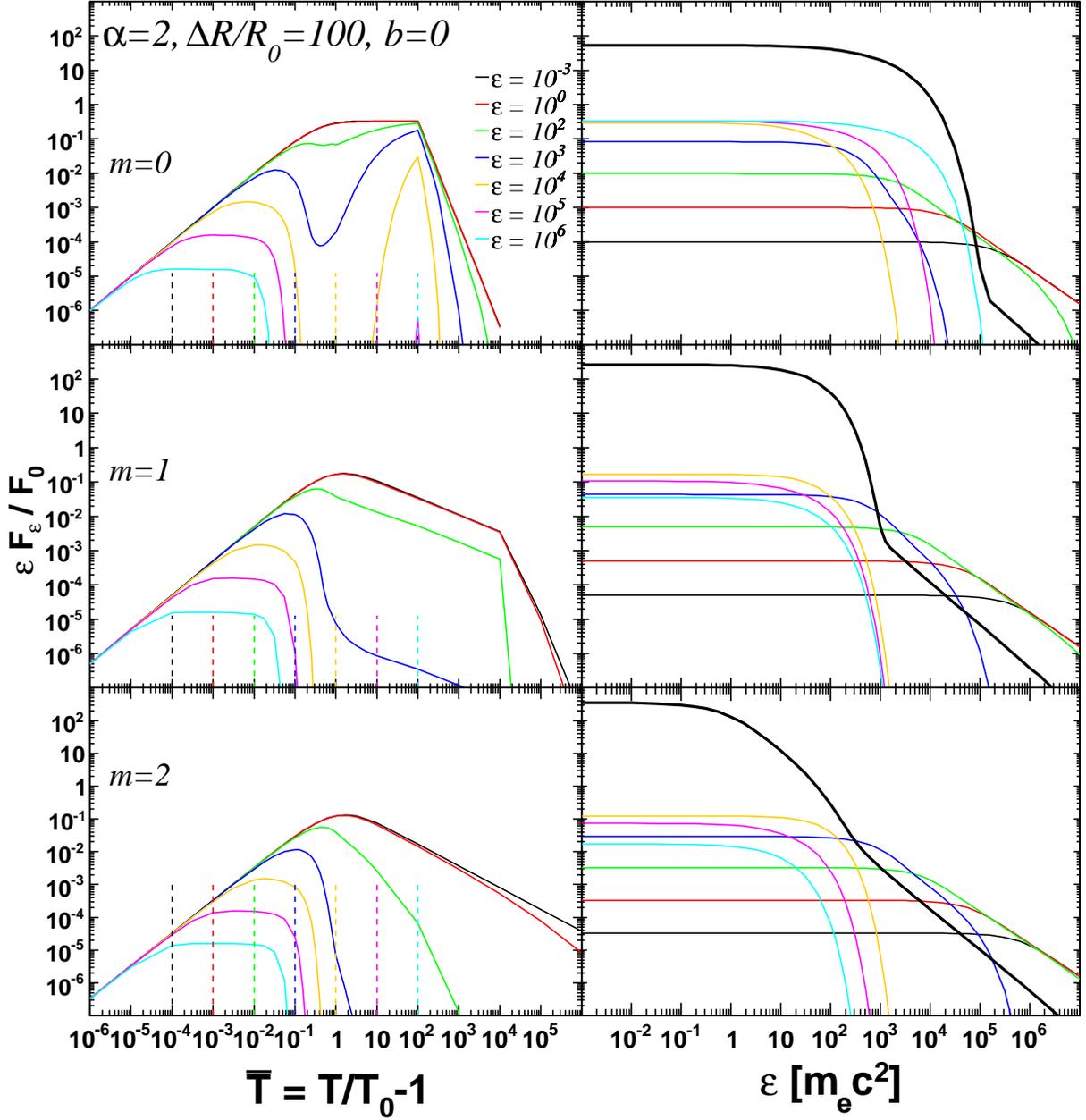}
\caption{Similar to Fig.~\ref{fig:6panel_DR}, with $b=0$, $\alpha=2$, $\tau_\star = 1$, 
but for a fixed $\Delta  R/R_0 = 100$ and varying $m$ where $\Gamma^2 \propto R^{-m}$.}
\label{fig:6panel_m}
\end{figure}

\begin{figure}[ht]
\centering
\includegraphics[scale=0.8]{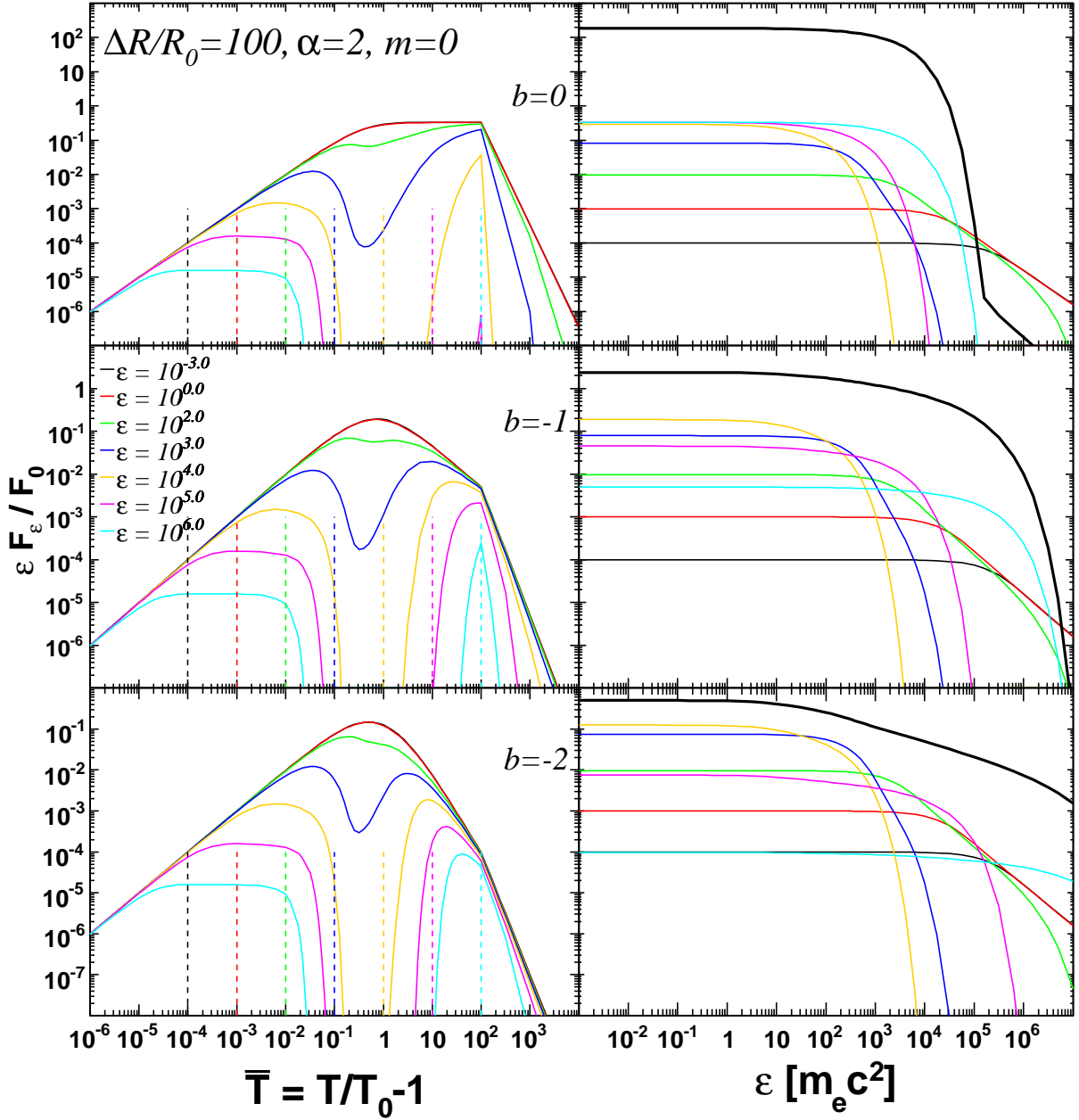}
\caption{Similar to Fig.~\ref{fig:6panel_DR}, with $m=0$, $\alpha=2$,
 $\tau_\star = 1$, but for a fixed $\Delta R/R_0 = 100$ and varying
 $b$ where the the spectral luminosity in the comoving frame of the
 shell scales as $L'_{\varepsilon'} \propto
 R^b(\varepsilon')^{1-\alpha}$.}
\label{fig:6panel_b}
\end{figure}

\begin{figure}[ht]
\centering
\includegraphics[scale=0.8]{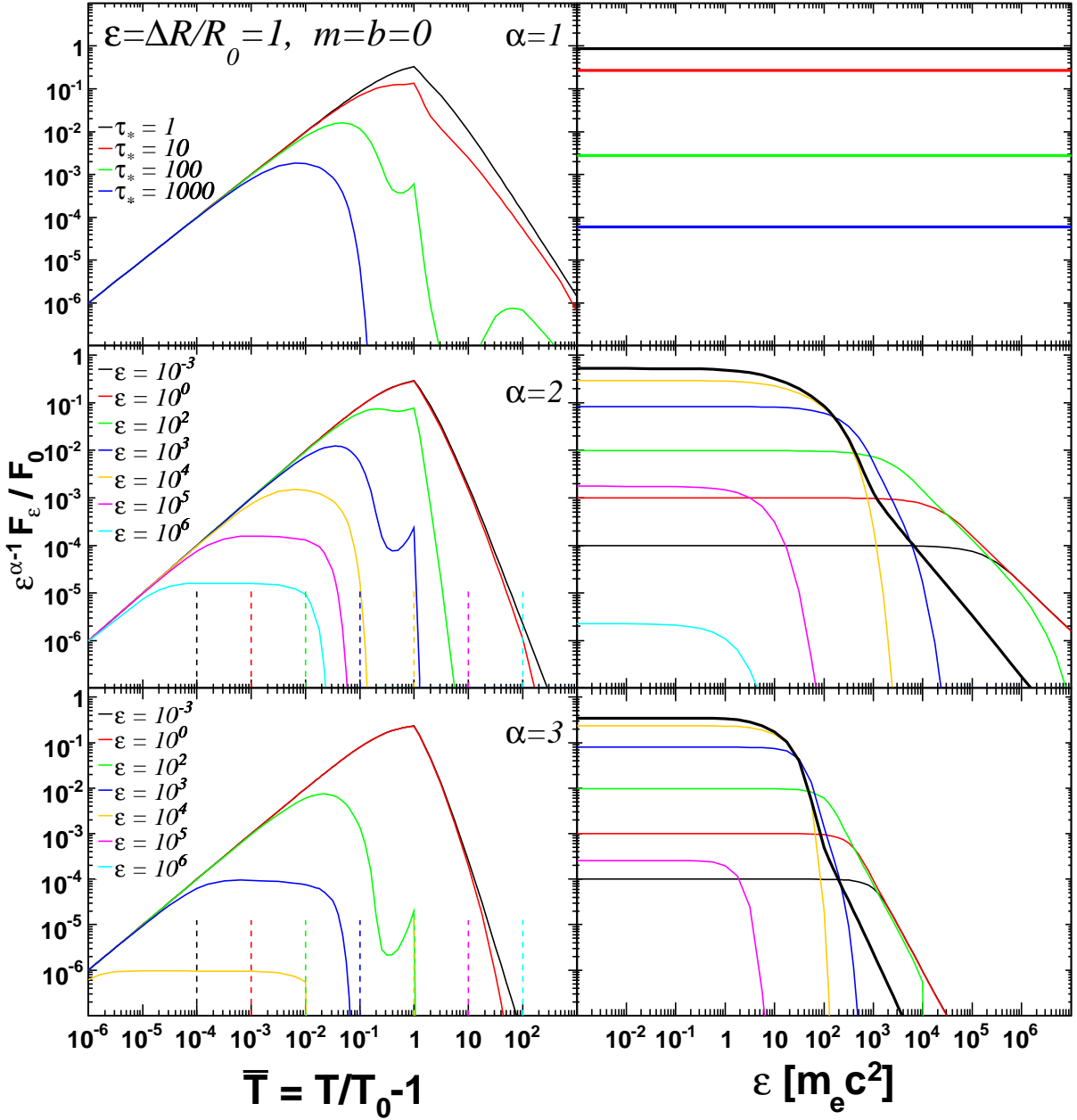}
\caption{Similar to Fig.~\ref{fig:6panel_DR}, 
with $b=m=0$, $\tau_\star = 1$, $\Delta R/R_0 = 1$
and  varying $\alpha$, where the the spectral luminosity in the comoving
  frame of the shell scales as $L'_{\varepsilon'} \propto
  R^b(\varepsilon')^{1-\alpha}$. The {\it middle panel} and {\it bottom panel} 
are for $\alpha=2$ and $3$, respectively.
The {\it top panel} is for $\alpha = 1$, 
for which $\tgg$ becomes independent of the photon energy
  $\varepsilon$ and therefore the spectrum is always a pure power law,
  $F_\varepsilon \propto \varepsilon^0$ and the flux depends only on
  time but not on the photon energy. For this reason we show light
  curve ({\it left panel}) and time integrated spectra ({\it right
  panel}) for different values of $\tau_\star$ (see
  Eq.~[\ref{tau_star}]).}
\label{fig:6panel_alpha}
\end{figure}

\begin{figure}[ht]
\centering
\includegraphics[scale=0.79]{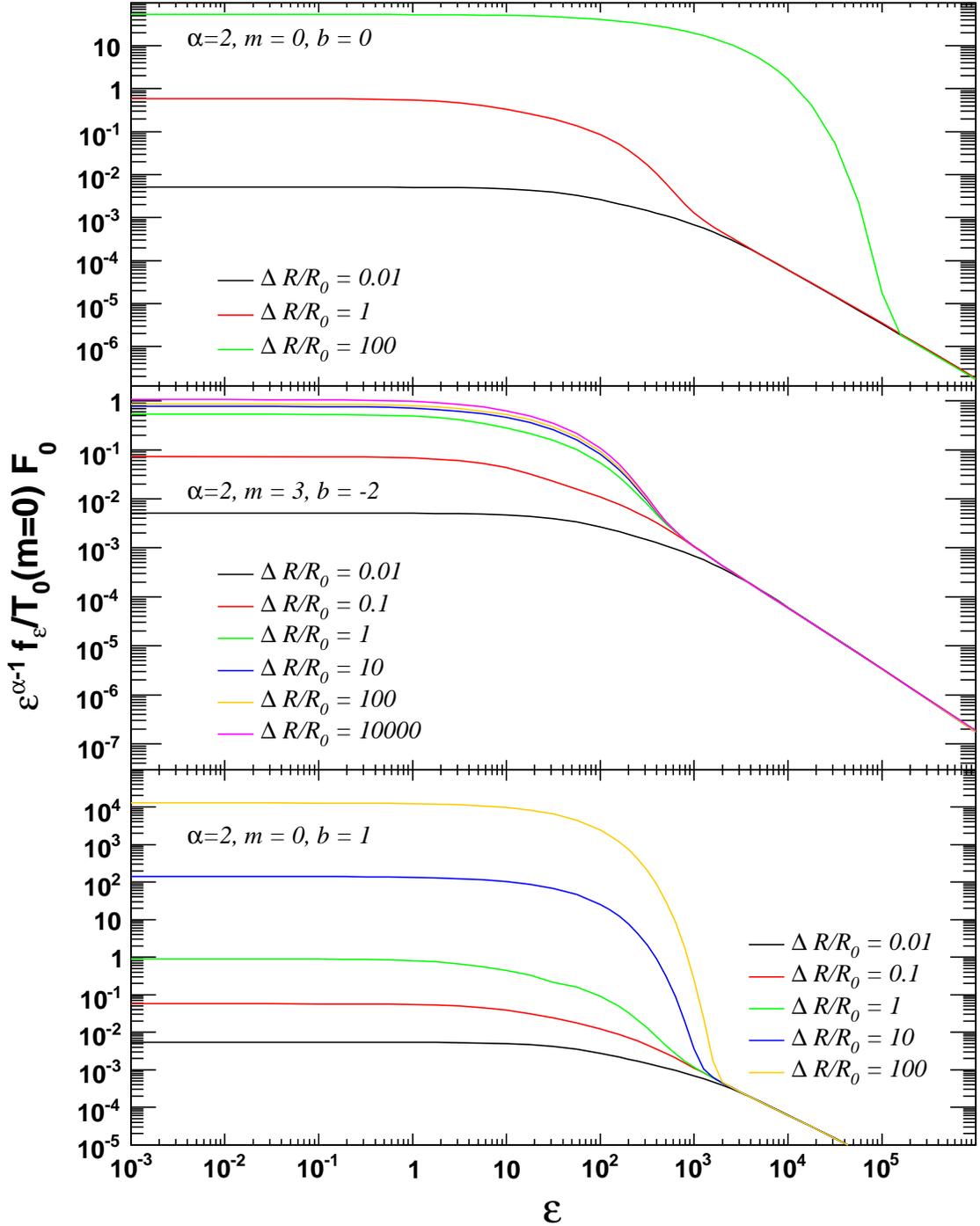}
\caption{The time integrated spectra for different values of $\Delta
  R/R_0$, while fixing the values of the other parameters. In
  the {\it top panel} $m = b = 0$, in the middle panel $m = 3$ and $b
  = -2$, while in the {\it bottom panel} $m = 0$ and $b = 1$. The
  values of the other model parameters in all the panels are $\alpha =
  2$ and $\tau_\star = 1$.}
\label{fig:int_spectra}
\end{figure}

Figures~\ref{fig:6panel_DR} -- \ref{fig:6panel_alpha} show light
curves and spectra for the semi-analytic model developed in the
preceding sections. We use fiducial parameter values of $m = b = 0$,
$\Delta R/R_0 = \tau_\star = 1$, and $\alpha = 2$, which are relevant
for the prompt gamma-ray emission in GRBs, and vary one parameter at a
time in order to see the effect of each model parameter more
clearly. When varying $m$ and $b$ (Figs.~\ref{fig:6panel_m} and
\ref{fig:6panel_b}, respectively) we use $\Delta R/R_0 = 100$ in order
to have a large enough range of emission radii so that the radial
dependence of the Lorentz factor and of the co-moving spectral
emissivity would have a significant effect on the light curves (for
$\Delta R/R_0 \ll 1$ the values of $m$ and $b$ hardly affect the light
curves). Figure~\ref{fig:int_spectra} shows the time integrated
spectra for several values of $\Delta R/R_0$, where each panel is for
a different set of values for the three parameters $(\alpha,\,m,\,b)$.
In order to ease the reading, Table~\ref{tab:params} summarizes
the various sets of parameters and the corresponding figures.

Figure \ref{fig:6panel_DR} shows the light curves for fixed values $m =
b = 0$, $\tau_\star = 1$, and $\alpha = 2$, while the various panels
correspond to different values of $\Delta R/R_0$ (of $0.01$, $1$, and
$100$, from top to bottom). At the lowest photon energies, well below
$\varepsilon_{1*}$ (which for the parameter values used here is $\sim
10^2$), opacity to pair production never becomes very significant, and
the light curves follow the behavior described in Eq.~(\ref{F_eps_T})
which is discussed in the preceding section. In this regime the
lightcurves are self similar in the sense that $\varepsilon^{\alpha-1}
F_\varepsilon$ is independent of $\varepsilon$ below
$\varepsilon_1(T)$. The different behavior for $\bar{T}_f \ll 1$ and
$\bar{T}_f \gg 1$ (where $\bar{T}_f = \Delta R/R_0$ for $m = 0$) that
appears in Eq.~(\ref{F_eps_T}) can clearly be seen by comparing the
upper and lower panels of Fig.~\ref{fig:6panel_DR}. For $\varepsilon
\gg \varepsilon_{1*}$, on the other hand, opacity to pair production
has a major effect on the light curves. In this regime the light
curves at $\bar{T} \ll 1$ follow Eq.~(\ref{F_eps_high}), showing a
pronounced constant flux plateau between $\bar{T}_{1i} \propto
\varepsilon^{1-\alpha}$, when the center of the image becomes
optically thick to pair production, and $\bar{T}_{1f} \sim
\bar{T}_{1i}^{1/(\alpha+1)}$, when the entire image becomes opaque,
followed by an exponential flux decay.  At $1 \lesssim \bar{T} <
\bar{T}_f$ the opacity does not vary drastically across the image and
may be described by a single value of $\epsilon_1(1 \ll \bar{T} <
\bar{T}_f) \propto \bar{T}^{(1-b-m\alpha/2)/[(m+1)(\alpha-1)]}$.  For
the parameter values used in Fig.~\ref{fig:6panel_DR}, $\varepsilon_1$
increases (linearly) with $\bar{T}$ in this range, and therefore the
opacity at a given $\varepsilon$ decreases with time, causing the
observed flux to increase with time until $\varepsilon_1$ sweeps
across $\varepsilon$ or until $\bar{T}_f$ is reached (whichever comes
first). At $\bar{T} > \bar{T}_f$ the situation is reversed, as the
observed emission comes from large angles relative to the line of
sight (``high-latitude" emission) and $\varepsilon_1$ decreases with
time.

We now turn to the photon energy spectrum. The instantaneous spectra
at $\bar{T} \ll 1$ follow the behavior described in
Eq.~(\ref{F_eps_smallTbar}). At very early times the exponential part
starts only at very high photon energies, making it very hard to
detect. When $\bar{T}\sim 1$ the intermediate power-law segment
disappears as $\varepsilon_{1i}$ and $\varepsilon_{1f}$ become nearly
equal (note that the low energy part of the curves appears flat in the
figures since we show $\varepsilon^{\alpha-1}F_\varepsilon$ which is
independent of $\varepsilon$ below $\varepsilon_{1}$). The time
integrated spectrum varies with the value of $\Delta R/R_0$. For
$\Delta R/R_0 \ll 1$ it consists of three power-law segments, as
described in Eq.~(\ref{time_int_spec2}). As $\Delta R/R_0$ increases,
the central power-law segment, at $\varepsilon_{1i}(\bar{T}_f) <
\varepsilon < \varepsilon_{1f}(\bar{T}_f)$, shrinks as
$\varepsilon_{1i}(\bar{T}_f)$ and $\varepsilon_{1f}(\bar{T}_f)$
approach each other, until it disappears for $\Delta R/R_0 \sim 1$
where $\bar{T}_f \sim 1$ and $\varepsilon_{1i}(\bar{T}_f) \sim
\varepsilon_{1f}(\bar{T}_f) \sim \varepsilon_{1*}$. For $\Delta R/R_0
\sim 1$ the time integrated spectrum is described by
Eq.~(\ref{spectral_break}), and consists of two power-law segments.
As $\Delta R/R_0$ increases above unity the time integrated spectrum
develops an exponential high-energy cutoff, while the power-law tail
at high energies becomes increasingly suppressed. This occurs since if
the flux at $1 < \bar{T} < \bar{T}_f$ does not drop faster than
$\bar{T}^{-1}$, which corresponds to $m(\alpha-2) < 2(b+1)$ (see
Eq.~[\ref{F_eps_T}]) as is indeed the case for the parameter values
used in Fig.~\ref{fig:6panel_DR}, then the time integrated flux is
dominated by contributions from $\bar{T} \sim \bar{T}_f \gg 1$ and
reflects the exponential cutoff of the instantaneous spectrum at that
time, which dominates over the high-energy power-law component that
arises from the superposition of the instantaneous spectra from
$\bar{T} \lesssim 1$.

Figures~\ref{fig:6panel_m} and \ref{fig:6panel_b} demonstrate the
effects of the two parameters $m$ and $b$. As discussed above, a large
value for $\Delta R/R$ (100) was chosen so that the radial dependence
of the Lorentz factor ($\Gamma^2 \propto R^{-m}$) and of the co-moving
spectral luminosity [$L'_{\varepsilon'} \propto
R^b(\varepsilon')^{1-\alpha}$] would have a large effect on the light
curves. For $\Delta R/R_0 \ll 1$ the values of $m$ and $b$ hardly
affect the light curves (since the emission takes place over a very
small range of radii in which both $\Gamma$ and $L'_{\varepsilon'}$
hardly vary). Figure~\ref{fig:6panel_m} also demonstrates the dependence
of $\bar{T}_f$ on $m$, where in the limit of $\Delta R/R_0 \gg 1$,
$\bar{T}_f \approx (\Delta R/R_0)^{m+1}$ (see Eq.~[\ref{T_bar_f}]).

As can be seen in Fig.~\ref{fig:6panel_m}, the power-law component of
the time integrated spectrum is largely independent of $m$, since it
originates from the superposition of the instantaneous spectra at
$\bar{T} \lesssim 1$, which are sampling a small range of emission
radii. The lower energy component, however, from the contribution of
the emission at times $1 < \bar{T} \lesssim \bar{T}_f$, is sensitive
to the value of $m$, since it sample a large range of emission radii.
For $m = 0$, $F_\varepsilon(1<\bar{T}<\bar{T}_f)$ is constant in time
(for the values of the other parameters that are used in
Fig.~\ref{fig:6panel_m}), while $\varepsilon_1(1<\bar{T}<\bar{T}_f)
\propto \bar{T}$, and both effects combine to produce a very
pronounced high-energy exponential cutoff. For $m = 1$,
$F_\varepsilon(1<\bar{T}<\bar{T}_f) \propto \bar{T}^{-1/2}$ while
$\varepsilon_1(1<\bar{T}<\bar{T}_f)$ is constant in time, which
results in a somewhat less pronounced, though still fairly large
high-energy exponential cutoff in the time integrated spectrum. For $m
= 2$, $F_\varepsilon(1<\bar{T}<\bar{T}_f) \propto \bar{T}^{-2/3}$
while $\varepsilon_1(1<\bar{T}<\bar{T}_f) \propto \bar{T}^{-1/3}$, so
that the time integrated spectrum in the range
$\varepsilon_1(\bar{T}_f) < \varepsilon < \varepsilon_1(\bar{T}=1)
\sim \varepsilon_{1*}$ is dominated by the contributions near the
time $T_1(\varepsilon)$ when $\varepsilon_1(\bar{T}_1) =
\varepsilon$. This results in a spectral slope of $\varepsilon
f_\varepsilon \propto \varepsilon^{-1}$ in the lower panel of
Fig.~\ref{fig:6panel_m}.

More generally, $F_{\varepsilon<\varepsilon_1(\bar{T})}(1 < \bar{T} <
\bar{T}_f) \sim \varepsilon^{1-\alpha} \bar{T}^{(2b-m\alpha)/[2(m+1)]}$
while $\varepsilon_1(1<\bar{T}<\bar{T}_f) \sim
\varepsilon_{1*}\bar{T}^{(1-b-m\alpha/2)/[(m+1)(\alpha-1)]}$, so that
when the flux is dominated by the contribution from $\bar{T} \sim
\bar{T}_1(\varepsilon)$, then the spectral slope of the time
integrated spectrum is given by
\begin{equation}\label{late_int_spec_slope}
\frac{d\log\varepsilon^{\alpha-1}f_\varepsilon}{d\log\varepsilon} =
\frac{(\alpha-1)\left[m(2-\alpha)+2(b+1)\right]}{2(1-b)-m\alpha}\ .
\end{equation}
This may be relevant if $\varepsilon_1(1<\bar{T}<\bar{T}_f)$ decreases
with $\bar{T}$, in which case this spectral slope is valid in the
range $\varepsilon_1(\bar{T}_f) < \varepsilon < \varepsilon_{1*}$.  It
may also be relevant if $\varepsilon_1(1<\bar{T}<\bar{T}_f)$ decreases
with $\bar{T}$, as discussed below.

In Fig.~\ref{fig:6panel_b}, the upper panel is identical to the upper
panel of Fig.~\ref{fig:6panel_m} and the lower panel of
Fig.~\ref{fig:6panel_DR}. In the middle panel
$F_\varepsilon(1<\bar{T}<\bar{T}_f) \propto \bar{T}^{-1}$ and
$\varepsilon_1(1<\bar{T}<\bar{T}_f) \propto \bar{T}^2$, while in the
bottom panel $F_\varepsilon(1<\bar{T}<\bar{T}_f) \propto \bar{T}^{-2}$
and $\varepsilon_1(1<\bar{T}<\bar{T}_f) \propto \bar{T}^3$. Both cases
result in a very pronounced exponential cutoff at very high photon
energies (which may be hard to detect), but show a shallow spectral
slope up to this exponential cutoff (which may be easier to detect).
This again results in the spectral slope given by
Eq.~(\ref{late_int_spec_slope}). However, in this case
$\varepsilon_1(1<\bar{T}<\bar{T}_f)$ increases with $\bar{T}$, and
therefore this spectral slope occurs in the range $\varepsilon_{1*} <
\varepsilon < \varepsilon_1(\bar{T}_f)$. This is valid, however, only
if indeed the time integrated flux in this spectral range is dominated
by the contribution from near $T_1(\varepsilon)$.  This is not valid
in the upper panel of Fig.~\ref{fig:6panel_b} (where it is dominated
by the contribution from $\bar{T} \sim \bar{T}_f$), and is only
marginally valid in the middle panel (where the contributions from all
the times in the range $T_1(\varepsilon) \lesssim \bar{T} \lesssim
\bar{T}_f$ are comparable). In the lower panel of
Fig.~\ref{fig:6panel_b} the flux in this spectral range is indeed
dominated by the contribution from $\bar{T} \sim
\bar{T}_1(\varepsilon)$, which results in a spectral slope of
$\varepsilon f_\varepsilon \propto \varepsilon^{-1/3}$ in this range.

In Fig.~\ref{fig:6panel_alpha}, which shows the effect of varying
$\alpha$, the top panel corresponds to $\alpha=1$, for which both the
flux and the optical depth become independent of $\varepsilon$. As a
result, we present for this case light curves for different values of
$\tau_*$, and the corresponding integrated spectra (which all have a
flat $f_\varepsilon$ and vary only in their normalization). For
$\alpha = 2$ ({\it middle panel}) and $\alpha = 3$ ({\it bottom
panel}), one can verify that the power laws on the middle and bottom
right panels have an index of approximately $4/3$ and
$5/2$ respectively, as expected from Eq.~(\ref{spectral_break}) after
rescaling by $\varepsilon^{\alpha-1}$.

Finally, Fig.~\ref{fig:int_spectra} illustrates the behavior of the
time integrated spectra, as discussed at the end of \S\
\ref{sec:analytic}. All the curves show a high energy power law tail
with an index of about $4/3$, as expected from
Eq.~(\ref{spectral_break}).  Moreover, given the rescaling by
$T_0(m=0)$ (which is independent of $m$) in
Fig.~\ref{fig:int_spectra}, it is easier to see that the time
integrated spectra become independent of $b$ and $m$ for $\Delta R/R_0
\ll 1$ (since in that limit, the same holds for the light curves and
instantaneous spectra). As discussed in the paragraph following
Eq.~(\ref{spectral_break}), the exponential cutoff is suppressed when
$m(\alpha-2)>2(b+1)$, as is the case on the middle panel only. In such
a case, the time integrated spectra are dominated by contributions near
$\bar{T}\sim 1$, and the effect of $\Delta R/R_0$ becomes negligible
for $\Delta R/R_0 \gg 1$, which explains the asymptotic behavior of
the spectra with increasing $\Delta R/R_0$. Finally, for very
impulsive sources, the intermediate power law segment in
Eq.~(\ref{time_int_spec2}) can be discerned, albeit with difficulty.

\section{Discussion}\label{sec:dis}

We have explored in great detail a model for the temporal and spatial
dependence of the opacity to pair production ($\gamma\gamma \to
e^+e^-$) in impulsive relativistic sources. Our simple, yet rich,
model features a thin spherical shell expanding ultra-relativistically
and emitting isotropically in its own rest frame within a finite range
of radii. Our two main results are the follwoing. First,
 while the instantaneous spectrum
(which is typically very hard to measure due to poor photon
statistics) has an exponential cutoff at high photon energies, the
time integrated spectrum over the duration of a flare or spike in the
light curve (which is easier to measure) has a power-law high-energy
tail. Second, photons above this spectral break in the time integrated
spectrum arrive mainly near the onset of the flare or spike in the
light curve. 

These two features provide a unique detectable signature of opacity to
pair production, making it easier to identify
observationally. Furthermore, these features are expected to be fairly
robust, even if the exact details (such as the exact change in the
spectral slope across the break, $\Delta \alpha$, or the exact shape
of the light curve at high photon energies above the spectral break)
may depend on the details of the model (such as the exact geometry,
which is assumed to be spherical in our model\footnote{In AGN, e.g., a
cylindrical geometry may be more appropriate. We intend to study such
a cylindrical geometry in a future work.}). The reason behind these
features is that in impulsive sources the photon field starts from
zero (or more realistically a non-zero value, which is still much
lower than that near the peak of the flare or spike in the light
curve) and builds-up with time, so that the optical depth to pair
production, $\tgg$, increases with time, and high energy photons can
escape mainly at early times while $\tgg$ is still below unity.

A source is considered impulsive for our purposes if the photon field
in the source and its vicinity changes considerably within the source
light crossing time. In this limit the time dependence of the photon
field and the resulting opacity to pair production is important. This
can naturally occur in relativistic sources, but is hard to produce in
non-relativistic sources (since it requires a relativistic signal in
order to turn the emission on or off on a time scale of the order of
the light crossing time of the source). In the opposite limit, where
the photon field hardly varies within the source light crossing time,
the photon field may be approximated as constant in time along the
trajectory of the photons, and can be evaluated at the time of
emission (this is considered a ``quasi-steady'' state). In our model,
the photon field approaches a quasi-steady state\footnote{Here, by
quasi-steady state, we mean that neglecting the time dependence of the
photon field would at most change the results by a factor of order
unity, but not qualitatively.} within a few light crossing times of
the emitting region ($\bar{T}_f > \bar{T} \gtrsim {\rm a\ few}$).
If the source is active for much longer times ($\bar{T}_f \gg 1
\Leftrightarrow \Delta R/R_0 \gg 1$), the time integrated spectrum
will usually be dominated by this late time quasi-steady emission, and
an exponential high-energy cutoff develops, while the high-energy
power-law tail becomes increasingly suppressed. For this reason, in
order for the source to be impulsive, the duration of the emission
should be at most comparable to the light crossing time of the source
($\bar{T}_f \lesssim 1 \Leftrightarrow \Delta R/R_0 \lesssim 1$).

We have considered a single, isolated emission episode which
corresponds to a single flare or spike in the observed light curve.
Furthermore, we have assumed no background photon field at the time
when the emission turns on. These are obviously idealized assumptions
and it is worth considering, at least qualitatively at this stage, the
modifications that may occur when these ideal conditions are not
satisfied. For the prompt emission or X-ray flares in GRBs the
background quasi-steady photon field is expected to be very low and
not contribute significantly to $\tgg$. This is probably also true for
Blazars of BL Lac type. In other types of Blazars (quasar hosted
Blazars), however, the external photon field, mainly due to emission
from the accretion disk and its scattered photons from the clouds in
the broad line region, is expected to provide the dominant
contribution to $\tgg$ in the source, even during flares
\citep{Sikora94}. In this case, our model would not be applicable,
since the external radiation field would prevent the escape of
high-energy photons near the onset of the spike, resulting in $\tgg$
that is largely independent of time.

Regarding the assumption that the flare/spike is isolated, in many
cases there are series of flares, so that except for the first flare
in the series for which our assumption should hold very well, for
consecutive flares the high energy photons could in principle pair
produce with photons emitted in previous flares.  This will be highly
suppressed if the time from the end of the previous flare is much
larger than its duration. Even if these two times are comparable, pair
production is still significantly suppressed since the relevant
photons can meet only further away from the source, where the photon
density is smaller and the angle between the directions of the photons
(in the lab frame) is smaller.  Such a proximity in time to a previous
flare/spike will still increase $\tgg$ to some degree, but this will
affect mainly the highest energy photons, with energies well above the
spectral break in the time integrated spectrum over the duration of
the flare/spike, which are relatively hard to detect due to the
smaller number of photons at such high energies. Therefore, in
practical terms, the differences from our idealized model are not
expected to be very large. This may even make it meaningful to
integrate the spectrum over many spikes/flares in order to increase
the photon statistics, in cases where the number of photons detected
in individual spikes is not large enough to enable a good spectral
analysis.

Other sources of opacity, such as scattering of photons on the pairs
that are produced, are also possible. The latter, however, is expected
to build-up in time on a comparable time scale to that of the opacity
to pair production that we study here. Therefore, it is not expected
to have a significant impact on our main conclusions. Opacity for
scattering on the electrons associated with the baryons or with
preexisting pairs within the outflow is also possible. However, it
will not greatly vary within a single dynamical time, and should also
affect lower energy photons (where the larger number of photons
enables a better spectral analysis). Furthermore, it is suppressed at
high photon energies due to the reduction in the cross section in the
Klein-Nishina regime.  Moreover, we find a rather unique combined
spectral and temporal signature for the opacity to pair production,
which could help distinguish between it and other sources of opacity.

In GRB afterglows the opacity to pair production is typically very low
and therefore not expected to be detectable in the GLAST energy range.
During the afterglow, after about one day, $L_{0,52} \sim
10^{-8}-10^{-7}$, $R \sim 10^{17}\;$cm corresponding to $R_{0,13} \sim
10^4$, and $\Gamma \sim 10$ corresponding to $\Gamma_{0,2} \sim
0.1$. According to Eq.~(\ref{eps_1s}), this implies a huge value of
$\varepsilon_{1*}$ ($\varepsilon_{1*}m_e c^2\sim 10^{15}-10^{16}\;$eV
for $\alpha \approx 2$). In practice the opacity would be even lower
than this, since the typical energy of the photons that would pair
produce with such high energy photons would be $\sim
\Gamma^2/\varepsilon_{1*}$ corresponding to $\nu \sim 10^{12}\;$Hz,
which is well below the assumed power law segment of the spectrum (so
that the number density of these low energy photons would in practice
be much lower than the default value according to our assumption of a
simple single power law spectrum). A possible exception to the very
low $\tau_{\gamma\gamma}$ during the afterglow may be the very early
afterglow in a stellar wind environment, near $T_{\rm dec}$ which is
of the order of seconds in this case. Typical parameters values there
are $R_{0,13} \sim 100$, $\Gamma_{0,2} \sim 1$, and $L_{0,52} \sim
0.1$, which might give $\varepsilon_{1*}m_e c^2$ as low as $\sim
100\;$GeV. Such values could be detected by GLAST, albeit with
difficulty.

For the internal shocks model, the GRB prompt emission occurs at a
much smaller radius compared to the afterglow, $R_{\rm GRB} \ll R_{\rm
dec}$. Furthermore, the luminosity of the prompt GRB emission is much
larger than that of the afterglow, and the Lorentz factor is higher.
Therefore, despite the higher Lorentz factor during the prompt GRB,
$\tgg$ is still much larger than during the afterglow.  In models
where the prompt emission occurs near the deceleration radius, $R_{\rm
GRB} \sim R_{\rm dec}$, the values of $\tgg$ in the very early
afterglow and in the prompt emission are comparable (perhaps somewhat
smaller in the early afterglow due to a smaller radiative efficiency),
but $\tgg$ is typically very low for both emission components
(i.e. the effects of opacity to pair production are not expected in
the GLAST energy range).

Once the spectral break in the time integrated spectrum over the
duration of a flare or spike in the light curve is observed in the
data, it can be used to constrain the values of the physical
parameters of the source, namely $\Gamma_0^{2\alpha}R_0$. A fit of our
model predictions to the data can in principle determine the values of
all the model parameters: $\alpha$, $m$, $b$, $\Delta R/R_0$,
$\tau_\star$ and $F_0$, which in turn determine $L_0$ (from $F_0$,
Eq.~[\ref{F_0}]) and $\Gamma_0^{2\alpha}R_0$ (from $\tau_\star$,
Eq.~[\ref{tau_star}]). In practice, however, the limited photon
statistics may render such a direct fit with such a large number of
free parameters impractical.  One way to overcome this problem is to
fix the values of some of the model parameters, e.g. $m = b = 0$ and
even $\Delta R/R_0 = 1$, if necessary.

A less accurate but less computationally demanding 
alternative is to fit the time integrated spectrum (over a flare or
spike in the light curve) to a parameterized function featuring a
smooth transition between two power laws,
\begin{equation}
f_\varepsilon = f_0\left[\left(
\frac{\varepsilon}{\varepsilon_{1*}}\right)^{-n(1-\alpha)}
+\left(\frac{\varepsilon}{\varepsilon_{1*}}
\right)^{-n(1-\alpha-\Delta\alpha)} \right]^{-1/n}\ ,
\end{equation}
where $n$ and $f_0$ determine the sharpness of the spectral break at
$\varepsilon_{1*}$ and its flux normalization, respectively, while
$f_{\varepsilon\ll\varepsilon_{1*}} \propto \varepsilon^{1-\alpha}$
and $f_{\varepsilon\gg\varepsilon_{1*}} \propto
\varepsilon^{1-\alpha-\Delta\alpha}$. Such a fit can determine both
the photon index, $\alpha$, and the photon energy $\varepsilon_{1*}
\approx \varepsilon_1(\bar{T}=1)$ of the spectral break in the time
integrated spectrum, as well as $F_0$. For $\Delta R/R_0 \lesssim 1$
and defining $\Delta T$ as the observed
variability time (in seconds), e.g. the observed FWHM of the flare or spike in the
light curve,  we have $f_0/\Delta T \approx \bar{T}_f
F_0\varepsilon_{1*}^{1-\alpha} \sim
F_0\varepsilon_{1*}^{1-\alpha}\Delta R/R_0$, which can be used
in Eq.~(\ref{F_0}) to determine $L_0$ :
\begin{equation}
L_0 \approx \pi d_L^2 \frac{f_0}{\Delta T}\,\frac{R_0}{\Delta R}
\left(\frac{1+z}{2}\right)^{\alpha-2}
\varepsilon_{1*}^{\alpha-1} = 1.3\times10^{51}
\,\frac{R_0}{\Delta R}\left(\frac{1+z}{2}\right)^{\alpha-2}
d_{L,z1}^2\frac{f_{0,-6}}{\Delta T}\,\varepsilon_{1*}^{\alpha-1}
\;{\rm erg\;s^{-1}}\ ,
\end{equation}
where $f_0 = 10^{-6}f_{0,-6}\;{\rm erg\;cm^{-2}}$, and $d_{L,z1}$ is
 $d_L$ in units of $d_L(z=1) \approx 2.05\times 10^{28}\;$cm (for
 standard cosmological parameters). It is hard to determine $\Delta
 R/R_0$ without a detailed fit to the model spectra (see
 Fig.~\ref{fig:int_spectra} for the dependence of the time integrated
 spectrum on $\Delta R/R_0$), and this is a price for the simplicity
 of this method and the use of simple analytic formulas rather than
 numerically evaluating the set of nested integrals in order to
 calculate our model predictions. One can either assume $\Delta R/R_0
 \approx 1$ or try to estimate its values by eye, guided by
 Fig.~\ref{fig:int_spectra}, if one wishes to avoid a direct fit to
 the model predictions.

\begin{figure}[t]
\centering \includegraphics[scale=0.67]{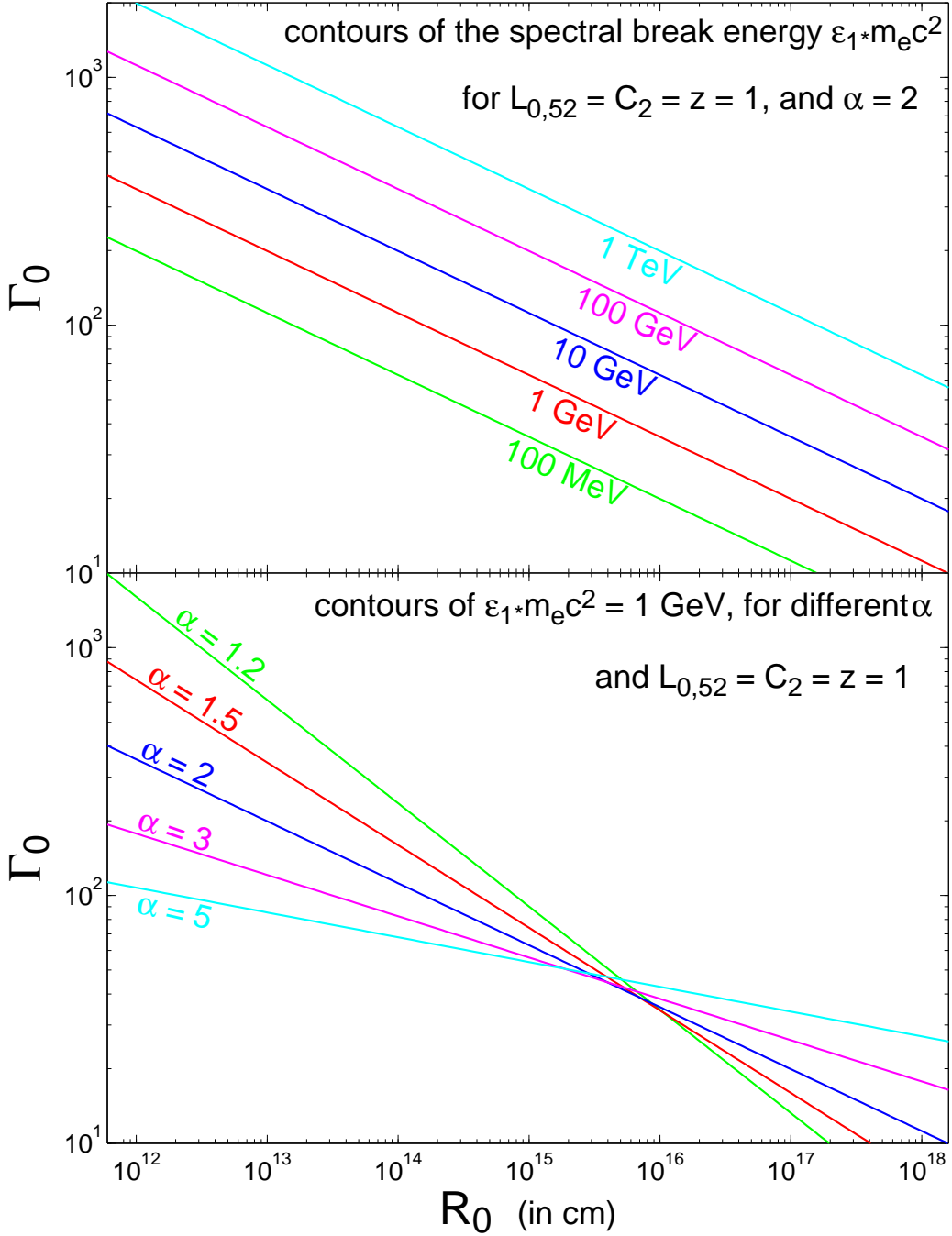}
\caption{Contour lines for of the photon energy $\varepsilon_{1*}m_e
c^2$ where the time integrated spectrum over a flare or spike in the
light curve steepens due to opacity to pair production, shown in the
$\Gamma_0 - R_0$ plane, according to Eq.~(\ref{eps_1s}). In the {\it
upper panel} $\varepsilon_{1*}m_e c^2$ is varied and $\alpha = 2$ is
fixed, while the {\it lower panel} $\alpha$ is varied and
$\varepsilon_{1*}m_e c^2 = 1\;$GeV is fixed. In both panels $L_{0,51}
= C_2 = z = 1$.}
\label{fig:eps1}
\end{figure}

The quantities $\alpha$, $L_0$, and $\varepsilon_{1*}$
may in turn be used to determine $\Gamma_0^{2\alpha}R_0$. In
order to do this in practice, we need to use Eq.~(\ref{tau_star}) and
the relation
\begin{align}
(1+z)\varepsilon_{1*} &\equiv (\tau_\star/C_\alpha)^{-1/(\alpha-1)}
= \left[249C_2(\alpha/2)^{5/3}10^{4(\alpha-2)}
L_{0,52}^{-1}(\Gamma_{0,2})^{2\alpha}
R_{0,13}\right]^{1/(\alpha-1)}\ ,
\\ \nonumber
\\ \label{eps_1s}
\varepsilon_{1*} m_e c^2 &= 
\frac{127\;{\rm MeV}}{(1+z)}\left[C_2\,(40.2)^{\alpha-2}
\left(\frac{\alpha}{2}\right)^{5/3}\frac{(\Gamma_{0,2})^{2\alpha}
R_{0,13}}{L_{0,52}}\right]^{1/(\alpha-1)}\ ,
\end{align} 
where $C_\alpha = 100C_2$ is a coefficient whose value is determined
numerically. The dependence of $\varepsilon_{1*}$ on $\Gamma_0$,
$R_0$, and $\alpha$ is demonstrated in Fig.~\ref{fig:eps1}.  Since a
test photon of dimensionless energy $\varepsilon_{1*}$ pair produces
primarily with photons of energy $\sim
\Gamma^2/(1+z)^2\varepsilon_{1*}$, and we fix the value of $L_0$
[i.e. the photon number density near $(1+z)\varepsilon = 1$], the
values of $\varepsilon_{1*}$ becomes almost independent of $\alpha$
near $\Gamma_0 \sim \sqrt{(1+z)\varepsilon_{1*}}$ (which is roughly
where the lines for the three values of $\alpha$ for the same value of
$\varepsilon_{1*}$ almost meet).

Eq.~(\ref{eps_1s}) can be inverted in order to obtain
\begin{equation}\label{physpar1}
(\Gamma_{0,2})^{2\alpha}R_{0,13} = 
C_2^{-1}40.2^{2-\alpha}(\alpha/2)^{-5/3}L_{0,52}
\left[\frac{(1+z)\varepsilon_{1*} m_e c^2}{127\;{\rm MeV}}\right]^{\alpha-1}\ .
\end{equation}
If one makes the additional assumption that $R_0 \sim \Gamma_0^2
c\Delta T/(1+z)$, which is valid for a large class of models, then
Eq.~(\ref{physpar1}) provides the following estimate for $\Gamma_0$:
\begin{equation}\label{Gamma_0gg}
\Gamma_0 \approx 100\,(1+z)^{\alpha/(2\alpha+2)}
\left[\frac{1.34}{C_2}\left(\frac{\alpha}{2}\right)^{-5/3}
\frac{L_{0,52}}{\Delta T}\right]^{1/(2\alpha+2)}
\left(\frac{\varepsilon_{1*} m_e c^2}
{5.1\;{\rm GeV}}\right)^{(\alpha-1)/(2\alpha+2)}\ .
\end{equation}
For GRBs, one may perform a consistency check for the assumption that
$R_0 \sim \Gamma_0^2 c\Delta T/(1+z)$ by comparing the value of
$\Gamma_0$ under this assumption from opacity to pair production
(Eq.~[\ref{Gamma_0gg}]) to the estimate for $\Gamma_0$ from the time,
$T_{\rm dec}$, of the onset of the afterglow emission,\footnote{This
  estimate is for the Lorentz factor of the outflow after the passage
  of the reverse shock, so it is close to that of the original
  outflow before it was decelerated by the reverse shock only as long
  as the reverse shock is at most mildly relativistic. For a highly
  relativistic reverse shock, the original Lorentz factor of the
  outflow can be much larger than this value.}
\begin{equation}
\Gamma_0(T_{\rm dec}) \approx \left[\frac{(3-k)E_{\rm iso}}{\pi A
    (2c)^{5-k}T_{\rm dec}^{3-k}}\right]^{1/2(4-k)}
= \left\{\begin{matrix}128\,E_{\rm iso,53}^{1/8}n_0^{-1/8}T_{\rm
    dec,2}^{-3/8} &\quad (k=0)\ , \cr\cr
131\,E_{\rm iso,53}^{1/4}A_\star^{-1/4}T_{\rm dec,0}^{-1/4} &\quad (k=2)\ ,
\end{matrix}\right.
\end{equation}
where $T_{\rm dec} = T_{\rm dec,0}\;{\rm s} = 100T_{\rm dec,2}\;{\rm
s}$, $E_{\rm iso} = 10^{53}E_{\rm iso,53}\;$erg is the isotropic equivalent
kinetic energy in the outflow, $\rho_{\rm ext} = A R^{-k}$ is the
external density, and is assumed to be a power law with radius,
which is $\rho_{\rm ext} = n m_p$ for a uniform external medium
($k=0$) of number density $n = n_0\;{\rm cm^{-3}}$, while $A = 5\times
10^{11}A_\star\;{\rm g\;cm^{-1}}$ for a stellar wind environment
($k=2$).

\acknowledgements We thank M. Peskin, M.~G. Baring, P. Coppi, and
L. Stawarz for useful discussions. 
This work is supported by the U.S. Department of Energy under 
contract number DE-AC02-76SF00515.
J.G. gratefully acknowledges a Royal Society Wolfson Research Merit Award.
 
\appendix

\section{Changes of variables in section \ref{sec:red}}\label{ap:varchanges}
\subsection{Change of variable from $(s,\mu_r)$ to $(R_e,R_t)$}
Since we integrate over $d\Omega_r = d\phi_r d\mu_r$ and the integrand
contains $|d\mu_e/d\mu_r|$, we can conveniently change variables from $\mu_r$
to $\tilde{R}_e$. It is straightforward to verify that, 
irrespective of the position of $R_t$ with 
respect to the upper bound of $R_e$, 
$\left|d\mu_e/d\mu_r\right|\;d\mu_r = 
(d\mu_e/d\tilde{R}_e)\;d\tilde{R}_e$, 
 with the integration over $\tre$ being performed from the smaller
 to the larger bound, and $d\mu_e/d\tilde{R}_e>0$ always. 
Furthermore, the perpendicular distance from the line of sight from
the center of the emitting sphere to the observer at infinity,
\begin{equation}\label{eq:rperp}
R_\perp \equiv R_{t,0}\sin\theta_{t,0} = R_t\sin\theta_t\ ,
\end{equation}
is constant along the trajectory of the test photon 
(see Fig.~\ref{fig:geom2}). Thus
\begin{equation}\label{ds_to_dR_t}
s = R_t\cos\theta_t - R_{t,0}\cos\theta_{t,0}\quad,\quad
ds = -\frac{R_\perp d\theta_t}{\sin^2\theta_t}
= \frac{R_\perp d\mu_t}{\left(1-\mu_t^2\right)^{3/2}}
= \frac{R_t dR_t}{\sqrt{R_t^2-R_\perp^2}} \approx R_{t,0}d\hat{R}_t\ ,
\end{equation}
where the last approximation holds since $R_\perp \leq R_{\perp,{\rm
max}} \approx R(R_{\perp,{\rm max}})/\Gamma(R_{\perp,{\rm max}}) =
{\cal O}(R_t/\Gamma) \ll R_t$ and the contribution from $R_t \ll
R(R_{\perp,{\rm max}})$ is negligible for $\Gamma \gg 1$.

\subsection{Integration over $d\phi_r$ and $d\ei$}\label{sec:integration}

The local photon field derived above is symmetric around the radial
direction (i.e. does not depend on $\phi_r$). As a consequence,
$\phi_r$ appears only in the function $\sigma^\star
\left[\chi(\varepsilon_t,\ei,\mu_{ti})\right](1-\mu_{ti})$, where
$\mu_{ti}$ is a function of $\cos{\phi_r}$ (see Eq.~[\ref{mu_ti}]).
Thus, we can write $\int_0^{2\pi}d\phi_r = 2\int_0^\pi d\phi_r$. Next,
we follow the insights of \citep{Stepney83} and \citep{Baring94} by
performing the change of variables $(\ei,\phi_r)\to(\chi,u)$, with
$\chi^2 = \varepsilon_t\ei u$ defined in Eq.~(\ref{chi}).  Defining
$\zeta_+ = \left[1-\cos{(\theta_r+\theta_t)}\right]/2$ and $\zeta_- =
\left[1-\cos{(\theta_r-\theta_t)}\right]/2$, where $\zeta_+ > \zeta_-$
for $(\theta_t,\theta_r)\in[0,\pi]$, Eq.~(\ref{mu_ti}) yields
$\int_0^{\pi} d\phi_r =2\int_{\zeta_-}^{\zeta_+} du
\left[(\zeta_+-\zeta_-)\sin\phi_r\right]^{-1} =
2\int_{\zeta_-}^{\zeta_+} du
\left[(\zeta_+-u)(u-\zeta_-)\right]^{-1/2}$.  Likewise
$\int_{2/\varepsilon_t}^{\infty}d\ei = 2\int_1^{\infty}\chi
d\chi(\varepsilon_t u)^{-1}$.  Eq.~(\ref{tau_gg_2bis}) now reads:
\begin{align}\nonumber
\tgg(\varepsilon_t,\theta_{t,0},R_{t,0}) &=
\frac{8\sigma_T}{(4\pi)^2 m_e c^3 R_{t,0}} \int_1^\infty
\frac{d\hat{R}_t}{\hat{R}_t^2} \int d\tre
\frac{\delta^3}{\tilde{r}^2}\cdot\frac{d\mu_e}{d\tre} \;
\\ \label{tau_gg_3bis}
&\quad\quad\quad\quad\quad\quad\quad
\times\int_{\zeta_-}^{\zeta_+} \frac{u
du}{\sqrt{(\zeta_+-u)(u-\zeta_-)}} \int_1^{\infty} 
\frac{d\chi}{\chi}\frac{\sigma^\star(\chi)}{\sigma_T}
L'_{\chi^2/\varepsilon_t\zeta\delta}(\tre)\ .\quad\quad
\end{align}

\subsection{$\displaystyle{L'_{\varepsilon'_i}(R_e) = 
L'_0 (\varepsilon'_i)^{1-\alpha}\times h(R_e/R_0)}$.}
\label{local_luminosity}

The specific luminosity in the co-moving frame is conveniently
parameterized as $L'_{\varepsilon'} =
L'_0(\varepsilon')^{1-\alpha}h(R_e/R_0)$ where $h(1) = 1$ is
normalized at $R_e/R_0 = \tre\hrt/\hrz = 1$. Similarly, we want to
parameterize the specific luminosity in the lab frame at $R_0$ as
$L_\varepsilon(R_0) \approx L_0\varepsilon^{1-\alpha}$, even though
the luminosity at a given radius is not really well defined, since the
Doppler factor also depends on the angle $\theta_{t,0}$ from the line
of sight. The normalization coefficients in the lab frame ($L_0$) and
in the co-moving frame ($L_0'$), are the specific luminosity at $R_0$
corresponding to a photon energy of $m_e c^2 \approx 511\;$keV in the
respective frames.  Since the typical value of the Doppler factor is
$\delta = \varepsilon/\varepsilon' \sim \Gamma$, and specifically
$\delta(R_0) \sim \Gamma(R_0) \equiv \Gamma_0$, these coefficients are
related by $L_0\varepsilon^{1-\alpha} \sim \Gamma_0
L'_0(\varepsilon')^{1-\alpha}$ and $L_0 \sim
\Gamma_0^{\alpha}L'_0$. Thus motivated, we use this relation as the
definition of $L_0$, $L_0 \equiv \Gamma_0^{\alpha}L'_0$.  Therefore,
$L'_{\varepsilon'_i}(R_e) = \Gamma_0^{-\alpha} L_0 (\ei')^{1-\alpha}
\times h(\tre\hrt/\hrz)$. It is convenient to express the optical
depth $\tgg$ in terms of $L_0$ which is approximately the observed
isotropic equivalent luminosity at an observed photon energy of
$511(1+z)^{-1}\;$keV near $\bar{T} \sim 1$ for $\Delta R \gtrsim
R_0$. For $\Delta R/R_0 \approx \bar{T}_f/(1+m) \ll 1$ (see
Eq.~[\ref{T_bar_f}]), the peak isotropic equivalent luminosity at
$511(1+z)^{-1}\;$keV is $\sim \bar{T}_f L_0$ and the corresponding
optical depth near the peak of the spike in the light curve at the
same photon energy is $\sim \bar{T}_f \tau_0$. Therefore, $L_0$ is
practically an observable quantity, making it convenient to work with.

Eq.~(\ref{tau_gg_3bis}) now becomes:
\begin{align}\nonumber 
\tgg(\varepsilon_t,\theta_{t,0},R_{t,0}) &= 
\frac{2\Gamma_0^{-\alpha}L_0\varepsilon_t^{\alpha-1} \sigma_T}
{(4\pi)^2 m_e c^3 R_{t,0}} \int_1^\infty \frac{d\hrt}{\hrt^2} 
\int d\tre \left(\frac{\delta^{2+\alpha}}{\tilde{r}^2}
\cdot\frac{d\mu_e}{d\tre}\right) h\left(\tre\frac{\hrt}{\hrz}\right)
\\ \label{tau_gg_4}
&\quad\times\int_{\zeta_-}^{\zeta_+} \frac{u^{\alpha} du}
{\sqrt{(\zeta_+-u)(u-\zeta_-)}} \times
 \frac{4}{\sigma_T}\int_1^{+\infty} d\chi \chi^{-(2\alpha-1)}
\sigma^\star(\chi) \ , 
\end{align}
From Eq.~(8) of \citep{Baring94} we can write, to a very good
approximation:
\begin{equation}
 \frac{4}{\sigma_T}\int_1^{+\infty} d\chi \chi^{-(2\alpha-1)}\sigma^\star(\chi) \sim \frac{7}{6\alpha^{5/3}}
\end{equation}
Making the change of variable $u\to t =
(u-\zeta_-)/(\zeta_+-\zeta_-)$, we also have \citep[see Eq.~15.3.1 in
][]{AbraSteg}:
\begin{equation}
 \int_{\zeta_-}^{\zeta_+}
\frac{u^{\alpha} du}{\sqrt{(\zeta_+-u)(u-\zeta_-)}} 
= \zeta_-^{\alpha}\int_0^1\frac{(1+\zeta t)^{\alpha}}{\sqrt{t(1-t)}}dt 
\equiv \frac{\zeta_-^{\alpha}}{\pi}\; {}_2F_1(-\alpha,0.5;1;-\zeta)
\end{equation}
where ${}_2F_1$ is a hypergeometric function and $\zeta =
(\zeta_+-\zeta_-)/\zeta_- > 0$. We define $H_\alpha(\zeta) \equiv
{}_2F_1(-\alpha,0.5;1;-\zeta)$, and notice that it is regularized by
the factor $\zeta_-^{\alpha}$ when $\zeta_-\to 0$.  Note that $\zeta$
is of order unity and $H_\alpha(\zeta)$ is a simple polynomial when
$\alpha$ is an integer \citep[see Eq.~15.4.1 in][and our
eqs.~(\ref{H_alpha1})-(\ref{H_alpha3}) for $\alpha$=1,2, and
3.]{AbraSteg}.

\section{Justification for the Approximations in Case~1}\label{ap:case1}

When calculating the photon field at the instantaneous location of the
test photon on case 1, where the test photons lags behind the shell,
$R_t < R_{\rm sh}(t_t)$, we have used an approximation for the value
of $\theta_r$, namely Eq.~(\ref{mu_r2}), which is valid for $\theta_r
\ll 1$ and break down for $\theta_r \sim 1$ which corresponds to
$1-\tre = {\cal O}(\Gamma_t^{-2})$. This is despite the fact that in
this case $\theta_r$ can assume any value between zero and $\pi$. The
justification for this convenient approximation is that the
contributions to the optical depth $\tgg$ from
$\theta_r \sim 1$, where our approximation breaks, is negligible
compared to the contribution from $\theta_r \ll 1$, where our
approximation is valid. In order to show this more explicitly, we
examine the dependence of the integrand in the integration over
$d\tre$ on the value of $\theta_r$ in the range
\begin{equation}\label{case1_contrib2}
\frac{1}{\Gamma_t} \ll \theta_r \ll 1 \quad\Longleftrightarrow\quad
\frac{1}{\Gamma_t^2} \ll \tilde{R}_{\rm sh} - \tre \approx 1-\tre \ll 1 \ ,
\end{equation}
which gives us a handle (up to factors of order unity) on its
dependence throughout the entire range of possible $\theta_r$
values. In this intermediate range of
$\theta_r$ values, $\tilde{R}_{\rm sh} - \tre \approx 1-\tre$ since
for case 1 $\Gamma_t^2(\tilde{R}_{\rm sh}-1) \lesssim {\rm a\ few}$,
and thus
\begin{equation}
  \tilde{r}^2 = \left(\tilde{R}_{\rm sh}-\tilde{R}_e\right)^2
+\frac{\left(\tilde{R}_{\rm sh}-\tilde{R}_e\right)
\left(\tilde{R}_{\rm sh}^{m+1}-\tilde{R}_e^{m+1}\right)}
{(m+1)\Gamma^2_t} + {\cal O}\left(\Gamma_t^{-4}\right) \approx
(\tilde{R}_{\rm sh}-\tre)^2 \approx (1-\tre)^2 \ .
\end{equation}
Likewise, Eq.~(\ref{dmu_e_dR}) yields 
\begin{align}
\frac{d\mu_e}{d\tilde{R}_e} &\approx \frac{1}{2\Gamma^2_t\tilde{R}_e^2}
\left[ \Gamma^2_t\left(\tilde{R}_{\rm sh}^2-1\right)
+\frac{1}{m+1}\left(1 + m\tilde{R}_e^{m+1}\right) - \tilde{R}_e^{m+2} \right]\\
&\approx \frac{f_m-\tre^{m+1}}{2(m+1)\Gamma_t^2\tre^2}
\approx \frac{f_m-1}{2(m+1)\Gamma_t^2}\ ,
\end{align}
Thus we see that in this intermediate range of $\theta_r$ values
$d\mu_e/d\tilde{R}_e$ is approximately constant, and is of order
$\Gamma_t^{-2}$. Besides, $H_\alpha$ is of order unity, $\zeta_-
\sim \theta_r^2$, and $\delta \sim \Gamma^{-1}\theta_r^{-2}$, so that
\begin{equation}\label{scaling_contrib2}
\frac{\delta^{\alpha+2}}{\tilde{r}^2}\,\frac{d\mu_e}{d\tre}\,
\zeta_-^{\alpha}H_\alpha(\zeta) \sim
\frac{1}{\Gamma_t^{4+\alpha}(1-\tre)^2\theta_r^4} 
\propto 
\frac{1}{(1-\tre)^2\theta_r^4}\ .
\end{equation}
\noindent Moreover, Eq.~(\ref{mu_r2}), which is valid for case 1 in the
limit $\theta_r \ll 1$ that applies in our intermediate regime,
implies that $(1-\tre)\theta_r^2$ is approximately constant in this
range of $\theta_r$ values given by Eq.~(\ref{case1_contrib2}).
Therefore, from Eq.~(\ref{scaling_contrib2}) we conclude that the
integrand in the integration over $d\tre$ is approximately constant
over this range in $\tre$, which is of interest here. Furthermore, the
integrand must still have a similar value, up to a factor of order
unity, even for $\theta_r \sim 1$, since the approximation of
$\theta_r \ll 1$ breaks only marginally, rather than very severely
(since $\theta_r$ cannot have values $\gg 1$). As the region where our
approximation breaks, $\theta_r \sim 1$, corresponds to $1-\tre =
{\cal O}(\Gamma_t^{-2}$), i.e. a range of the order of $\Gamma_t^{-2}$
in $\tre$, which is much smaller than the range over which our
approximation is valid, and is also much smaller than the range in
Eq.~(\ref{case1_contrib2}), we conclude that the contribution to the
integral from $\theta_r \sim 1$ can safely be neglected.

\section{Properties of the Photon Field in Case~3}

By changing the integration variable from $\mu_r$ to $\tre$ we
eliminated the need to express $\tilde{R}_e$ as a function of $\mu_r$,
and to calculate the minimal value $\nu_r$ which corresponds to
$\theta_{r,{\rm max}}$.  Nevertheless, this is still interesting in
terms of the properties of the local photon field, so it is given in
this appendix. Each
value of $\mu_r$ may correspond to two different values of
$\tilde{R}_e$, one at the front and one at the back of the equal
arrival time surface of photons to the point
$(R_t,t_t)$. Eq.~(\ref{mu_r3}) can be re-written as
\begin{align}\nonumber
\tilde{R}_e^{m+2}-\left[\tilde{R}_{e,{\rm max}}^{m+1}
+ 2(m+1)\Gamma_t^2(1-\mu_r)\right]\tilde{R}_e +
2(m+1)\Gamma_t^2(1-\mu_r) &= \\
\tilde{R}_e^{m+2}
- 2(m+1)\Gamma_t^2\left(\frac{ct_t}{R_t}-\mu_r\right)\tilde{R}_e 
+ 2(m+1)\Gamma_t^2(1-\mu_r) &= 0\ .
\end{align}
For $m = 0$ this becomes a second order equation with the solutions
\begin{align}\nonumber
\tilde{R}_e &= \Gamma_t^2\left(\frac{ct_t}{R_t}-1\right)
\left[1\pm\sqrt{1-\frac{2\Gamma_t^2(1-\mu_r)}
{\Gamma_t^4\left(ct_t/R_t-1\right)}}\,\right]
\\
&= \frac{\tilde{R}_{e,{\rm max}}+\left(\Gamma_t\theta_r\right)^2}{2}
\left\{1\pm\sqrt{1-\frac{4\left(\Gamma_t\theta_r\right)^2}
{\left[\tilde{R}_{e,{\rm max}}
+\left(\Gamma_t\theta_r\right)^2\right]^2}}\,\right\}\ ,
\end{align}
where $\theta_{r,{\rm max}}$ may be obtained by the condition of a
single solution, i.e. that the expression in the square root vanishes.
This implies
\begin{equation}
\left(\Gamma_t\theta_{r,{\rm max}}\right)^2 = 
\left(2-\tilde{R}_{e,{\rm max}}\right)\left[1-\sqrt{1-\left(
\frac{\tilde{R}_{e,{\rm max}}}{2-\tilde{R}_{e,{\rm max}}}\right)^2}\right]\ ,
\end{equation}
where we chose the root of the equation which corresponds to the
familiar result of $\Gamma_t\theta_{r,{\rm max}} \approx
\tilde{R}_{e,{\rm max}}/2$ for $\tilde{R}_{e,{\rm max}} \ll 1$.
 
More generally, $\mu_{r,{\rm min}} = \cos\theta_{r,{\rm max}}$ may be
found by the condition that $d\mu_r/d\tilde{R}_e = 0$. Using
Eq.~(\ref{dmu_r3_dR}) this results in
\begin{equation}
(m+1)\tilde{R}_e^{m+2} - (m+2)\tilde{R}_e^{m+1} 
+ \tilde{R}_{e,{\rm max}}^{m+1} = 0\quad,\quad
\left(\Gamma_t\theta_{r,{\rm max}}\right)^2 =
\left[\tilde{R}_e(\theta_{r,{\rm max}})\right]^{m+2}\ .
\end{equation}
Alternatively, one can use the latter relation, which is obtained by
substituting $d\mu_r/d\tilde{R}_e = 0$ from Eq.~(\ref{dmu_r3_dR}) into
Eq.~(\ref{mu_r3}), to obtain an explicit equation for $\theta_{r,{\rm
max}}$,
\begin{equation}
(m+1)\left(\Gamma_t\theta_{r,{\rm max}}\right)^2
-(m+2)\left(\Gamma_t\theta_{r,{\rm max}}\right)^{2(m+1)/(m+2)}
+\tilde{R}_{e,{\rm max}}^{m+1} = 0 \ .
\end{equation}

\section{The Scaling of $\tau_{\gamma\gamma}$ with $\bar{T}$ }
\label{tau_Tbar}

It is instructive to explicitly derive the scaling of
$\tau_{\gamma\gamma} = \tau_0(\varepsilon_t,\hrt){\cal F}(x)$ with
$\bar{T}$, in the three regimes $1 \gg \bar{T} < \bar{T}_f$, $1 \ll
\bar{T} < \bar{T}_f$, and $\bar{T} \gg \bar{T}_f$. The only time
dependence of $\tau_0$ on $\bar{T}$ is through $\hrz =
y^{-1}(T/T_0)^{-1/(m+1)}$ (see Eq.~[\ref{tau_0}]), so that
$\tau_0\propto (1+\bar{T})^{[b-1+\alpha m/2]/(m+1)}$.

\subsection{$1\gg\bar{T}<\bar{T}_f$}

For $1\gg\bar{T}<\bar{T}_f$, $\tau_0$ is thus approximately constant
and the time dependence of $\tau_{\gamma\gamma}$ is dominated by the
the time dependence of ${\cal F}(x)$, which we now consider in more
detail. First, the maximal value of the emission angle $\theta_{t,0}$
and correspondingly of $x = (\gamma_{t,0}\theta_{t,0})^2$ along the
EATS-I is given by
\begin{equation}
T - T_0 = \frac{R_0\theta_{t,0}^2}{2c} = T_0(m+1)x_{\rm max} 
\quad \Longleftrightarrow \quad 
x_{\rm max}(\bar{T}) = \frac{\bar{T}}{(m+1)}\ .
\end{equation}
This result holds in general, and can be readily obtained by noticing
 the $x_{\rm max}$ always corresponds to $y_{\rm min} =
 (T/T_0)^{-1/(m+1)}$, and substituting the latter in
 Eq.~(\ref{Rhat_x}).  Therefore, $x \leq x_{\rm max} \ll 1$ for
 $\bar{T} \ll 1$ and we are always in case 3. Second, it is
 straightforward to show that
\begin{equation}
f_m(\hrt)\left(\frac{\hrz+\Delta\hat{R}}{\hrt}\right)^{-(m+1)} =
\left(\frac{1+\bar{T}}{1+\bar{T_f}}\right)\frac{1+x(m+1)(1-\hrt^{-1})}
{1+x(m+1)} < \frac{1+\bar{T}}{1+\bar{T_f}}\ ,
\end{equation}
so that for $\bar{T} \leq \bar{T}_f$ we always have 
\begin{equation}\label{re3}
\tilde{R}_{e,{\rm max}} = \tilde{R}_{e,3} = f_m(\hrt)^{1/(m+1)} = 
\hrt^{-1}\left[1+x(m+1)\left(1-\hrt^{-1}\right)\right]^{1/(m+1)}\ ,
\end{equation}
 and
\begin{equation}\label{Fxapp}
{\cal F}(x) = \int_1^\infty d\hrt
\int_{\hrz/\hrt}^{\tilde{R}_{e,3}} d\tre \ {\cal I}(\hrt,\tre,x)\ ,
\end{equation}
where ${\cal I}(\hrt,\tre,x)$ is given in Eq.~(\ref{I}), and
\begin{equation}
\tilde{R}_{e,{\rm min}} = \frac{\hrz}{\hrt} = 
\hrt^{-1}\left[\frac{1+(m+1)x}{1+\bar{T}}\right]^{1/(m+1)}\ .
\end{equation}
Keeping terms to first order in $\bar{T}$ (and $x$), the range of
$\tilde{R}_e$ value that is being integrated over in Eq.~(\ref{Fxapp})
is
\begin{equation}
\Delta \tilde{R}_e = \tilde{R}_{e,{\rm max}} - \tilde{R}_{e,{\rm min}} 
\approx \frac{\bar{T}}{(m+1)\hrt} - \frac{x}{\hrt^2} 
= {\cal O}(\bar{T}) \ll 1\ .
\end{equation}
The integrand includes in several places the expression
\begin{equation}
f_m(\hrt)-\tre^{m+1} = \tilde{R}_{e,{\rm max}}^{m+1}-\tre^{m+1}
\approx \frac{\tilde{R}_{e,{\rm max}}^{m}}{(m+1)}
\left(\tilde{R}_{e,{\rm max}}-\tre\right) \ll 1\ ,
\end{equation}
which is either comparable to or much smaller than $1-\tre$, which
also appears in the integrand, thus defining different regimes. The
relevant ratio to compare to unity is
\begin{equation}\label{ratio1}
\max\left(\frac{\tilde{R}_{e,{\rm max}}-\tre}{1-\tre}\right) = 
\frac{\Delta\tre}{1-\tilde{R}_{e,{\rm min}}} = 
\frac{1-\tilde{R}_{e,{\rm min}}}{1-\tilde{R}_{e,{\rm max}}}-1 \approx
\left(\hrt-1\right)^{-1}\left(\frac{\bar{T}}{m+1}-\frac{x}{\hrt}\right)\ ,
\end{equation}
which measures both the fractional change in $1-\tre$ and the minimal
value of its ratio to $\tilde{R}_{e,{\rm max}}-\tre$. 

For $x = x_{\rm max} = \bar{T}/(m+1)$ this ratio is $x_{\rm max}/\hrt
\leq \bar{T}/(m+1) \ll 1$ so that $1-\tre$ is both approximately
constant and much larger than $\tilde{R}_{e,{\rm max}}-\tre \sim
f_m(\hrt)-\tre^{m+1}$. Therefore, the only term that varies
significantly with $\tre$ in the inner integrand is
$\bar{\zeta}_-^{\,\alpha} H_\alpha(\zeta)$. For $\zeta \gg 1$, $H_\alpha(\zeta)
\sim \zeta^\alpha$ so that $\bar{\zeta}_-^{\,\alpha} H_\alpha(\zeta) \sim
(\bar{\zeta}_-\zeta)^\alpha \propto \bar{T}^{\alpha/2}(\tilde{R}_{e,{\rm
max}}-\tre)^{\alpha/2}$ where the integration over $(\tilde{R}_{e,{\rm
max}}-\tre)^{\alpha/2}$ results in a factor of $\bar{T}^{1+\alpha/2}$
so that altogether the inner integral is $\propto
\bar{T}^{\alpha+1}$. The outer integral is of the form $\int_1^\infty
d\hrt g(\hrt) = {\rm const}$. For $\zeta \lesssim 1$, $H_\alpha(\zeta) \sim 1$
and $\bar{\zeta}_-^{\,\alpha} H_\alpha(\zeta) \sim
\bar{\zeta}_-^{\,\alpha} \sim
(\Gamma_t\theta_r-\Gamma_t\theta_t)^{2\alpha}$ which consists of a sum
of terms of the form $\bar{T}^{a}(\tilde{R}_{e,{\rm
max}}-\tre)^{\alpha-a}$ that upon integration are $\propto
\bar{T}^{\alpha+1}$. Thus, 
\begin{equation}
{\cal F}(x_{\rm max}) \propto \bar{T}^{\alpha+1}\ . 
\end{equation}
Note that in this case most of the contribution
to the optical depth comes from $\hrt \lesssim 2$ or $\Delta \hrt \sim
1$. 

For $x = 0$, $\theta_t = \zeta = 0$ so that $H_\alpha(\zeta) =
1$. Furthermore, the ratio in Eq.~(\ref{ratio1}) becomes larger than
unity for $\hrt - 1 < \bar{T}/(m+1)$, and in this regime
$f_m(\hrt)-\tre^{m+1} \sim \tilde{R}_{e,{\rm max}}-\tre \sim 1 - \tre$
so that $\bar{\zeta}_-$ is roughly constant and the inner integrand
scales as $(1-\tre)^{-1}$, which upon integration scales linearly with
$\bar{T}$,
\begin{align}\nonumber
\int_1^{1+x_{\rm max}}d\hrt\,g(\hrt)\int_{(1-x_{\rm
    max})/\hrt}^{1/\hrt}\frac{d\tre}{(1-\tre)} &\approx
    g(1)\int_0^{x_{\rm max}}d(\hrt-1)\ln\left[\frac{(\hrt-1)+x_{\rm
    max}}{(\hrt-1)}\right]
\\
 &= g(1)(2\ln 2)x_{\rm max} \propto \bar{T}\ .
\end{align}
For $\hrt - 1 \gg x_{\rm max} = \bar{T}/(m+1)$, the approximation
discussed in the previous paragraph apply, and this part of the
integration over $\hrt$ does not contribute significantly to the total
optical depth, so that
\begin{equation}
{\cal F}(x = 0) \propto \bar{T}\ .
\end{equation}
Physically, the lack of significant contribution to the optical depth
from $\hrt - 1 \gg x_{\rm max} = \bar{T}/(m+1)$ may be understood
since the maximal value of $\theta_r$ (which corresponds to $R_e =
R_0$) starts to decrease significantly,
\begin{align}
\max[(\Gamma_0\theta_r)^2] &\approx \frac{\bar{T}}{(m+1)}\hrt^{-2}
\left[\hrt-1+\frac{\bar{T}}{(m+1)}\right]^{-1} 
\nonumber \\
&\approx
\left\{\begin{matrix}1 & \hrt-1 \ll \frac{\bar{T}}{(m+1)}\ , \cr\cr
\frac{\bar{T}}{(m+1)(\hrt-1)\hrt^2} \ll 1 & \hrt-1 \gg
\frac{\bar{T}}{(m+1)}\ ,
\end{matrix}\right.
\end{align}
which suppresses the opacity to pair production.

\subsection{$1 \ll \bar{T} < \bar{T}_f$}

For $1 \ll \bar{T} < \bar{T}_f$, we have $R_{t,0} \gg R_0$, so that
$\hrz/\hrt\ll 1$ and may effectively be taken as zero. Furthermore,
$R_{e,{\rm max}} < R_L(\bar{T}) < R_L(\bar{T}_f) = R_0 + \Delta R$
(since $\bar{T} < \bar{T}_f$) so that $\tilde{R}_{e,2} = 1$ and
$\tilde{R}_{e,3} = f_m(x,\hrt)^{1/(m+1)}$ is given by
Eq.~(\ref{re3}), and Eq.~(\ref{2int}) now reads
\begin{equation}
{\cal F}(x) = \int_1^{\hat{R}_{2}(x)} d\hrt
\int_0^1 d\tre \ {\cal I}(\hrt,\tre) 
 \; + \; \int_{\hat{R}_{2}(x)}^\infty d\hrt
\int_0^{f_m(x,\hrt)^{1/(m+1)}} d\tre \ {\cal I}(\hrt,\tre)\ . 
\end{equation}
In this regime neither the boundaries of integration nor the
integrand, ${\cal I}$, depend on $\bar{T}$. As a consequence, the
dependence of $\tau_{\gamma\gamma}$ in this regime is only through
$\tau_0$, and we have
\begin{equation}
\tgg(1 \ll \bar{T} < \bar{T_f}) \approx \tau_0(\bar{T}){\cal F}(x)
\propto \bar{T}^{[b-1+\alpha m/2]/(m+1)}\ .
\end{equation}

\section{On the definition of the optical depth}\label{factor_onehalf}

We start with the explicitly Lorentz invariant expression for the
differential interaction rate of two particles, denoted '1' and '2',
colliding with respective momenta $\vec{p}_1$ and $\vec{p}_2$, as
given in Eq.~(24a) of \citet{Weaver76} :
\begin{equation}
 R_{12}(\vec{p}_1,\vec{p}_2) \equiv 
\frac{n_1(\vec{p}_1)n_2(\vec{p}_2)(1-\vec{\beta}_1\cdot\vec{\beta}_2)
\left[(p_1\cdot p_2)^2 - m_1^2m_2^2c^4\right]^{1/2}}
{p_1\cdot p_2}c\sigma\ ,
\label{eq:rate}
\end{equation}
where $p_1$, $p_2$ are the four-momenta of particles '1' and '2',
respectively, $m_1$ and $m_2$ are their masses, $n_1(\vec{p}_1)$ and
$n_2(\vec{p}_2)$ their phase-space density and $\sigma$ is the
generalized Lorentz-invariant cross-section, usually computed in the
center of momentum frame. In Eq.~(\ref{eq:rate}), we have explicitly
written the dependence of $R_{12}$ on the momenta, which is missing in
\citet{Weaver76}, in order to distinguish it with the total
interaction rate $<R_{12}>$. The latter results from an integration
over the phase spaces of both particles (see eqs.~(2),(27) in
\citep{Weaver76}) :
\begin{equation}
 <R_{12}> = \frac{1}{1+\delta_{12}}
\int\int R_{12}(\vec{p}_1,\vec{p}_2) d^3\vec{p}_1 d^3\vec{p}_2 \ .
\label{eq:totalrate}
\end{equation}
In Eq.~(\ref{eq:totalrate}), the Kronecker symbol $\delta_{12}$ is 1
if the two particles are identical and 0 otherwise.
It accounts for the fact that, for identical particles, 
the double intergration counts twice each pair of interacting particles.

Now, we define $R_{12}(\vec{p}_1)$ as the interaction rate of a {\em
given} particle '1' of momentum $p_1$.  It writes $R_{12}(\vec{p}_1) =
\int R_{12}(\vec{p}_1,\vec{p}_2) d^3\vec{p}_2$, without a Kronecker
symbol because there cannot be any double counting when there is no
double integration. Specializing now to $\gamma\gamma$-interactions,
the interaction rate $R_{\gamma\gamma}(\vec{p}_1)$ is equal to the
decrease in $n_1$ per unit time :
$dn_1(\vec{p}_1)/dt=-R_{\gamma\gamma}(\vec{p}_1)$.  Defining the
optical depth of a particle of type 1 and momentum $\vec{p}_1$ as the
corresponding attenuation per unit length :
$dn_1/n_1\equiv-\tau(\vec{p}_1)ds$, where $ds$ is an element of
trajectory of particle 1, we obtain :
\begin{equation}
 \tau_{\gamma\gamma}(\vec{p}_1) \equiv R_{\gamma\gamma}(\vec{p_1})/c n_1 
= \int n_2(\vec{p}_2)(1-\vec{\beta}_1\cdot\vec{\beta}_2)
\sigma d^3\vec{p}_2 \ ,
\label{eq:tau}
\end{equation}
where in the last equality we made use of $m_1=m_2=0$ in
Eq.~(\ref{eq:rate}).  We thus re-derived Eq.~(\ref{dtau_gg}) (in
integral form), and showed that there is no factor 1/2 involved
because the computation of the optical depth does not warrant a double
integration over the phase space of both particles. Because they
compute the total reaction rates and not the optical depth,
\citet{Weaver76} and \citet{Stepney83} do have this factor.

Another source of confusion arises from the fact that in their seminal
paper, \citet{Gould67} specialized to an isotropic distribution for
particles 2, which brings up a factor 1/2 due only to the
normalization of the integration over $\cos\theta$. In other words,
introducing $dn\equiv n_2(\vec{p}_2)d^3\vec{p}_2 =
(1/2)n(\epsilon)d\epsilon\sin\theta d\theta$ in Eq.~(\ref{eq:tau})
immediately yields their Eq.~(7).

\begin{table}[ht]
\centering
\begin{tabular}{|l|l|l|}
\hline
notation & definition & Eq./\S\\
\hline
$\varepsilon \equiv E_{\rm ph}/m_ec^2$ &
observed photon energy normalized by the electron rest energy&\S\;\ref{sec:model}\\
$t,R,\theta$ & spherical coordinates (time, radius from the source, polar angle)& ---\\
$R_0,\Delta R$ & onset radius and range of the emission episode & \S\;\ref{sec:model}\\
$R_{\rm sh}(t),\Gamma,\Gamma_0\equiv \Gamma(R_0)$ 
& radius and bulk Lorentz factor of the emitting shell& (\ref{t_beta})\\
$m \equiv 2\,\frac{d\log\Gamma}{d\log R}$ & 
power law index of $\Gamma^2$ with radius $R$ &
\S\;\ref{sec:model}\\
$L_0 \equiv \Gamma_0^\alpha L'_0$ & roughly: observed isotropic
equivalent luminosity at $R_0$ \& $\varepsilon = 1$ & (\ref{tau_gen})\\
$\alpha \equiv -\frac{d\log N_{\rm ph}}{d\log E_{\rm ph}}$ & 
photon index at large photon energies & \S\;\ref{sec:model}\\
$b \equiv \frac{d\log L'_{\varepsilon'}}{d\log R}$ & 
power law index of co-moving spectral luminosity with radius & 
\S\;\ref{sec:model}\\
$t_t, R_t, \theta_t, \Gamma_t \equiv \Gamma(R_t)$ 
& test photon spherical coordinates and shell Lorentz factor at $R_t$& \S\;\ref{sec:tau}\\
$t_0,R_{t,0},\theta_{t,0},\Gamma_{t,0}$ 
& initial test photon spherical coordinates and Lorentz factor & (\ref{T}),(\ref{theta})\\
$R_e,\theta_e$ & emission radius and polar angle of 
interacting photon&\S\;\ref{sec:tau}\\
$T_0, T$ & arrival times of first and subsequent photons at the observer & (\ref{R_L}),(\ref{T})\\
$R_L(T), R_{e, {\rm max}}$ & maximal radius of emission along 
the EATS-I and EATS-II& (\ref{R_L})\\
$\varepsilon_1$ & dimensionless photon energy at which 
$\tau_{\gamma\gamma}(\varepsilon_1) = 1$ & \S\;\ref{sec:analytic}\\
$s$ & path length along the test photon trajectory & (\ref{dtau_gg})\\
$R_\perp$ & 
distance of test photon from the line of sight to the origin & 
(\ref{eq:rperp})\\
$\theta_{ti}$ & angle between directions of test
photon and interacting photon & (\ref{chi})\\
$\varepsilon_t$, $\varepsilon_i$ & dimentionless test/interacting
photon energies in the lab frame
& (\ref{dtau_gg})\\
$\mu \equiv \cos\theta$ & cosine of angle $\theta$ & ---\\
$\chi \equiv \sqrt{\frac{\varepsilon_t\ei(1-\mu_{ti})}{2}}$ & 
dimentionless photon energy in the center of momentum frame & 
(\ref{chi})\\
$\zeta \equiv (1-\mu_{ti})/2$ & convenient integration variable & 
\S\;\ref{sec:integration}\\
$r$ & interacting photon emission to 
test photon intersection distance & \S\;\ref{sec:tau}\\
$\theta_r$ & angle of an interacting photon relative to the radial direction & 
\S\;\ref{sec:tau}\\
$\delta \equiv (1+z)\varepsilon/\varepsilon'$ & Doppler factor between
the co-moving and lab frames & (\ref{doppler})\\
$f_m$ & useful quantity & (\ref{f_m})\\
$\tau_*$ & typical optical depth at $\varepsilon = 1$ on a
dynamical time ($\bar{T}_f > \bar{T} \sim 1$) & (\ref{tau_star})\\
$\tau_0,{\cal F}(x)$ & explicit analytic and integral parts of the optical depth & 
(\ref{tau_0})\\
$x \equiv (\Gamma_{t,0}\theta_{t,0})^2$ & 
rescaled emission angle squared & (\ref{Rhat_x})\\
$y \equiv R_{t,0}/R_L(T)$ & emission radius rescaled to the maximum radius 
on an EATS-I & (\ref{F1})\\
$\hat{R} \equiv R/R_{t,0}$ & radius rescaled to a given test photon emission
radius & \S\;\ref{sec:red}\\
$\tilde{R} \equiv R/R_t$ & radius rescaled to the instantaneous test photon
radius & \S\;\ref{sec:red}\\
$\bar{T} \equiv T/T_0-1$ & arrival time of photons rescaled to the earliest 
arrival time & \S\;\ref{sec:analytic}\\
$\bar{\delta} \equiv \delta/\Gamma_t,\bar{\mu}_e \equiv \Gamma^2_t\mu_e$ & rescaled Lorentz 
factor and cosine of the emission angle& 
(\ref{rescaled1})\\
$\bar{\zeta}_- \equiv \Gamma^2_t\zeta_-$ & rescaled integration variable &
(\ref{rescaled2})\\
$Y \equiv \frac{y-y_{\rm min}}{y_{\rm max}-y_{\rm min}}, Y_*$ & 
rescaled variable $y$, and  value at which ${\cal F}$ changes its behavior& (\ref{Y_Ystar})\\ 
\hline
\end{tabular}
\caption{Notation and definition of some quantities used throughout
  this work.}
\label{tab:basic}
\end{table}

\begin{table}[ht]
\centering
\begin{tabular}{|c|c|c|c|c|}
\hline
$\alpha$ & $m$ & $b$ & $\log_{10}(\Delta{R}/R_0)$ & Figures\\
\hline
2&0&0&$-2,-1,0,1,2$&\ref{fig:6panel_DR}, \ref{fig:int_spectra}\\
2&0&1&$-2,-1,0,1,2$&\ref{fig:int_spectra}\\
2&3&-2&$-2,-1,0,1,2,4$&\ref{fig:int_spectra}\\
2&$0,1,2$&0&2&\ref{fig:6panel_m}\\
2&0&$-2,-1,0$&2&\ref{fig:6panel_b}\\
$2,3,4$&0&0&0&\ref{fig:6panel_alpha} \\
\hline
\end{tabular}
\caption{The different sets of parameters for which results are shown
in this work.}
\label{tab:params}
\end{table}

\end{document}